\documentclass[review]{elsarticle}

\usepackage{graphicx}
\usepackage{amssymb}

\usepackage[margin=2.5cm]{geometry}
\usepackage{graphicx}
\usepackage{dcolumn}
\usepackage{bm}
\usepackage{braket}
\usepackage{graphicx}
\usepackage{epstopdf}
\usepackage{mwe}    
\usepackage{subfig}
\usepackage[T1]{fontenc}
\usepackage{caption}
\usepackage{amssymb}
\usepackage{amsmath}
\usepackage{booktabs}
\usepackage{tabularx}
\usepackage{multirow}
\usepackage{floatrow}
\usepackage{xcolor}
\usepackage[font=normalsize]{caption}

\newcommand{\tensor}[1]{\bm{\mathsf{#1}}} 

\journal{Journal Name}

\begin{document}

\begin{frontmatter}


\title{Central Moment Lattice Boltzmann Method on a Rectangular Lattice}


\author{Eman  Yahia}
\ead{eman.yahia@ucdenver.edu}
\author{Kannan N. Premnath}
\ead{kannan.premnath@ucdenver.edu}

\address{\normalsize Department of Mechanical Engineering\\ College of Engineering, Design and Computing\\ University of Colorado Denver\\ 1200 Larimer street, Denver, Colorado 80217 , U.S.A}



\begin{abstract}
Simulating inhomogeneous flows with different characteristic scales in different coordinate directions using the collide-and-stream based lattice Boltzmann methods (LBM) can be accomplished efficiently using rectangular lattice grids. We develop and investigate a new rectangular central moment LBM based on non-orthogonal moment basis and involving multiple relaxation times. The equilibria to which the central moments relax under collision in this approach are obtained from matching with those corresponding to the continuous Maxwell distribution. A Chapman-Enskog analysis is performed to derive the correction terms to the second order moment equilibria involving the grid aspect ratio and velocity gradients that restores the isotropy of the viscous stress tensor and eliminates the non-Galilean invariant cubic velocity terms of the resulting hydrodynamical equations. A special case of this rectangular formulation involving the raw moments is also constructed. The resulting schemes represent a considerable simplification, especially for the transformation matrices and isotropy corrections, and improvement over the existing lattice Boltzmann schemes based on raw moments on rectangular lattice grids that use orthogonal moment basis. Numerical validation study of both the proposed rectangular LBMs for a variety of benchmark flows are performed that show good accuracy at various grid aspect ratios. The ability of our proposed schemes to simulate flows at relatively lower grid aspect ratios and higher Reynolds numbers than considered in prior approaches is demonstrated. Furthermore, simulations reveal the superior stability characteristics of the rectangular central moment LBM over that based on raw moments in handling shear flows at lower viscosities and/or higher characteristic velocities. In addition, computational advantages of using our rectangular LB formulation in lieu of that based on the square lattice is shown.

\end{abstract}




\end{frontmatter}



\section{Introduction} \label{sec:1}

The lattice Boltzmann (LB) method~\cite{mcnamara1988use,higuera1989boltzmann,he1997theory} has been receiving a remarkable interest as a promising computational fluid dynamics (CFD) technique. It is a kinetic method that evolves the distribution functions due to the effects of collisions, which are often represented by relaxation model under certain symmetry, isotropy and conservation constraints (e.g.,~\cite{qian1992lattice,lallemand2000theory,d2002multiple,karlin1999perfect}) and due to their streaming along the particle characteristic directions. The features and applications of this mesoscopic computational technique have been discussed in various reviews~\cite{benzi1992lattice,chen1998lattice,yu2003viscous,aidun2010lattice,sharma2020current} and monographs~\cite{succi2001lattice,guo2002lattice,kruger2017lattice}. For simulating inhomogeneous fluid motion, such as those involving boundary layer flows with different scales for variations in different coordinate directions, or flows in domains where the spatial extent of one of the directions is considerably shorter than the others, i.e., characterized by geometric anisotropy such as in sheetlike porous media, it is highly desirable to use nonuniform grids for enabling efficient simulations. However, the symmetry constraints and the coupling of the particle velocity and coordinate space discretizations restrict the use of uniform grids, e.g., square lattice in two-dimensions (2D). To address this issue, broadly, the following two types of modifications to the LBM have been considered based on (a) the decoupling the discretizations of the velocity space and the spatial coordinate space and (b) the rectangular lattice grid to naturally accommodate the inhomogeneity in flows. In the first category, the resulting LBM does not maintain the lock-step advection during the streaming step, and either involves interpolations (e.g.,~\cite{he1996some}) or the use of traditional discretizations such as finite volume or finite element schemes, which leads to a more complicated approach with attendant additional overhead (e.g.,~\cite{xi1999finite, chen1998volumetric, peng1998lattice,li2004least}). The second category maintains the perfect-shift advection during the streaming step that incurs relatively low numerical dissipation, an important numerical advantage, and is the focus of this work. However, to recover the inherent isotropy of the viscous stress tensor in LB simulations using such rectangular lattices require making certain modifications to the algorithm. Thus, the prior LB schemes on rectangular lattice grids, which was inspired from an early work~\cite{koelman1991simple}, can be further classified according to the following modifications made: (i) designing the collision step with sufficient degrees of freedom and parametrization of the relaxation rates to maintain isotropy~\cite{bouzidi2001lattice,zhou2012mrt,zong2016designing}, (ii) extend the lattice with additional particle velocities~\cite{hegele2013rectangular} (iii) the use of extended moment equilibria to correct for isotropy~\cite{peng2016lattice,peng2016hydrodynamically}, and (iv) use of coordinate and velocity transformations and counteracting source terms~\cite{wang2019simulating}. Categorizing from a different consideration, such rectangular lattice-based LB algorithms use either the single relaxation time (SRT) collision model~\cite{koelman1991simple,hegele2013rectangular,peng2016lattice,wang2019simulating} or the multiple relaxation time (MRT) collision operator~\cite{bouzidi2001lattice,zhou2012mrt,zong2016designing,peng2016hydrodynamically}, involving the relaxation of the distribution functions or the raw moments, respectively.

Recognizing that the earlier SRT scheme on a rhombic lattice does not have the additional degrees of freedom~\cite{koelman1991simple}, Ref.~\cite{hegele2013rectangular} proposed another rectangular SRT-LBM using additional particle velocities whose equilibria, involving their weights and scaling factors, obtained via solving a quadrature problem, and validated for the Taylor-Green vortex flow using moderate grid aspect ratios (defined in the next section). Note that using additional particle velocities adds to the computational overhead and may complicate the implementation of the boundary conditions. A different rectangular SRT-LB scheme with extended equilibrium distribution functions was proposed in Ref.~\cite{peng2016lattice}, which was, however, found to be stable only if the grid aspect ratio is above $0.3$. Recently, Ref.~\cite{wang2019simulating} adopted a different approach by introducing artificial source terms obtained via a coordinate/velocity transformation, which was found to be severely limited to using grid aspect ratio is larger than $0.5$. Moreover, none of the above rectangular LB formulations are flexible enough to adjust the shear and bulk viscosities independently.

On the other hand, Ref.~\cite{bouzidi2001lattice} presented the first MRT-LB formulation on a two-dimensional nine velocities (D2Q9) rectangular lattice grid by introducing coupling between various relaxation parameters and the grid aspect ratio via a linear stability analysis. However, as shown later in Ref.~\cite{zong2016designing}, this scheme is not able to completely recover the isotropy of the macroscopic fluid flow equations. A different rectangular MRT-LB approach which maintains the transformation matrix independent of the grid aspect ratio~\cite{zhou2012mrt} was found to exhibit similar spurious behavior. More recently, via an inverse design analysis based on the Chapman-Enskog expansion~\cite{chapman1990mathematical}, Ref.~\cite{zong2016designing} introduced a rectangular MRT-LB method with an additional adjustable parameter that determines the relative orientation in the energy-normal stress subspace, which can be adjusted to completely eliminate the anisotropy. However, the resulting scheme appears to be quite complicated in specifying such an additional parameter as a function of the speed of sound and the grid aspect ratio, and with stable results achieved only for the grid aspect ratio above 0.2. Later, inspired by the lattice kinetic scheme~\cite{inamuro2002lattice}, Ref.~\cite{peng2016lattice} presented a consistent MRT-LBM on a rectangular grid in which the equilibrium moments are extended to include the stress components, which are designed in such a way as to restore the isotropy of the recovered hydrodynamical equations. However, the guidance for setting up the associated free parameters to recover the physically correct transport coefficients seems involved. While this scheme showed good agreement with benchmark results, results on its numerical stability at relatively low viscosities or large Reynolds numbers at different grid aspect ratios were not reported. Moreover, all the existing MRT-LB schemes on rectangular lattice grids involve raw moments, where the moment basis are orthogonalized via a Gram-Schmidt orthogonalization; however, it has recently been demonstrated that the orthogonalization can couple the evolution of the higher order moments to those of the lower moments thereby impacting their numerical stability characteristics~\cite{dubois2015lattice}. In the context of the rectangular lattice grid, the use of such an orthogonal moment basis also results in unwieldy expressions for the transformation matrices dependent on the lattice grid ratio, which compromises their implementation. Moreover, the existing rectangular LB schemes do not eliminate the cubic velocity errors arising from aliasing effects on the D2Q9 lattice.

A significant improvement over the standard MRT-LB methods based on raw moments is to consider performing relaxation of central moments to their equilibria under collision~\cite{geier2006cascaded}. Here, the central moments are obtained from the distribution functions based on the peculiar velocity and naturally preserves the Galilean invariance of all the moments independently supported by the lattice. The central moment equilibria are generally constructed via a matching principle based on the continuous Maxwell distribution function. As a result, when compared to the standard SRT-LB and MRT-LB schemes, whose equilibria generally involve fluid velocity terms truncated up to the second order, the central moment LB methods involve higher order fluid velocity terms, which support their enhanced stability characteristics. As discussed in Ref.~\cite{geier2015cumulant}, the method can be constructed using different moment basis, including those based on non-orthogonal moments. Recently, the central moment LB method has been further extended, improved and applied to variety of flowing systems (see e.g.,~\cite{asinari2008generalized,premnath2009incorporating,premnath2011three,ning2016numerical,dubois2015lattice,Rosis16,sharma2017new,elseid2018cascaded,hajabdollahi2018symmetrized,Hajabdollahi201897,chavez2018improving,fei2018modeling,HAJABDOLLAHI2018838,hajabdollahi2019cascaded,adam2019numerical,hajabdollahi2021central,adam2021cascaded}). Moreover, the numerical investigations in Refs.~\cite{ning2016numerical,dubois2015lattice,chavez2018improving,adam2021cascaded} demonstrated the superior stability characteristics of the central moment LB schemes. It should, however, be mentioned here that the central moment LBM has so far been developed only for square lattice grids in 2D and cubic lattice grids in 3D.

From the above, we can now summarize the main drawbacks of the existing LB schemes on rectangular lattice grids as follows. They are generally constructed using orthogonal moment basis, their raw moment equilibria contain terms only up to the second order in fluid velocity with several free parameters requiring cumbersome guidance involved for their specifications and with no corrections for the cubic velocity error terms due to aliasing effects, with attendant complicated expressions for the corrections terms eliminate the grid anisotropy and for the transformation matrices dependent on the grid aspect ratios. These features render such schemes with relatively narrow stability range and compromising their accuracy, computational efficiency, and implementation. All these limitations will be addressed in this work by constructing and investigating a new rectangular central moment LBM~\cite{yahiaAPSDFD2017,yahiaAPSDFD2018}. A non-orthogonal moment basis will be used in this regard. Moreover, we will also present a special case of this approach involving rectangular non-orthogonal raw moment MRT-LBM, which represents a simplification and improvement over other existing MRT-LB schemes on rectangular lattice grids. The rectangular non-orthogonal raw moment and central moment LB versions developed in this paper will be referred to as the RNR-LBM and RC-LBM, respectively. Consistency of our new rectangular LB formulation with the Navier-Stokes (NS) equations will be demonstrated via a Chapman-Enskog analysis and through which the correction terms to the second order moments involving the grid aspect ratio and velocity gradients that fully restore the isotropy of the hydrodynamical behavior will be identified. The use of non-orthogonal moment basis leads to a considerable simplification of such correction terms and associated transformation matrices, with a more efficient implementation along with robust numerical features as it avoids the spurious coupling of moments due to orthogonalization. The moment equilibria are constructed by matching with those obtained from the continuous Maxwellian and thereby involving higher order fluid velocity terms and without many free parameters.  Moreover, unlike other previous rectangular LB schemes, our approach also eliminates the non-Galilean invariant (GI) cubic velocity errors arising due to aliasing effects in the D2Q9 lattice. Numerical validation study of both the proposed rectangular LB schemes for a variety of benchmark flow problems will be performed to demonstrate their accuracy. Moreover, the superior numerical stability of the rectangular central moment LB formulation, i.e., RC-LBM, will be shown for handling a wide range of grid aspect ratios and at low viscosities or higher Reynolds numbers, and its computational effectiveness over that based on the square lattice will also be demonstrated. While the method is developed and studied here in 2D, it allows extension to three-dimensions.

This paper is organized as follows. In Section~\ref{sec:2}, we present a Chapman-Enskog analysis of the non-orthogonal moment LB formulation on a rectangular D2Q9 lattice grid, identify the correction terms that restore the isotropy and eliminate the non-GI cubic velocity terms and show consistency to the NS equations. A rectangular raw moment LB (RNR-LB) algorithm is constructed based on this analysis. Then, in Secs.~\ref{sec:3} and \ref{sec:4}, a rectangular raw moment LB (RNR-LB) and central moment LB (RC-LB) using the correction terms derived in Sec.~\ref{sec:2} will be developed. A numerical validation study of both RNR-LBM and RC-LBM for a variety of benchmark fluid flow problems are performed in Sec.~\ref{sec:5}. A comparative study involving numerical stability at different grid aspect ratios demonstrating the improvements with using the RC-LBM will be presented in Sec.~\ref{sec:6}. Finally, the main conclusions of this investigation are summarized in Sec.~\ref{sec:7}.

\section{Chapman-Enskog Analysis using Non-orthogonal Moment Basis on a Rectangular Lattice: Isotropy Corrections, Hydrodynamical Equations, and Local Expressions for the Strain Rate Tensor} \label{sec:2}
\subsection{Basis vectors, transformation matrix, moment equilibria and definition of corrections}
\begin{figure}[t]
\centering
\includegraphics[width=0.4\linewidth]{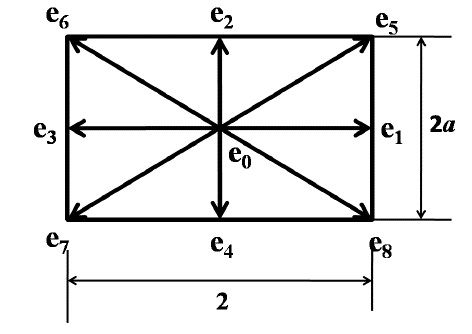}
\caption{Two dimensional-nine velocities rectangular lattice grid.}
\label{fig:1} 
\end{figure}

The two dimensional nine velocities lattice (D2Q9) representing the rectangular lattice grid considered in this study is shown in Figure \ref{fig:1}.
The rectangular lattice grid is parameterized by the \emph{grid aspect ratio}, $a$ defined as the ratio between the space steps in the $y$ and $x$ coordinate directions, $\Delta y$ and $\Delta x$, respectively, i.e., $a=\Delta y/ \Delta x$. The particle velocities $\bm{e}_i$, where $i={x,y}$, can be written as follows:
\begin{equation}\label{eq:1}
{\bm{e}_i} =
  \begin{cases}
     (0,0)      & \quad i =0\\
     ( \cos (i-1) \; \pi/2 , \; \; a \sin (i-1) \pi/2)c   & \quad  i =1-4 \\
     (\cos (2i-9) \;  \pi/4 , \;  a \sin (2i-9) \pi/4)c   & \quad  i =5-9,
  \end{cases}
\end{equation}
where $c$ is the lattice speed in the $x$ coordinate direction given by $c = \Delta x/\Delta t$ with $\Delta t$ being the time step. The Cartesian components of the particle velocities can then be listed in terms of the grid aspect ratio $a$ as
\begin{subequations}\label{eq:2}
\begin{align}
\ket{e_{x} }  = {\begin{bmatrix} \;0  &  1  &  0  &  -1  &   0  &  1  &  -1  &  -1  &   1 \; \end{bmatrix}}^\dagger, \\
 \ket{e_{y}}    = {\begin{bmatrix}\;0  &  0  &  a  &   0  &  -a  &  a  &   a  &  -a  &  -a \;  \end{bmatrix}}^\dagger.
\label{eq:2}
\end{align}
\end{subequations}
Here and in what follows, we use the 'ket' operator $ \ket{\cdot} $ notation to indicate a column vector of any variable defined for the lattice velocity set, while $\dagger$ refers to the transpose operation. The combination of the monomials of the form $\ket{e_x^m e_y^n}$, where $m$ and $n$ are integers, then defines the following set of natural or non-orthogonal basis vectors for the D2Q9 lattice:~\cite{geier2006cascaded,premnath2009incorporating}
\begin{equation} \label{eq:3}
\tensor{T}= \Big[\; \ket{1}, \ket{e_{x}}, \ket{e_{y}}, \ket{e_{x}^2+e_{y}^2},  \ket{e_{x}^2-e_{y}^2}, \ket{e_{x} e_{y}}, \ket{e_{x}^2 e_{y}}, \ket{e_{x} e_{y}^2}, \ket{e_{x}^2 e_{y}^2} \Big],
\end{equation}
where $\ket{1}$ is given by
\begin{equation}
\ket{1}={\begin{bmatrix} \;1  &  1  &  1  &  1  &   1  &  1  &  1  &  1  &   1 \; \end{bmatrix}}^\dagger.
\end{equation}
In Eq.~(\ref{eq:3}), the diagonal components of the basis vectors of the second order moments $\ket{e_{x}^2}$ and $\ket{e_{y}^2}$ are equivalently rearranged to isolate its trace or the isotropic part, which will be related to the bulk viscosity, from the other components, which are related to the shear viscosity so that both of these transport coefficients can be adjusted independently.

We note here that in the existing moment based LB formulations on rectangular lattice grids, an orthogonal moment basis is used to define the collision operator. As pointed out in Ref.~\cite{dubois2015lattice}, the orthogonalization introduces coupling of the higher order moments with the lower order moments under collision that can reduce the numerical stability range. Moreover, it can lead to cumbersome expressions for the transformation matrices and attendant isotropy corrections for the rectangular lattice. Hence, in this work, we employ the simpler non-orthogonal vector basis that plays a crucial rule in enhancing the numerical features of our approach. It may be noted such a basis was recently employed to develop a compact LB formulation for local vorticity computation scheme~\cite{hajabdollahi2020local}. Thus, the transformation matrix $\tensor{T}$ that maps the distribution functions from the velocity space to the moment space is established from Eq.~(\ref{eq:3}) for the rectangular lattice, which reads as
\begin{equation}\label{eq:4}
\tensor{T}=
  \begin{bmatrix}
   1 &  1 &  1 &  1 &  1 &  1 &  1  &  1  &  1\\
  0  &  1 &  0 &  -1 &  0 &  1 &  -1  &  -1  &  1\\
  0  &  0  & a  & 0  &  -a   & a & a & -a &   -a \\
  0  &  1  & a^2  & 1   &  a^2   & h_1 & h_1 & h_1 &   h_1 \\
  0  &  1  & -a^2  & 1   & -a^2   & h_2 & h_2 & h_2 &   h_2 \\
  0  &  0  & 0  & 0  &  0   & a  & -a & a &  -a \\
  0  &  0  & 0  & 0  &  0   & a  & a & -a &  -a \\
  0  &  0  & 0  & 0  &  0   & a  & -a^2 & -a^2 &  a^2 \\
  0  &  0  & 0  & 0  &  0   &  a^2 &  a^2 &  a^2 &  a^2\
\end{bmatrix},
\end{equation}
\\
where $h_1=1+a^2$, and $h_2=1-a^2$. Subsequently, the distribution functions in the velocity space $\mathbf{f}$, the equilibrium distribution functions
$\mathbf{f^{eq}}$, and the sources due to any applied body force $\mathbf{S}$ can be projected onto the moment space through the transformation matrix  \tensor{T} as
\begin{equation}
\mathbf{m}= \tensor{T}\;\mathbf{f}, \qquad \mathbf{m^{eq}}=  \tensor{T} \mathbf{f^{eq}}, \qquad \mathbf{\Phi}= \tensor{T}\;\mathbf{S}, \label{eq:tranformationrawmoments}
\end{equation}
where $\mathbf{f} ={\big( f_0, f_1, f_2, ...., f_8 \big)}^\dagger$, $\mathbf{f^{eq}} = {\big(f^{eq}_0, f^{eq}_1, f^{eq}_2, ...., f^{eq}_8 \big)}^\dagger$ and $\mathbf{\mathbf{S}} = {\big( S_0, S_1, S_2, ...., S_8 \big)}^\dagger$. The use of rectangular lattice introduces anisotropy in the viscous stress tensor given in terms of the grid aspect ratio $a$, which is related to the second order moment non-equilibrium moments and needs to be corrected for via appropriate counteracting correction terms. Since by definition, the second order non-equilibrium raw moments are identical to those of the central moments, for the purpose of performing a Chapman-Enskog (C-E) analysis and deriving the appropriate correction terms, it suffices to consider the simpler raw moment based lattice Boltzmann equation (LBE), i.e., MRT-LBE, which can be expressed as
\begin{equation}\label{eq:5}
 \mathbf{f} (\bm{x}+\mathbf{e}\Delta t, t+\Delta t)- \mathbf{f} (\bm{x},t) =  \tensor{T^{-1}}
 \Big[  \tensor{\hat{\Lambda}} \;  \left(\; \mathbf{m}^{eq}-\mathbf{m} \;\right)\Big] + \frac{1}{2} [\mathbf{S} (\bm{x},t) + \mathbf{S}(\bm{x}+\mathbf{e}\Delta t, t+\Delta t)]\Delta t.
\end{equation}
Here, the first term on the right hand side (RHS) of this equation (Eq.~(\ref{eq:5})) represents the changes under collision as a result of the various raw moments relaxing to their corresponding equilibria at rates given in terms of the relaxation matrix $ \tensor{\hat{\Lambda}} = diag \; (\omega_0  ,\; \omega_1,\;   \omega_2, \omega_3, .....,\omega_8)$, where $\omega_j$ ($j=0,1,...8$) are the relaxation parameters, with the changes mapped back into the velocity space via the $\tensor{T^{-1}}$ operator. On the other hand, the second term on the RHS of Eq.~(\ref{eq:5}) represents the effect of the body force via the source term, which is discretized using the trapezoidal rule. By applying the standard variable transformation $\mathbf{\overline{f}}=\mathbf{f} -\frac{1}{2} \mathbf{S}\Delta t$, the implicitness in this term can be removed. After dropping the 'overbar' symbol to simplify notation, then Eq.~(\ref{eq:5}) simplifies to~\cite{premnath2009incorporating}
\begin{equation}\label{eq:6}
 \mathbf{f} (\bm{x}+\mathbf{e}\Delta t, t+\Delta t)- \mathbf{f} (\bm{x},t) =  \tensor{T^{-1}}
 \Big[  \tensor{\hat{\Lambda}} \;  \left(\; \mathbf{m}^{eq}-\mathbf{m} \;\right) + \left(\tensor{I} - \frac{\tensor{\hat{\Lambda}}}{2}\right)  \mathbf{\Phi}\Delta t \Big].
\end{equation}
The raw moments of the distribution functions $f_i$, their equilibria $f_i^{eq}$, and the source terms $S_i$ used in the above can be defined as
\begin{equation} \label{eq:7}
\eta_{x^m y^n}=\sum_{i=0}^{8} f_i e_{ix}^m e_{iy}^n, \qquad \eta^{eq}_{x^m y^n}=\sum_{i=0}^{8} f^{eq}_i e_{ix}^m e_{iy}^n, \qquad \sigma_{x^m y^n}=\sum_{i=0}^{8} S_i e_{ix}^m e_{iy}^n,
\end{equation}
where $(m+n)$ refers to the order of the moment. Based on these and Eq.~(\ref{eq:tranformationrawmoments}), the 9-dimensional vectors of the raw moments of the distribution functions $\mathbf{m}$, their equilibria $\mathbf{m}^{eq}$ and the source terms $\mathbf{\Phi}$ used in Eq.~(\ref{eq:6}) can be enumerated as
\begin{subequations} \label{eq:9}
\begin{eqnarray}
\mathbf{m}&=&\left(m_{0}, m_{1}, m_{2},\ldots,m_{8}\right)^{\dag}\nonumber\\
&=&\left({\eta}_{0}, {\eta}_{x}, {\eta}_{y}, {\eta}_{xx+yy}, {\eta}_{xx-yy}, {\eta}_{xy}, {\eta}_{xxy}, {\eta}_{xyy}, {\eta}_{xxyy}\right)^{\dag},\\
\mathbf{m}^{eq}&=&\left(m_{0}^{eq}, m_{1}^{eq}, m_{2}^{eq},\ldots,m_{8}^{eq}\right)^{\dag}\nonumber\\
&=&\left({\eta}_{0}^{eq}, {\eta}_{x}^{eq}, {\eta}_{y}^{eq}, {\eta}_{xx+yy}^{eq}, {\eta}_{xx-yy}^{eq}, {\eta}_{xy}^{eq}, {\eta}_{xxy}^{eq}, {\eta}_{xyy}^{eq}, {\eta}_{xxyy}^{eq}\right)^{\dag},\\
\mathbf{\Phi}&=&\left(\Phi_{0}, \Phi_{1}, \Phi_{2},\ldots, \Phi_{8}\right)^{\dag}\nonumber\\
&=&\left(\sigma_{0}, \sigma_{x}, \sigma_{y}, \sigma_{xx+yy}, \sigma_{xx-yy}, \sigma_{xy}, \sigma_{xxy}, \sigma_{xyy}, \sigma_{xxyy}\right)^{\dag}.
\end{eqnarray}
\end{subequations}

Then, with the aim of removing the anisotropy in the emergent hydrodynamical equations arising from use of the rectangular lattice and to eliminate the non-Galilean invariant cubic velocity terms due to the aliasing effects on the D2Q9 lattice, we now extend the components of the equilibrium moments $\mathbf{m}^{eq,r}$ defined for the square lattice, by including two types of correction terms perturbed by the time step $\Delta t$ (which will also serve as a small parameter in the C-E expansion later) as follows:
\begin{eqnarray} \label{eq:10}
\mathbf{m}^{eq}&=& \mathbf{m}^{eq,r}+\Delta {t} \;  \mathbf{m}^{eq,s} + \Delta {t} \;  \mathbf{m}^{eq,G},
\end{eqnarray}
where $\mathbf{m}^{eq,s}$ represents the equilibrium moment correction needed to eliminate the deviation from isotropy caused by the use of the rectangular lattice and $\mathbf{m}^{eq,G}$ represents the additional correction required for removing the non-GI terms. Accordingly, we define the components of each of them as follows:
\begin{subequations} \label{eq:11}
\begin{eqnarray}
\mathbf{m}^{eq,r}&=&\left(m^{eq,r}_{0}, m^{eq,r}_{1}, m^{eq,r}_{2},\ldots, m^{eq,r}_{8}\right)^{\dag}\nonumber\\
&=&\left(\eta_0^{eq,r}, \eta_x^{eq,r}, \eta_y^{eq,r}, \eta_{xx+yy}^{eq,r}, \eta_{xx-yy}^{eq,r}, \eta_{xy}^{eq,r}, \eta_{xxy}^{eq,r}, \eta_{xyy}^{eq,r}, \eta_{xxyy}^{eq,r}\right)^{\dag},\\
\mathbf{m}^{eq,s}&=&\left(0, 0, 0,m_3^{eq,s}, m_4^{eq,s}, m_5^{eq,s}, 0, 0, 0 \right)^{\dag},\\
\mathbf{m}^{eq,G}&=&\left(0, 0, 0,m_3^{eq,G}, m_4^{eq,G}, m_5^{eq,G}, 0, 0, 0 \right)^{\dag}.
\end{eqnarray}
\end{subequations}
Notice that in the above that non-zero correction terms suffice only for the second order moments (with indices 3, 4 and 5), which are related to the viscous stress tensor. In Eq.~(\ref{eq:10}) and (\ref{eq:11}), the components of the raw moment equilibria for the square lattice $\mathbf{m}^{eq,r}$ follow via the binomial transformation of the corresponding central moment equilibria obtained via matching the respective continuous central moments of the Maxwell distribution function, which read as~\cite{premnath2009incorporating}
\begin{eqnarray}\label{eq:12}
 m^{eq,r}_0&=& \eta^{eq,r}_0=  \rho,\nonumber\\
 m^{eq,r}_1&=& \eta^{eq,r}_x=  \rho  u_x,\nonumber\\
 m^{eq,r}_2&=&\eta^{eq,r}_y= \rho  u_y,\nonumber\\
 m^{eq,r}_3&=&\eta^{eq,r}_{xx+yy}=  2 \rho c_s^2+ \rho  (u_x^2+ u_y^2 ),\nonumber\\
 m^{eq,r}_4&=&\eta^{eq,r}_{xx-yy}=  \rho   (u_x^2- u_y^2 ),\nonumber\\
 m^{eq,r}_5&=&\eta^{eq,r}_{xy}=  \rho  u_x   u_y,\nonumber\\
 m^{eq,r}_6&=&\eta^{eq,r}_{xxy}=  c_s^2  \rho  u_y+ \rho  u_x^2 u_y,\nonumber\\
 m^{eq,r}_7&=&\eta^{eq,r}_{xyy}=  c_s^2  \rho  u_x+ \rho  u_x   u_y^2,\nonumber\\
 m^{eq,r}_8&=&\eta^{eq,r}_{xxyy}=  \rho  c_s^4 + 3 \rho   c_s^2  (u_x^2+ u_y^2 )+ \rho u_x^2 u_y^2,
\end{eqnarray}
where $\rho$ is the density and $\bm{u}=(u_x,u_y)$ is the velocity of the fluid, and $c_s$ is the speed of sound, which is a free parameter and will be related to the transport coefficients via a C-E analysis later. It may be noted that unlike the prior rectangular MRT-LB schemes that use equilibrium moments with fluid velocity terms only up to the second order, the above moment equilibria (Eq.~(\ref{eq:12})) involve the use of higher order velocity terms arising naturally via the use of the matching principle noted above, which preserves the GI of the moments independently supported by the lattice aside from those subjected to the aliasing effects. In addition to avoiding the orthogonalization in defining the moment basis, consideration of such more refined equilibria is expected to yield a more robust rectangular LB formulation. The components of the source moments that would yield consistency with the NS equations are given by~\cite{premnath2009incorporating}
\begin{eqnarray}\label{eq:13}
\Phi_0 &=& \sigma_{0}=0,\nonumber\\
\Phi_1 &=& \sigma_{x}=F_x,\nonumber\\
\Phi_2 &=& \sigma_{y}=F_y,\nonumber\\
\Phi_3 &=& \sigma_{xx+yy}=2(F_x u_x+F_y u_y),\nonumber\\
\Phi_4 &=& \sigma_{xx-yy}=2(F_x u_x-F_y u_y),\nonumber\\
\Phi_5 &=& \sigma_{xy}=F_xu_y+F_yu_x,\nonumber\\
\Phi_6 &=& \sigma_{xxy}=F_yu_x^2+2F_xu_xu_y,\nonumber\\
\Phi_7 &=& \sigma_{xyy}=F_xu_y^2+2F_yu_yu_x,\nonumber\\
\Phi_8 &=& \sigma_{xxyy}=2(F_xu_xu_y^2+F_yu_yu_x^2),
\end{eqnarray}
where $\bm{F}=(F_x,F_y)$ is the local body force applied to the fluid.

\subsection{Chapman-Enskog Analysis: Derivation of isotropy corrections and recovery of NS equations}
In order to derive the explicit expressions for the two types of corrections $ \mathbf{m}^{eq,s}\;$ and $\mathbf{m}^{eq,G}\;$ appearing in Eq.~(\ref{eq:10}), we shall now perform the Chapman-Enskog multi-scale expansion \cite{chapman1990mathematical}, which would then allow us to recover the NS equations from the corresponding rectangular LB scheme (Eq.~(\ref{eq:6})). In this regard, the approach used in Refs.~\cite{premnath2009incorporating} and~\cite{Hajabdollahi201897} will be adopted. First, we expand the moment $\mathbf{m}$ about its equilibria $\mathbf{m}^{(0)}$ by including the non-equilibrium effects as a perturbation and also the time derivative $\partial_t$ via a multiscale time expansion as
\begin{equation} \label{eq:14}
\mathbf{m} =\mathbf{m}^{(0)}+\epsilon \;   \mathbf{m}^{(1)} + \epsilon^2 \;   \mathbf{m}^{(2)}+ \ldots,\qquad \partial_t= \partial_{t_0}+\epsilon \partial_{t_1} + \epsilon^2 \partial_{t_2}+\ldots,
 \end{equation}
where $ \epsilon$ is a small perturbation bookkeeping parameter set equal to the time increment $\epsilon=\Delta t$. Substituting these expansions into rectangular MRT-LBE with non-orthogonal moment basis (Eq.~(\ref{eq:6})) and successively equating terms of the same order of $\epsilon$ on each side of this equation, we get
\begin{subequations}
\begin{eqnarray}
 \centering
&O (\epsilon^0 ):  \mathbf{m}^{(0)} =  \mathbf{m}^{eq,r},  \label{eq:15a},\\
&O (\epsilon^1 ): D_{t_0} \mathbf{m}^{(0)} =  - \tensor{\hat{\Lambda}} \; \mathbf{m}^{(1)} + \tensor{\hat{\Lambda}}\;(\;\mathbf{m}^{eq,s} + \mathbf{m}^{eq,G}\;) +\mathbf{\Phi},  \label{eq:15b} \\
&O (\epsilon^2 ) : {\partial_t}_1 \; \mathbf{m}^{(0)} +  D_{t_0} \;\left( \tensor{I} - \frac{\tensor{\hat{\Lambda}}} {2} \right) \mathbf{m}^{(1)} + D_{t_0} \; \frac { \tensor{\hat{\Lambda}} } {2} \left( \mathbf{m}^{eq,s}+ \mathbf{m}^{eq,G}\right)=  - \tensor{\hat{\Lambda}} \; \mathbf{m}^{(2)}, \label{eq:15c}
\end{eqnarray}
\end{subequations}
where $D_{t_0}$ is the streaming operator involving the fastest time scale $t_0$ and defined by $D_{t_0}^{(0)}={\partial_t}_0 +\hat{E_i}  \partial_i$, and $\tensor{\hat{E}_i}$ is given by $\tensor{\hat{E}_i}= \big( \tensor{T} \;( \mathbf{e_i} \; \tensor{I})\tensor{T^{-1}} \big)$. Notice that the modifications to the moment equilibria, i.e., the corrections $\mathbf{m}^{eq,s}$ and $\mathbf{m}^{eq,G}$ appear in the equation at first order in $\epsilon$. Rewriting Eqs.~(\ref{eq:15b})and (\ref{eq:15c}) in the long form, they can be respectively expressed as
\begin{eqnarray} \label{eq:16a}
& \partial_{t_0} \mathbf{m}^{(0)} + \partial_x \tensor{\hat{E}_x} \mathbf{m}^{(0)} + \partial_y  \tensor{\hat{E}_y}  \mathbf{m}^{(0)} = \mathbf{\Phi} - \tensor{\hat{\Lambda}} \mathbf{m}^{(1)}+ \tensor{\hat{\Lambda}} (\mathbf{m}^{eq,s} + \mathbf{m}^{eq,G}),
\end{eqnarray}
\begin{eqnarray} \label{eq:16b}
& \partial_{t_1}\mathbf{m}^{(0)}\;+ \partial_{t_0}(\tensor{I}-\frac{\tensor{\hat{\Lambda}} }{2}) \; \mathbf{m}^{(1)} \; +\partial_x \tensor{\hat{E_x}} (\tensor{I} -\frac{\tensor{\hat{\Lambda}} }{2})\; \mathbf{m}^{(1)}  \;+ \partial_y \tensor{\hat{E_y}} (\tensor{I} -\frac{\tensor{\hat{\Lambda}} }{2})\; \mathbf{m}^{(1)}
+\partial_{t_0} \; \frac{\tensor{\hat{\Lambda}} }{2} \; ( \mathbf{m}^{eq,s}+\mathbf{m}^{eq,G}) \nonumber\\
&+ \partial_x \; \frac{\tensor{\hat{\Lambda}} }{2}\; ( \mathbf{m}^{eq,s}+\mathbf{m}^{eq,G)})+ \partial_y \; \frac{\tensor{\hat{\Lambda}} }{2} \; ( \mathbf{m}^{eq,s}+\mathbf{m}^{eq,G})  = -\tensor{\hat{\Lambda}}\mathbf{m}^{(2)}. \label{eq:19b}
\end{eqnarray}

As shown in Eq.~(\ref{eq:15a}), the zeroth moment $\mathbf{m}^{(0)}$ is just the equilibrium moment for the square lattice $\mathbf{m}^{eq,r}$ defined in Eq.~(\ref{eq:12}). Thus, using $\mathbf{m}^{(0)}$ via Eq.~(\ref{eq:12}) into the $O(\epsilon)$ Eq.~(\ref{eq:16a}), its leading components, i.e., up to the second order moments, which are relevant to recovering the hydrodynamical equations are enumerated as
\begin{subequations}
\begin{eqnarray}\label{eq:17a}
 &\partial_{t_0}\rho + \partial_x \rho u_x + \partial_y \rho u_y = 0,
\end{eqnarray}
\begin{eqnarray}\label{eq:17b}
 &\partial_{t_0}\rho u_x + \partial_x (\rho c_s^2  + \rho u_x^2) + \partial_y (\rho u_x u_y) = F_x,
\end{eqnarray}
\begin{eqnarray}\label{eq:17c}
&\partial_{t_0}\rho u_y + \partial_x (\rho u_x u_y)+ \partial_y (\rho c_s^2 + \rho u_y^2) = F_y,
\end{eqnarray}
\begin{eqnarray}\label{eq:17d}
& \partial_{t_0}\left( 2 \rho c_s^2 + \rho \left(u_x^2+ u_y^2\right) \right)+  \partial_x \big\{(c_s^2+1)\rho u_x + \rho u_x u_y^2 \big\} + \underline{\partial_y \big\{(c_s^2+a^2)\rho u_y + \rho u_x^2 u_y \big\}} = \nonumber \\
& - \omega_3\;  m_3^{(1)}+ \omega_3 \; (m_3^{eq,s}+ m_3^{eq,G})+ 2(F_x u_x+F_y u_y),
\end{eqnarray}
\begin{eqnarray}\label{eq:17e}
& \partial_{t_0}(\rho u_x^2-\rho u_y^2)+  \partial_x \big\{(1-c_s^2)\rho u_x - \rho u_x u_y^2)\big\} + \underline{\partial_y \big\{(c_s^2- a^2) \rho u_y + \rho u_x^2 u_y \big\}} = \nonumber  \\
&- \omega_4 \; m_4^{(1)} + \omega_4 \; (m_4^{eq,s}+ m_4^{eq,G} )+2(F_x u_x-F_y u_y) ,
\end{eqnarray}
\begin{eqnarray}\label{eq:17f}
&\partial_{t_0}( \rho u_x u_y)+  \partial_x \big\{c_s^2 \rho u_y + \rho u_x^2 u_y \big\} + \partial_y \big\{c_s^2\rho u_x + \rho u_x u_y^2\big\}= \nonumber  \\
&- \omega_5 \; m_5^{(1)} + \omega_5 \; (m_5^{eq,s}+ m_5^{eq,G} )+F_xu_y+F_yu_x.
\end{eqnarray}
\end{subequations}
Observe that Eqs.~(\ref{eq:17a})-(\ref{eq:17c}), which correspond to the evolution of the conservative moments (i.e., density and the two momentum components), are independent of lattice geometry. On the other hand, the influence of the deviation from isotropy due to the presence of the grid aspect ratio $a$ appears as expected in the equations for the evolution of the second order moments (see the underlined terms in Eqs.~(\ref{eq:17d})-(\ref{eq:17f})). Similarly, we list the leading three relevant components of the $O(\epsilon^2)$  Eq.~(\ref{eq:16b}), which are required to complete the evolution of the density and momentum components at the next slower time scale $t_1$ to derive the NS equations, as follows:
\begin{subequations}
\begin{eqnarray}
&\partial_{t_1}\rho=0,
\label{eq:18a}\\&
\partial_{t_1}\left(\rho u_x\right)+\partial_x \left[\frac{1}{2}\left(1-\frac{\omega_3}{2}\right) m_3^{(1)}+\frac{1}{2}\left(1-\frac{\omega_4}{2}\right)m_4^{(1)}\right]
+\partial_y \left[\left(1-\frac{\omega_5}{2}\right)m_5^{(1)}\right]\nonumber \\&
+\partial_x\left[\frac{\omega_3}{4} (m_3^{eq,s}+m_3^{eq,G})+\frac{\omega_4}{4} (m_4^{eq,s}+m_4^{eq,G}) \right]+
\partial_y\left[\frac{\omega_5}{2} (m_5^{eq,s}+m_5^{eq,G})\right]=0,
\label{eq:18b}\\&
\partial_{t_1}\left(\rho u_y\right)+\partial_x \left[\left(1-\frac{\omega_5}{2}\right) m_5^{(1)}\right]+\partial_y \left[\frac{1}{2}\left(1-\frac{\omega_3}{2}\right)m_3^{(1)}-\frac{1}{2}\left(1-\frac{\omega_4}{2}\right) m_4^{(1)}\right] \nonumber \\&
+\partial_x\left[\frac{\omega_5}{2} (m_5^{eq,s}+m_5^{eq,G})\right]+\partial_y\left[\frac{\omega_3}{4} (m_3^{eq,s}+m_3^{eq,G})-\frac{\omega_4}{4} (m_4^{eq,s}+m_4^{eq,G})\right]=0.
\label{eq:18c}
\end{eqnarray}
\end{subequations}

To proceed further in deriving the correction terms, we need to first obtain the expressions for the non-equilibrium moments $m_3^{(1)}$, $m_4^{(1)}$ and $m_5^{(1)}$, which are needed in Eqs.~(\ref{eq:18b}) and (\ref{eq:18c}) to establish the consistency with the NS equations after eliminating the anisotropy and non-GI terms. Such second order non-equilibrium moment components follow from Eqs.~(\ref{eq:17d})-(\ref{eq:17f}) after substituting for the temporal derivatives of the momentum in terms of the spatial derivative terms (via Eqs.~(\ref{eq:17b}) and (\ref{eq:17c})) and subsequently simplifying them (see Ref.~\cite{HAJABDOLLAHI2018838} for details). Then, we get the following results:
\begin{eqnarray}\label{eq:19}
&m_3^{(1)} = \; \; - \frac{1}{\omega_3} \;  \Big\{ (- c_s^2 +1) \rho \;   \partial_x u_x  +  (- c_s^2 +a^2) \rho \;   \partial_y  u_y  +  (-3 c_s^2 +1) u_x \;   \partial_x \rho +  (-3 c_s^2 +a^2)  u_y \;   \partial_y \;  \rho  \nonumber \\&
-3 \rho \; (u_x^2  \;  \partial_x \;  u_x + u_y^2  \; \partial_y \; u_y ) \Big \} +  (m_3^{eq,s} + m_3^{eq,G}),
\end{eqnarray}
\begin{eqnarray}\label{eq:20}
&m_4^{(1)} =\; \; - \frac{1}{\omega_4} \;  \Big\{ (1 - c_s^2 ) \rho  \partial_x u_x  +  ( c_s^2 - a^2) \rho   \partial_y  u_y +  (-3 c_s^2 +1) u_x \;   \partial_x \rho -  (-3 c_s^2 + a^2)  u_y \;   \partial_y \;  \rho
\nonumber \\&
-3 \rho \; (u_x^2  \;  \partial_x \;  u_x - u_y^2  \; \partial_y \; u_y ) \Big \} +  (m_4^{eq,s} + m_4^{eq,G}),
\end{eqnarray}
\begin{eqnarray} \label{eq:21}
&m_5^{(1)} =\; \; - \frac{1}{\omega_5} \; c_s^2  \rho \;  (  \partial_y u_x  +   \partial_x  u_y ) +  (m_5^{eq,s} + m_5^{eq,G}).
\end{eqnarray}
Note that these last three equations contain the error terms related to the anisotropy terms (dependent on the grid aspect ratio) and the non-GI cubic velocity terms as well as the correction terms whose forms are yet to be determined. The next step is to combine the conserved moments equations related to temporal variations using the scale $t_0$ (Eqs.~(\ref{eq:17a})-(\ref{eq:17c})) with $\epsilon$ times the corresponding equations involving the time scale $t_1$ (Eqs.~(\ref{eq:18a})-(\ref{eq:18c})). Then, taking into account that $\partial_t={\partial_t}_0 + \epsilon  {\partial_t}_1 $, we arrive at the following hydrodynamical equations for the evolution of the density and the components of the momentum fields:
\begin{subequations}\label{eq:22}
 \begin{eqnarray}\label{eq:22a}
 &\partial_{t}\rho u_y + \partial_x (\rho u_x u_y)+ \partial_y (\rho c_s^2 + \rho u_y^2) = 0,
 \end{eqnarray}
 \begin{eqnarray}\label{eq:22b}
 &\partial_{t}\left(\rho u_x\right)+\partial_x \left(c_s^2 \rho+\rho u_x^2\right)+\partial_y \left(\rho u_xu_y\right) = F_x-\epsilon\partial_x \left[\frac{1}{2}\left(1-\frac{\omega_3}{2}\right){m}_3^{(1)}+\frac{1}{2}\left(1-\frac{\omega_4}{2}\right)m_4^{(1)}\right]
 -\epsilon\partial_y\left[\left(1-\frac{\omega_5}{2}\right)m_5^{(1)}\right]  \nonumber \\&
 -\epsilon\partial_x\left[\frac{\omega_3}{4}m_3^{eq,s}+\frac{\omega_4}{4} m_4^{eq,s}\right]
 -\epsilon\partial_y\left[\frac{\omega_5}{2} m_5^{eq,s}\right] -\epsilon\partial_x\left[\frac{\omega_3}{4}m_3^{eq,G}+\frac{\omega_4}{4} m_4^{eq,G}\right]-\epsilon\partial_y\left[\frac{\omega_5}{2} m_5^{eq,G}\right],
\end{eqnarray}
\begin{eqnarray}\label{eq:22c}
&\partial_{t}\left(\rho u_y\right)+\partial_x \left(\rho u_xu_y\right)+\partial_y \left(c_s^2 \rho+\rho u_y^2\right) = F_y-\epsilon\partial_x\left[\left(1-\frac{\omega_5}{2}\right)m_5^{(1)}\right]-\epsilon\partial_y\left[\frac{1}{2}
\left(1-\frac{\omega_3}{2}\right) m_3^{(1)}-\frac{1}{2}\left(1-\frac{\omega_4}{2}\right)m_4^{(1)}\right]
\nonumber \\&
-\epsilon\partial_x\left[\frac{\omega_5}{2} m_5^{eq,s)}\right]
-\epsilon\partial_y\left[\frac{\omega_3}{4} m_3^{eq,s}-\frac{\omega_4}{4} m_4^{eq,s}\right]
-\epsilon\partial_x\left[\frac{\omega_5}{2} m_5^{eq,G)}\right]
-\epsilon\partial_y\left[\frac{\omega_3}{4} m_3^{eq,G}-\frac{\omega_4}{4} m_4^{eq,G}\right].
 \end{eqnarray}
\end{subequations}
In the above three equations (Eqs.~(\ref{eq:22a})-(\ref{eq:22c})), we then substitute for the non-equilibrium moments $m_3^{(1)}$, $m_4^{(1)}$ and $m_5^{(1)}$ given in Eqs.~(\ref{eq:19})-(\ref{eq:21}), respectively. Then, we isolate the error terms and the counteracting correction terms from the desired fluid flow equations represented by the NS equations. This leads to the constraint that $(\mbox{correction term})_j+(1-\omega_j/2)(\mbox{error term})_j=0$, where $j=3,4$ and $5$ and $(\mbox{error term})_j$ are the terms that deviate from the target NS equations (see Ref.~\cite{HAJABDOLLAHI2018838} for details). For example, the aliasing effects on the standard D2Q9 lattice for the third order longitudinal moments, i.e., $\sum_\alpha f_\alpha e_{\alpha i}^3=\sum_\alpha f_\alpha e_{\alpha i}$, where $i \in \{x,y\}$, lead to the cubic velocity errors $-\frac{3\rho}{\omega_3} (u_x^2 \partial_x  u_x + u_y^2  \partial_y u_y )$ in the non-equilibrium moment $m_3^{(1)}$ in Eq.~(\ref{eq:19}) and $-\frac{3\rho}{\omega_4}(u_x^2  \partial_x u_x - u_y^2 \partial_y u_y )$ in the non-equilibrium moment $m_4^{(1)}$ in Eq.~(\ref{eq:20}). These are then eliminated by the counteracting correction terms in the second order extended moment equilibria $m_3^{eq,G}$ and $m_4^{eq,G}$, respectively, determined by the above constraint. Following this strategy, we can then determine the expressions for the isotropy correction terms $m_j^{eq,s}$ and the GI correction terms $m_j^{eq,G}$ for $j=3,4$ and $5$, which read
\begin{eqnarray}\label{eq:23}
&m_3^{eq,s} = \left( \frac{1}{\omega_3}- \frac{1}{2} \right)\Big\{  (-3 c_s^2 +1) \rho \partial_x u_x  +  (-3 c_s^2 +a^2) \rho\partial_y  u_y +  (-3 c_s^2 + 1) u_x   \partial_x \rho  +  (-3 c_s^2 + \; a^2)  u_y \partial_y \rho \Big \},
\end{eqnarray}
\begin{eqnarray}\label{eq:24}
&m_4^{eq,s} =\left( \frac{1}{\omega_4}- \frac{1}{2} \right)\; \Big\{  (-3 c_s^2 + 1) \rho \partial_x u_x  +  (3 c_s^2 - a^2) \rho \partial_y  u_y  +  (-3 c_s^2 +1) u_x \partial_x \rho  +  (3 c_s^2  - \; a^2)  u_y \;   \partial_y \rho \Big \},
\end{eqnarray}
\begin{eqnarray}\label{eq:25}
&m_3^{eq,G} = -3\rho\left( \frac{1}{\omega_3}- \frac{1}{2} \right)\left(u_x^2 \partial_x u_x + u_y^2 \partial_y u_y  \right),
\end{eqnarray}
\begin{eqnarray}\label{eq:26}
&m_4^{eq,G} = -3\rho\left( \frac{1}{\omega_4}- \frac{1}{2} \right) \left(u_x^2 \partial_x u_x - u_y^2 \partial_y u_y \right),
\end{eqnarray}
\begin{eqnarray} \label{eq:27}
&m_5^{eq,s} = 0, \; \; \; \; m_5^{eq,G} = 0.
\end{eqnarray}
The use of Eqs.~(\ref{eq:23})-(\ref{eq:27}), which are among the key results of this work expressing the required corrections terms, in Eqs.~(\ref{eq:22a})-(\ref{eq:22c}) then implies the rectangular MRT-LBE is consistent with the fluid dynamics with isotropic viscous stress tensor represented by the NS equations given by
\begin{eqnarray}
&\partial_t \rho + \bm{\nabla}\cdot \bm{j} = 0,
\label{eq:29}
\end{eqnarray}
\begin{eqnarray}
&\partial_t j_x+\bm{\nabla}\cdot \left(\bm{j}u_x\right)= -\partial_x p+\partial_x\left[\nu (2\partial_x j_x-\bm{\nabla}\cdot\bm{j})+\zeta (\bm{\nabla}\cdot\bm{j}) \right]
+\partial_y\left[\nu (\partial_x j_y+\partial_y j_x) \right]+F_x,
\label{eq:30}
\end{eqnarray}
\begin{eqnarray}
&\partial_t j_y+\bm{\nabla}\cdot \left(\bm{j}u_x\right)= -\partial_y p+\partial_x\left[\nu (\partial_x j_y+\partial_y j_x) \right]+ \partial_y\left[\nu (2\partial_y j_y-\bm{\nabla}\cdot\bm{j})+\zeta (\bm{\nabla}\cdot\bm{j}) \right]
+F_y,
\label{eq:31}
\end{eqnarray}
where $p=\rho c_s^2$ is the pressure, $\bm{j}=\left( j_x, j_y \right)= \left( \rho u_x, \rho u_y\right)$ is the momentum, $ \nu $ is the shear kinematic viscosity, $ \zeta $ is the bulk kinematic viscosity, respectively, which are written as a function of the relaxation parameters of the second order moments, i.e., $\omega_j$, where $j=3,4$ and $5$ as
\begin{eqnarray}
\nu &=& c_s^2\left( \frac{1}{\omega_4}- \frac{1}{2}\right)\Delta t= c_s^2\left( \frac{1}{\omega_5}- \frac{1}{2}\right)\Delta t, \label{eq:32}\\
\zeta &=& c_s^2\left(\frac{1}{\omega_3}- \frac{1}{2}\right)\Delta t. \label{eq:33}
\end{eqnarray}
The relaxation parameters for the higher order moments $\omega_j$, where $j=6,7$ and $8$ influence numerical stability (see e.g.,~\cite{ning2016numerical}) and are set to unity in this work. It may be noted that the correction terms given in Eqs.~(\ref{eq:23}) and (\ref{eq:24}) for the diagonal components of the second order moment equilibria are dependent on the grid aspect ratio $a$ and the speed of sound $c_s$, both of which are free parameters of our formulation, with the latter adjusted based on the choice of the former to maintain numerical stability. Equation~(\ref{eq:27}) implies that no additional corrections are necessary for the off-diagonal second order moment equilibria for the rectangular LB formulation. Moreover, the transport coefficients given in Eqs.~(\ref{eq:32}) and (\ref{eq:33}) are not parameterized by the grid aspect ratio, and maintain the simple expressions applicable for the square lattice. It may be noted that, as a special case, when we set $a=1$ and the speed of sound $c_s=1/\sqrt{3}$, the previous results for the square lattice are recovered (see e.g.,~\cite{HAJABDOLLAHI2018838}) and the isotropy corrections (Eqs.~(\ref{eq:23}) and (\ref{eq:24})) vanish.

\subsection{Local expressions for strain rate tensor in terms for rectangular lattice}
The diagonal parts of the strain rate tensor, i.e., $\partial_x u_x$ and $\partial_y u_y$ appear in the equilibria correction terms (Eqs.~(\ref{eq:23})-(\ref{eq:26}). These along with the off-diagonal component $(\partial_y u_x+\partial_x u_y)$ can be obtained locally in terms of the following second-order non-equilibrium moments:
\begin{subequations}
\begin{eqnarray}
m_3^{(1)}&=& m_3 - m_3^{eq,r}= \eta_{xx+yy} -  m_3^{eq,r}, \label{eq:34}\\
m_4^{(1)}&=& m_4 - m_4^{eq,r}= \eta_{xx-yy} -  m_4^{eq,r}, \label{eq:35}\\
m_5^{(1)}&=& m_5 - m_5^{eq,r}= \eta_{xy} - m_5^{eq,r}, \label{eq:36}
\end{eqnarray}
\end{subequations}
where the $m_j^{eq,r}$ ($j=3,4$ and $5$) are given in Eq.~(\ref{eq:12}). By substituting Eqs.~(\ref{eq:23})-(\ref{eq:27}) in Eqs.~(\ref{eq:19})-(\ref{eq:21}) and using Eqs.~(\ref{eq:34})-(\ref{eq:36}), rearranging and solving for $\partial_x u_x$, $\partial_y u_y$ and $(\partial_y u_x+\partial_x u_y)$ and simplifying the resulting expressions, we then get the following local expressions for the strain rate tensor applicable for a rectangular lattice:
\begin{subequations}
\begin{eqnarray}
\partial_x u_x &=& \frac{C_2 \Big[ \eta_{xx+yy} - m_3^{eq,r} -B_3 \partial_x \rho -B_4 \partial_y \rho \Big] - B_2 \Big[ \eta_{xx-yy} - m_4^{eq,r} -C_3 \partial_x \rho -C_4 \partial_y \rho \Big]}{\big[C_2 B_1 - C_1  B_2 \big]},\label{eq:37} \\
\partial_y u_y&=& \frac{C_1 \Big[ \eta_{xx+yy} - m_3^{eq,r}-B_3 \partial_x \rho -B_4 \partial_y \rho \Big] - B_1 \Big[\eta_{xx-yy} - m_4^{eq,r} -C_3 \partial_x \rho -C_4 \partial_y \rho \Big]}{\Big[C_1 B_2 - C_2  B_1 \Big]},\label{eq:38} \\
\partial_x u_x  + \partial_y u_y &=& \frac{ \left[ \eta_{xy} -m_5^{eq,r} \right]}{ D_1 }.\label{eq:39}
\end{eqnarray}
\end{subequations}
The density gradients $\partial_x \rho$ and $\partial_y \rho$ appearing in the above are computed using a central difference approach. The coefficients in Eqs.~(\ref{eq:37})-(\ref{eq:39}), $B_1$, $B_2$, $B_3$, $C_1$, $C_2$, $C_3$ and $D_1$ are defined as a function of the model parameters, viz., the lattice speed of sound $c_s$ and the grid aspect ratio $a$ as
\begin{eqnarray*} \label{eq:40}
B_1 &=& \rho \left(\frac{-2 c_s^2}{\omega_3} - \frac{-3 c_s^2 +1}{2}+\frac{3 u_x^2}{2}\right),\quad \; \; B_3 =  \left( \frac{3 c_s^2 -1}{2}\right) \; u_x,  \\
B_2 &=& \rho \left(\frac{-2 c_s^2}{\omega_3} - \frac{-3 c_s^2 + a^2}{2}+\frac{3 u_y^2}{2}\right),\quad B_4 =  \left( \frac{3 c_s^2 - a^2}{2}\right) \; u_y,\\
C_1 &=& \rho \left(\frac{-2 c_s^2}{\omega_4} - \frac{-3 c_s^2 +1}{2}+\frac{3 u_x^2}{2}\right),\quad \; \; C_3 =  \left(\frac{3 c_s^2  -1}{2}\right) \; u_x,\\
C_2 &=& \rho \left(\frac{2 c_s^2}{\omega_4} - \frac{3 c_s^2 -a^2}{2}-\frac{3 u_y^2}{2}\right),\quad \; \; \; \; \; C_4 =  \left( \frac{-3 c_s^2  +a^2}{2}\right) \; u_y,\\
D_1 &=& -\frac{\rho c_s^2}{\omega_5}.
\end{eqnarray*}

\section{Rectangular Raw moment Lattice Boltzmann Method using Non-orthogonal Moment Basis (RNR-LBM)}\label{sec:3}
We will now discuss the implementation of the RNR-LB algorithm based on Eq.~(\ref{eq:6}). For ease of presentation, a matrix-vector representation will be used, while in the actual calculations, the matrix products should not be implemented, but carried out in their component forms by optimizing the operations involved.
\begin{itemize}

\item Compute pre-collision raw moments
\begin{equation*}
\mathbf{m}=\tensor{T}\mathbf{f},
\end{equation*}
where $\tensor{T}$ is given in Eq.~(\ref{eq:4}) and the elements of $\mathbf{f}$, i.e., $f_j$ are at time level $t$, i.e., $f_j=f_j(\bm{x},t)$.

\item Compute post-collision raw moments: Relaxation under collision including sources for body forces
\begin{equation*}
\widetilde{m}_j = m_j+\omega_j(m_j^{eq}-m_j)+(1-\omega_j/2)\Phi_j\Delta t, \quad j=0,1,\ldots 8.
\end{equation*}
Here, following Eq.~(\ref{eq:10}), the extended moment equilibria for the rectangular lattice are computed as $m_j^{eq}=m_j^{eq,r}+\Delta t(m_j^{eq,s}+m_j^{eq,G})$, where $m_j^{eq,r}$, $m_j^{eq,s}$, and $m_j^{eq,G}$ are given in Eqs.~(\ref{eq:11}), (\ref{eq:12}), (\ref{eq:23})-(\ref{eq:27}). The required velocity gradients in the equilibria corrections are computed locally using Eqs.~(\ref{eq:37}) and (\ref{eq:38}). The source terms $\Phi_j$ are obtained from Eq.~(\ref{eq:13}).

\item Compute post-collision distribution functions
\begin{equation*}
\widetilde{\mathbf{f}}=\tensor{T}^{-1}\widetilde{\mathbf{m}},
\end{equation*}
where the inverse transformation matrix $\tensor{T}^{-1}$ mapping from raw moments to distribution functions is given in Eq.~(\ref{eq:Tinverse}) in~\ref{sec:appendix1}.

\item Perform streaming of distribution functions
\begin{equation*}
f_j(\bm{x},t+\Delta t)=\widetilde{f}_j(\bm{x}-\bm{e}_j\Delta t,t).
\end{equation*}

\item Update hydrodynamic fields
\newline
Based on $f_j(\bm{x},t+\Delta t)$ at the new time level $t+\Delta t$ from the step above, the hydrodynamic fields are updated via their moments as
\begin{equation*}
\rho =\sum_{j=0}^{8} f_j, \quad     \rho \bm{u} =\sum_{j=0}^{8} f_j \bm{e}_j + \frac{1}{2} \mathbf{F}\Delta t.
\end{equation*}

\end{itemize}

\section{Rectangular Central moment Lattice Boltzmann Method using Non-orthogonal Moment Basis (RC-LBM)} \label{sec:4} \par
A more general rectangular LB scheme can be constructed using central moments in a moving frame of reference based on the local fluid velocity. Thus, we will now define the discrete central moments of the distribution functions, their equilibria and the source terms of order $(m+n)$ as
\begin{subequations}
\begin{eqnarray}
 \eta^c_{x^m y^n}&=& \sum_{i=0}^{8} f_i (e_{ix} -u_x)^m  (e_{iy} -u_y)^n, \label{eq:43} \\
 \eta^{c,eq}_{x^m y^n}&=& \sum_{i=0}^{8} f_i^{eq}(e_{ix} -u_x)^m  (e_{iy} -u_y)^n,\label{eq:44}\\
 \sigma ^c_{x^m y^n}&=& \sum_{i=0}^{8} S_i (e_{ix} -u_x)^m  (e_{iy} -u_y)^n. \label{eq:45}
\end{eqnarray}
\end{subequations}
Here, and in what follows, the superscript `c' is used to denote central moments of a given quantity. For the natural moment basis independently supported by the D2Q9 lattice given in Eq.~(\ref{eq:3}), we will then list the vectors $\mathbf{m}^c$, $\mathbf{m}^{c,eq}$ and $\mathbf{\Phi}^c$, which enumerate the corresponding components of the central moments as
\begin{subequations}
\begin{eqnarray}
 \mathbf{m}^c&=& {\big( \eta_0^c,\eta_x^c,\eta_y^c,\eta_{xx+yy}^c,\eta_{xx-yy}^c, \eta_{xy}^c ,\eta_{xxy}^c, \eta_{xyy}^c,\eta_{xxyy}^c  \big)}^\dagger,\label{eq:46} \\
  \mathbf{m}^{c,eq}&=& {\big( \eta_0^{c,eq},\eta_x^{c,eq},\eta_y^{c,eq}, \eta_{xx+yy}^{c,eq},\eta_{xx-yy}^{c,eq}, \eta_{xy}^{c,eq} ,\eta_{xxy}^{c,eq}, \eta_{xyy}^{c,eq},\eta_{xxyy}^{c,eq} \big)}^\dagger,\label{eq:47} \\
  \mathbf{\Phi}^c&=& {\big( \sigma_0^{c},\sigma_x^{c}, \sigma_y^{c}, \sigma_{xx+yy}^{c}, \sigma_{xx-yy}^{c}, \sigma_{xy}^{c} , \sigma_{xxy}^{c}, \sigma_{xyy}^{c}, \sigma_{xxyy}^{c} \big)}^\dagger.  \label{eq:48}
\end{eqnarray}
\end{subequations}
In addition, analogously, we can define the components of the discrete equilibrium central moments of the regular square lattice as $\eta^{c,eqr}_{x^m y^n}$ as part of the vector $\mathbf{m}^{c,eqr}$. They can be obtained by matching the corresponding continuous central moments of the Maxwell distribution function~\cite{geier2006cascaded,premnath2009incorporating}, and the components of the central moments of the sources can be prescribed
to yield consistency with the NS equations~\cite{premnath2009incorporating}. Thus,
\begin{subequations}
\begin{eqnarray}
\mathbf{m}^{c,eqr}&=& \left(\eta_0^{c,eqr}, \eta_x^{c,eqr}, \eta_y^{c,eqr}, \eta_{xx}^{c,eqr}, \eta_{yy}^{c,eqr}, \eta_{xy}^{c,eqr} , \eta_{xxy}^{c,eqr}, \eta_{xyy}^{c,eqr}, \eta_{xxyy}^{c,eqr}\right)^{\dag},\nonumber \label{eq:centralmomenteqMaxwellian}\\
&=&\left(0, 0, 0,\rho c_s^2, \rho c_s^2, 0, 0, 0, \rho c_s^4 \right)^{\dag}, \\
\mathbf{\Phi}^{c}&=& \left(\sigma_0^{c}, \sigma_x^{c}, \sigma_y^{c}, \sigma_{xx}^{c}, \sigma_{yy}^{c},\sigma_{xy}^{c}, \sigma_{xxy}^{c}, \sigma_{xyy}^{c}, \sigma_{xxyy}^{c}\right)^{\dag},\nonumber \\
&=&\left(0, F_x, F_y, 0, 0, 0, 0, 0, 0 \right)^{\dag}.\label{eq:centralmomentforcing}
\end{eqnarray}
\end{subequations}
The components of the central moment equilibria $\mathbf{m}^{c,eq}$ for the rectangular lattice can be constructed from those of the square lattice $\mathbf{m}^{c,eqr}$ by correcting for the grid anistropy and the non-GI terms, i.e.,
\begin{equation}\label{eq:centralmomenteqcorrection}
\mathbf{m}^{c,eq}=\mathbf{m}^{c,eqr}+\Delta {t} \;  \mathbf{m}^{eq,s} + \Delta {t} \;  \mathbf{m}^{eq,G},
\end{equation}
As seen earlier, such corrections are related to the non-equilibrium second order moments involving the viscous stress tensor. Since by construction, the non-equilibrium second order central moments are identical to those of raw moments, $\mathbf{m}^{eq,s}$ and $\mathbf{m}^{eq,G}$ in Eq.~(\ref{eq:centralmomenteqcorrection}) are the same as to those given in Eqs.~(\ref{eq:11}), (\ref{eq:12}), (\ref{eq:23})-(\ref{eq:27}). Hence, the central moment equilibria applicable for the rectangular lattice read as
\begin{eqnarray}\label{eq:49}
\mathbf{m}^{c,eq} &=& \left[ {\begin{array}{*{20}{l}} m_0^{c,eq}\\ m_1^{c,eq}\\  m_2^{c,eq}\\ m_3^{c,eq}\\ m_4^{c,eq}\\ m_5^{c,eq}\\ m_6^{c,eq}\\ m_7^{c,eq}\\ m_8^{c,eq} \end{array}} \right]
=\left[ {\begin{array}{*{20}{l}} \rho\\ 0\\  0\\ 2 \rho c_s^2\\ 0\\ 0\\ 0\\ 0\\ \rho c_s^4 \end{array}} \right] + \Delta t \left[ {\begin{array}{*{20}{l}} 0\\ 0\\  0\\  m_3^{eq,s}+ m_3^{eq,G}\\ m_4^{eq,s}+ m_4^{eq,G}\\ m_5^{eq,s}+ m_5^{eq,G}\\ 0\\ 0\\ 0 \end{array}} \right].
\end{eqnarray}
In formulating the RC-LBM, we need to map between raw moments and cental moments, which can be represented as follows:
\begin{equation}
\mathbf{m}^c= \tensor{\mathcal{F}} \mathbf{m},\quad \mathbf{m} = \tensor{\mathcal{F}}^{-1}\mathbf{m}^c,
\end{equation}
where $\tensor{\mathcal{F}}$ is the frame transformation matrix which converts the set of raw moments into central moments. $\tensor{\mathcal{F}}$ forms a lower triangular matrix involving the fluid velocity components $u_x$ and $u_y$ and can be obtained via enumerating the binomial transforms for the finite set of moments. Its inverse $\tensor{\mathcal{F}}^{-1}$ follows readily from an interesting property of $\tensor{\mathcal{F}}$. These are briefly discussed in~\ref{Sec:appendix2} and the elements of $\tensor{\mathcal{F}}$ and $\tensor{\mathcal{F}}^{-1}$ are provided in Eqs.~(\ref{eq:53}) and (\ref{eq:55}), respectively.

Then, based on the above considerations, the rectangular central moment (RC)-LBE involving central moment relaxations under collision and forcing can be written as
\begin{equation}\label{eq:RC-LBE}
 \mathbf{f} (\bm{x}+\mathbf{e}\Delta t, t+\Delta t)- \mathbf{f} (\bm{x},t) =  \tensor{T^{-1}}\tensor{\mathcal{F}^{-1}}
 \Big[  \tensor{\hat{\Lambda}} \;  \left(\; \mathbf{m}^{c,eq}-\mathbf{m}^c \;\right) + \left(\tensor{I} - \frac{\tensor{\hat{\Lambda}}}{2}\right)  \mathbf{\Phi}^c\Delta t \Big].
\end{equation}

The RC-LB algorithm implementing Eq.~(\ref{eq:RC-LBE}) is given as follows:
\begin{itemize}

\item Compute pre-collision raw moments
\begin{equation*}
\mathbf{m}=\tensor{T}\mathbf{f},
\end{equation*}
where $\tensor{T}$ is given in Eq.~(\ref{eq:4}). Hence, this step is the same as that given in the previous section.

\item Compute pre-collision central moments
\begin{equation*}
\mathbf{m}^c=\tensor{\mathcal{F}}\mathbf{m},
\end{equation*}
where $\tensor{\mathcal{F}}$ is given in Eq.~(\ref{eq:53}).

\item Compute post-collision central moments: Relaxation under collision including sources for body forces
\begin{equation*}
\widetilde{m}^c_j = m_j^c+\omega_j(m_j^{c,eq}-m_j^c)+(1-\omega_j/2)\Phi_j^c\Delta t, \quad j=0,1,\ldots 8.
\end{equation*}
Here, the extended central moment equilibria $m_j^{c,eq}$ are given in Eq.~(\ref{eq:49}) and the central moments of sources $\Phi_j^c$ follow from Eq.~(\ref{eq:centralmomentforcing}).

\item Compute post-collision raw moments
\begin{equation*}
\widetilde{\mathbf{m}}=\tensor{\mathcal{F}^{-1}}\widetilde{\mathbf{m}}^c,
\end{equation*}
where $\tensor{\mathcal{F}^{-1}}$ is given in Eq.~(\ref{eq:55}).

\item Compute post-collision distribution functions
\begin{equation*}
\widetilde{\mathbf{f}}=\tensor{T}^{-1}\widetilde{\mathbf{m}},
\end{equation*}
where $\tensor{T}^{-1}$ is provided in Eq.~(\ref{eq:Tinverse}).

\item Perform streaming of distribution functions
\begin{equation*}
f_j(\bm{x},t+\Delta t)=\widetilde{f}_j(\bm{x}-\bm{e}_j\Delta t,t).
\end{equation*}

\item Update hydrodynamic fields
\begin{equation*}
\rho =\sum_{j=0}^{8} f_j, \quad     \rho \bm{u} =\sum_{j=0}^{8} f_j \bm{e}_j + \frac{1}{2} \mathbf{F}\Delta t.
\end{equation*}

\end{itemize}

\section{Results and Discussion} \label{sec:5}
We will now present a numerical validation study of the new rectangular LB formulations, i.e., RNR-LBM and RC-LBM, for a variety of canonical flow problems using different grid aspect ratios and characteristic flow parameters.

\subsection {Steady flow driven by a body force between two parallel plates}  \par
First, we perform simulations of the flow between two parallel plates separated by a distance of $H$ and subjected to a constant body force $F_x$ imposed in the direction of fluid flow using both RNR-LBM and RC-LBM. The analytical solution for this problem is given by $u(y)= U[1-(y-H/2)^2/(H/2)^2]$, where the normal coordinate $y$ is measured from the bottom plate and $U$ is the maximum velocity occurring at the midway location between the channel defined by $U = F_x (H/2)^2/{2 \nu}$. Here, $\nu$ is the shear kinematic viscosity, which as written in Eq.~(\ref{eq:32}), is related to the relaxation parameters $\omega_3$ and $\omega_4$ associated with the relaxation of the second order moments. For brevity, here and in what follows, we express these two parameters in terms of a relaxation time $\tau$ given by $\tau = 1/\omega_4=1/\omega_5$, which will be used to adjust the desired shear viscosity. All the other relaxation parameters in both RNR-LBM and RC-LBM, are set to unity in the simulations presented in the following. Periodic boundary conditions are employed along the flow directions and no-slip boundary conditions are imposed at the walls using the standard half-way bounce back scheme. For the purpose of making comparisons, we define a characteristic Reynolds number as $\mbox{Re}=UH/\nu$.

\begin{figure}[H]
    \centering
    \captionsetup{justification=centering}
   \includegraphics[scale=0.7]{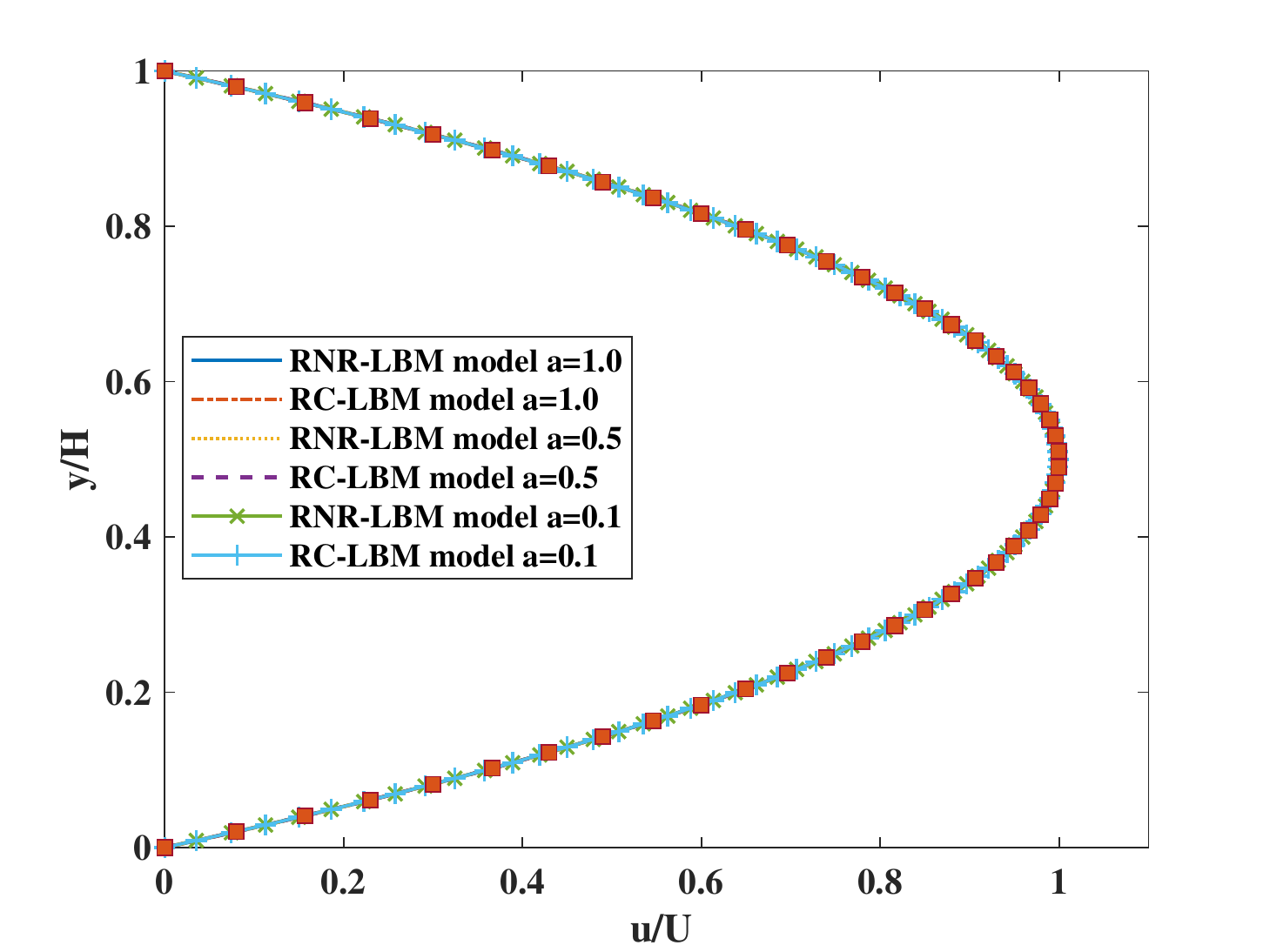}
    \caption{Comparison of computed velocity profiles simulated using the RNR-LBM and RC-LBM with the grid aspect ratios of $a=1.0, 0.5$, and $0.1$ against the analytical solution for 2D channel flow at $Re=50$.}
    \label{fig:2}
\end{figure}

A tabulation of the model parameters used to simulate this flow problem is provided in Table~\ref{tab:1}. We employ different mesh resolutions as the grid aspect ratio $a$ is varied within a range of $\{1.0, 0.8, 0.5, 0.3, 0.1\}$. It may be noted that the geometric anisotropy of the rectangular lattice grid increases as the grid aspect ratio decreases.
\begin{table}[h]
\small
\centering
\captionsetup{justification=centering}
\caption{Parameters used in the simulations of 2D channel flow at different lattice grid aspect ratios $a=0.8, 0.5, 0.3$, and $0.1 $ with a constant Reynolds number $\mbox{Re}=50$.}
\begin{tabular}{|c c c c c c c|}
\hline
 $a$ &   $ N_x \times N_y$ & $F_x$ & U &$ c_s^2 $ & $\tau$ & $\nu$ \\
\hline\hline
 0.8 & $100 \times 125$ & $9.16 \times 10^{-7}$ & 0.0336 & 0.3333 & 0.6& 0.03333\\
 0.5 & $100 \times 200$ & $2.11 \times 10^{-7}$ & 0.016 & 0.16 &  0.6 &0.016\\
 0.3 & $100 \times 300$ & $7.42\times 10^{-9}$ & 0.0.003 & 0.03 & 0.6 & 0.003\\
 0.1 & $ 50 \times 500$  & $1.7 \times 10^{-9}$  & 0.001 & 0.005 & 0.6& 0.0005\\

\hline
\end{tabular}
\label{tab:1}
\end{table}
All simulations are performed at a constant Reynolds number of $Re=50$ using a fixed relaxation time $\tau=0.6$. Figure~\ref{fig:2} shows a comparison between the computed velocity profiles obtained using both RNR-LBM and RC-LBM using the grid aspect ratios of $a=1.0, 0.5$, and $0.1$ against the analytical solution. Excellent agreement between the computed results and the steady state exact solution can be seen.

\begin{figure}[H]
  \centering
  \captionsetup{justification=centering}
  \includegraphics[scale=0.6]{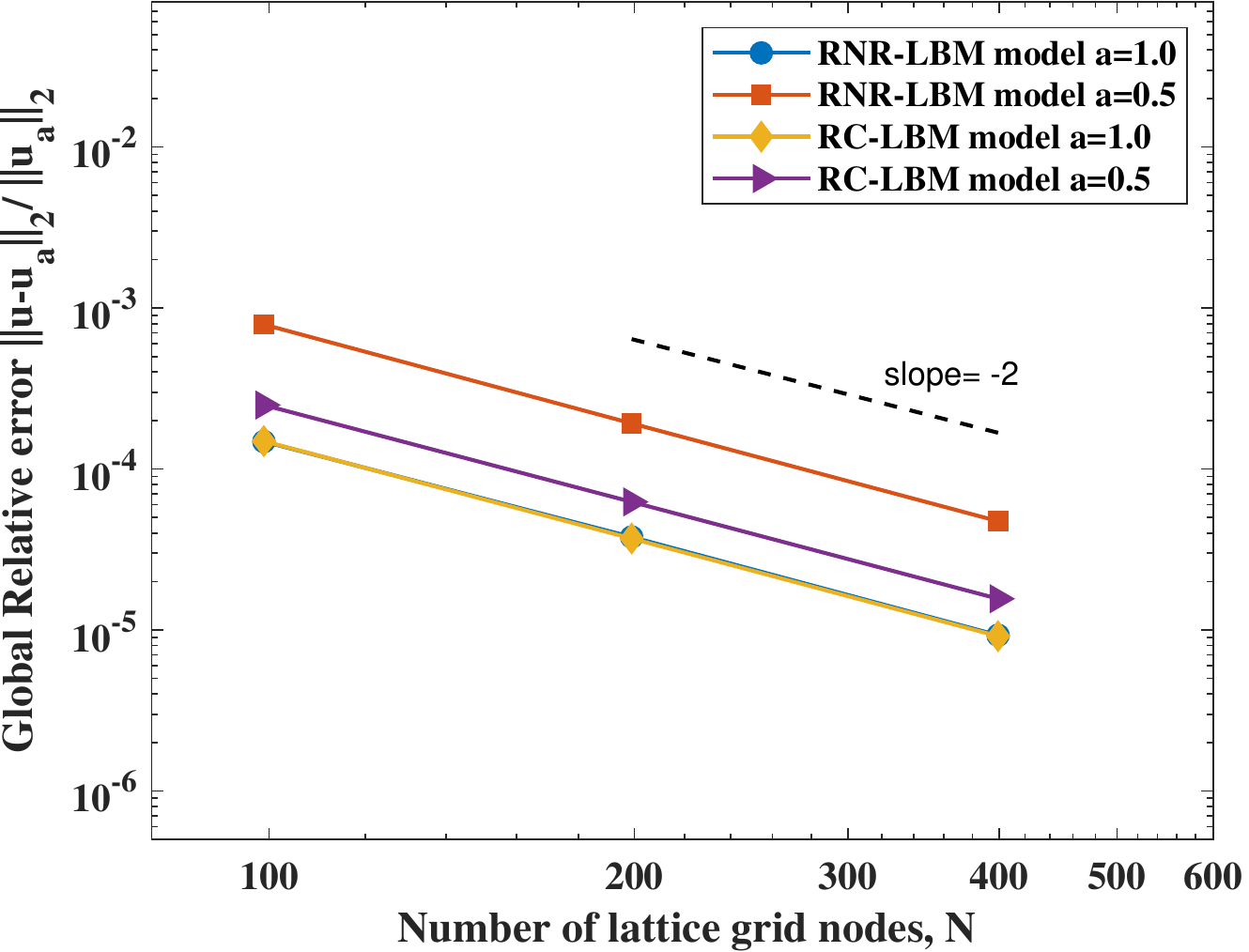}
  \caption{Relative global errors resulting from using the RNR-LBM and RC-LBM at various grid resolutions for 2D channel flow at $\mbox{Re=50}$.}
 \label{fig:3}
\end{figure}

Then, in order to illustrate the order of accuracy of the rectangular LB formulations under grid refinement, we define a global relative error in terms of the second norm of the difference between the computed values of the velocity and the analytical solution as $ \sum_{i} ||{u_{c,i}}-u_{a,i}||_2/ \sum_{i} ||u_{a,i} ||_2$, where $u_c$ and $u_a$ are the computed results and analytical solution, respectively, and the summation is carried out for the entire domain. Figure \ref{fig:3} illustrates the variation of the relative global errors at resolutions of $N = 100, 200, 300$ and $400$ in the $y$ direction by fixing the number of nodes in the $x$ to be $100$ obtained using RNR-LBM and RC-LBM with two grid aspect ratios of $a=1.0$ and $0.5$. The errors for $a=0.5$ are relatively larger than that of $a=1.0$ since the former introduces additional truncation errors dependent on the grid aspect ratio in the higher order moments. It can be seen that both the rectangular LB formulations exhibit a second order grid convergence rate. This second order accuracy of both these schemes under the usual diffusive scaling is consistent with the general property of the standard LB discretization and not dependent on the collision model used. However, the central moment based RC-LBM is found to result in significantly lower magnitudes of global errors when compared the raw moment based RNR-LBM. Since the RC-LBM performs the collision step in a moving frame of reference relative to the local fluid velocity involving the relaxation of the central moments to their equilibria, it is Galilean invariant for all the moments supported independently by the lattice. This results in smaller errors than that for the raw moments based RNR-LBM, whose higher order moments are subjected to additional truncation errors dependent on the fluid velocity, which scales with the spatial discretization under diffusive scaling. Hence, the RC-LBM is found to be more accurate than the RNR-LBM.
This is further evident from Table~\ref{tab:2}, which demonstrates the RC-LBM yields better accuracy than RNR-LBM for simulating this flow at the same grid ratio of $a=0.5$ with various choices of the number of grid nodes in the wall normal direction. In particular, the magnitudes of the relatively global errors are found to decrease by a factor greater than $3$ with the use of RC-LBM when compared to RNR-LBM.
\begin{table}[h]
\small
\centering
\captionsetup{justification=centering}
\caption{Comparison of the relative global errors resulting from using RNR-LBM and RC-LBM using $a=0.5$ with $N=100,200$, and $400$ grid nodes in the wall normal direction for 2D channel flow at $\mbox{Re=50}$.}
\begin{tabular}{c c c}
\hline
\hline
 Grid resolution &   Relative global error &   Relative global error\\
   & RNR-LBM  &  RC-LBM \\
 \hline\hline
  $100$ & $7.90 \times 10^{-4}$ &$2.50 \times 10^{-4}$ \\
 $200$ & $6.25 \times 10^{-4}$ & $1.92 \times 10^{-4}$ \\
 $400 $& $4.74\times 10^{-5}$ & $1.56\times 10^{-5}$ \\

\hline
\end{tabular}
\label{tab:2}
\end{table}

\subsection {Transient shear driven flow between two parallel plates}
\begin{figure}[H]
    \centering
    \captionsetup{justification=centering}
   \includegraphics[scale=0.7]{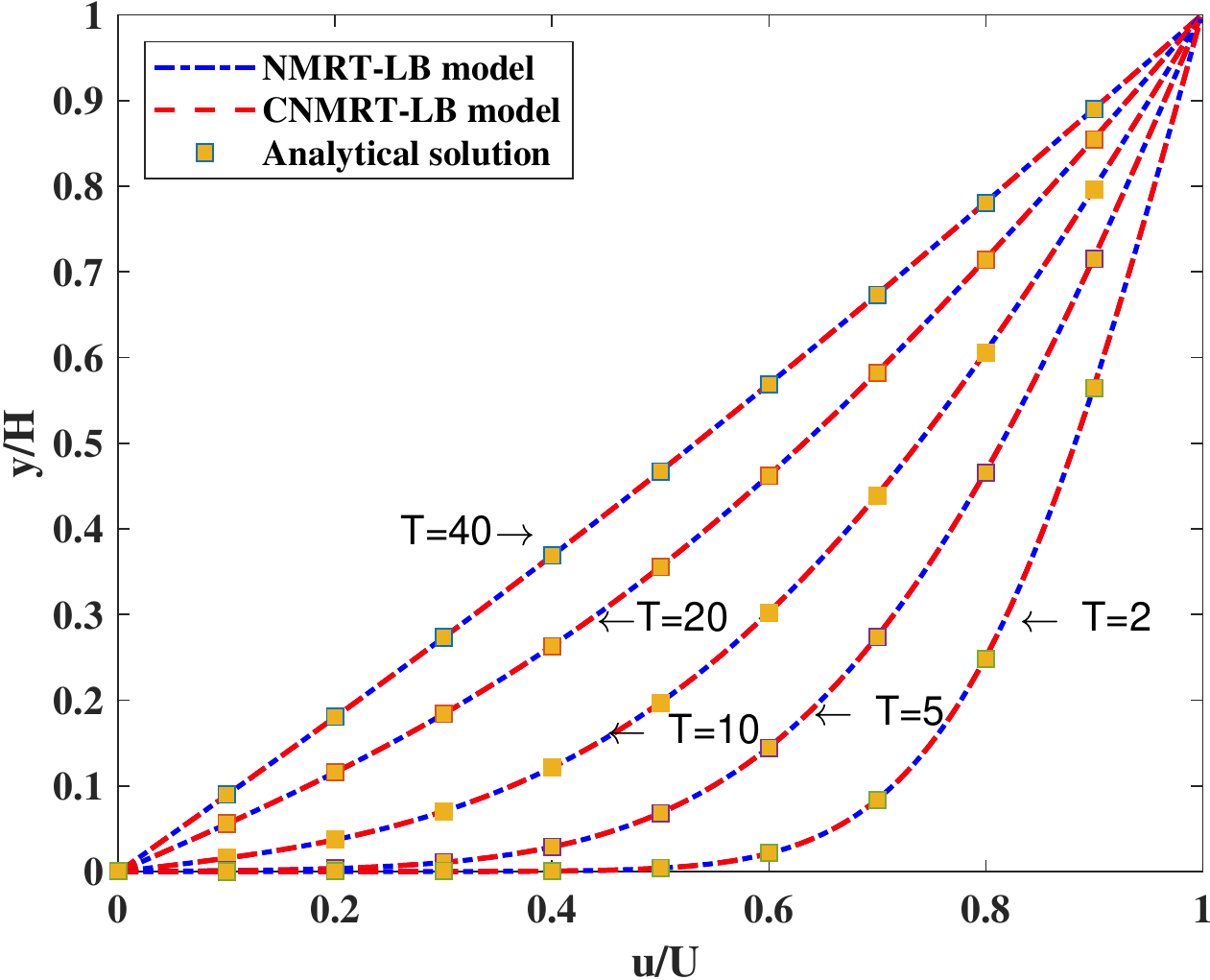}
    \caption{Comparison of the instantaneous velocity profiles computed using RNR-LBM and RC-LBM at a grid aspect ratio of $a=0.1$ against the analytical solution at time instants of $T=2, 5, 10,20$ and $40$ for 2D shear driven flow between two parallel plates.}
    \label{fig:4}
 \end{figure}

The second test problem is a transient flow between parallel plates with a spacing of $H$ driven by shear due to the motion of the upper plate with a constant velocity $U$. The time-dependent analytical solution of the velocity profile $u(y,t)$ generated between the two plates is given by~\cite{pozrikidis2011introduction}
\begin{equation*}
 u(y,t)= U \frac{y}{H} - \frac{2U}{\pi} \sum_{n=1}^{\infty} \frac{1}{n} \exp{\left[-\frac{n^2 \pi^2 \nu t}{H^2}\right]}\sin
 \left[ n\pi \left(1-y/H\right)\right],
\end{equation*}
where the wall-normal coordinate distance $y$ is measured from the lower plate. The characteristic time scale for this problem is $T^*=H^2/\nu$ and thus the dimensionless time $T$ can be defined as $T=t/T^*$. We consider a grid resolution of $50\times 500$ with a grid aspect ratio $a=0.1$ to resolve the
domain, in which the fluid is initially at rest and the upper plate is set in motion with a velocity $U=0.02$. The relaxation time $\tau$ is chosen to
be $0.8$. As in the previous case, periodic boundary conditions are employed in the flow direction. The motion of the top wall is accounted for in the no-slip boundary condition via applying the momentum augmented half-way bounce back scheme. In the derivation of the halfway bounce back scheme for the rectangular grid, since the momentum appears in the moment equilibria, when it is mapped back to the distribution functions via the inverse of the transformation matrix, i.e. $\tensor{T}^{-1}$, the resulting formulas will be parameterized by the grid aspect ratio $a$, in addition to the plate velocity $U$ and the speed of sound $c_s$. That is, if $\bm{x}_f$ is a fluid node nearest to the wall and the opposite particle directions are represented by $\bm{e}_{\overline{i}}=-\bm{e}_i$, then the incoming particle distribution functions $f_i$ are obtained from the outgoing post-collision particle distribution functions $\widetilde{f}_{\overline{i}}$ by $f_i(\bm{x}_f,t+\Delta t)=\widetilde{f}_{\overline{i}}(\bm{x}_f,t)-(f^{eq}_{\overline{i}}(\rho_w,U))-f^{eq}_i(\rho_w,U)$, where $\rho_w$ is the density of fluid at the wall. Evaluating these for the incoming directions associated with the top plate, i.e., $i=\{4,7,8\}$ and using $\mathbf{f^{eq}} =\tensor{T^{-1}}\mathbf{m^{eq}}$ based on the wall conditions, we get
\begin{subequations}
\begin{eqnarray} \label{eq:57}
 f_4(\bm{x}_f, t+\Delta{t})&=&f_2(\bm{x}_f,t),\label{eq:57} \\
 f_7(\bm{x}_f, t+\Delta{t})&=&f_5(\bm{x}_f,t) -\frac{ \rho_w c_s^2  U}{2 a^2},\label{eq:58} \\
 f_8(\bm{x}_f, t+\Delta{t})&=&f_6(\bm{x}_f,t)+\frac{\rho_w c_s^2 U}{2 a^2}.\label{eq:59}
\end{eqnarray}
\end{subequations}
These expressions, which parameterize the influence of the rectangular lattice via $a$, are simpler than those presented in Ref.~\cite{zong2016designing} due to the use of non-orthogonal moment basis and the construction of the equilibria directly from the Maxwell distribution function via matching without the use of free parameters in this work. Based on these and performing simulations by considering the fluid to be initially at rest, Fig.~\ref{fig:4} presents the instantaneous velocity profiles $u(y,t)$ computed using the RNR-LBM and RC-LBM at various dimensionless time instants $T$ of 2, 5, 10, 20 and 40, along with the analytical solution plotted at all these time instants. Very good agreement between the two rectangular LB schemes and the time-dependent analytical solution are found. The temporal development of the velocity field leading to a steady state linear profile is well reproduced by both RNR-LBM and RC-LBM.

\subsection {Pulsatile flow between two parallel plates driven by a periodic body force}\par
\begin{figure}[H]

    \centering
    \captionsetup{justification=centering}
    \includegraphics[scale=0.7]{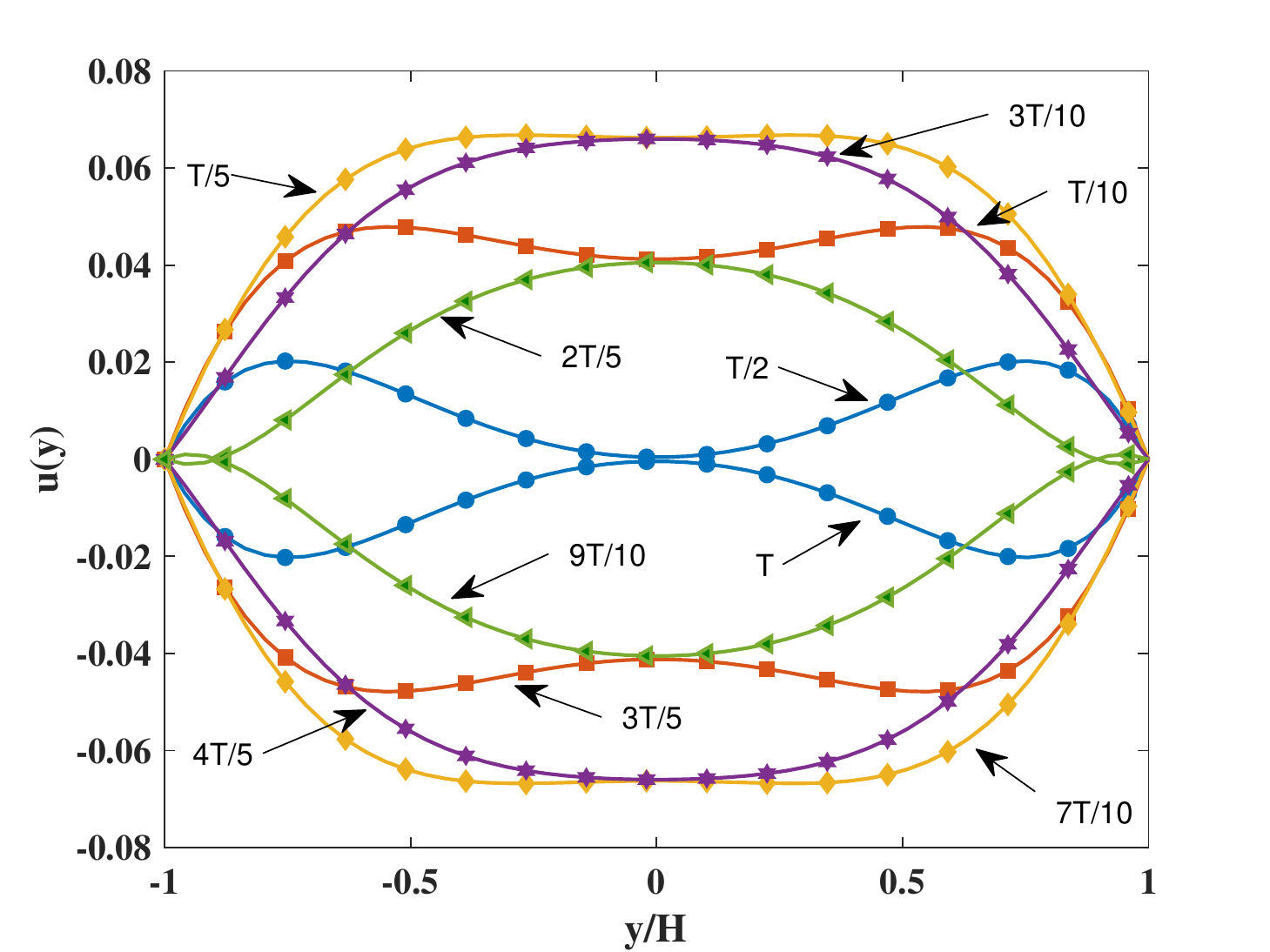}
    \caption{Comparison of the velocity profiles computed using the RC-LBM (lines) with a grid aspect ratio $a=0.125$ with the analytical solution (open symbols) for 2D Womersley flow at different instants within a time period $T=40,000$ and $\mbox{Wo}=4.25$.}
    \label{fig:5}
 \end{figure}

In addition, we now examine our rectangular LB approach for their validation involving flow bounded by two parallel plates with a separation distance of $H$ and subjected to a sinusoidally time varying body force $F_x(t)$, viz., for the classical Womersley flow. The imposed body force is represented as $F_x=F_m \cos \varpi t$, where $\varpi$ is the angular frequency given by $\varpi=2\pi/T$ with $T$ being the time period and $F_m$ is the peak amplitude of the force. The analytical solution for this flow reads as
\begin{equation*}
u_x(y)= \mathcal{R} \Big[ \frac{F_m i}{\varpi} (1- e^{i \varpi t}) \frac{\cos \frac{\beta y}{H}}{\cos \beta} \Big],
\end{equation*}
where $\mathcal{R}[\cdot]$ implies taking the real part of the expression, $\beta = \sqrt{-i\mbox{Wo}^2}$ and $\mbox{Wo}$ is the Womersley number is related to the ratio of the viscous flow time scale and the time scale of imposed force variations, i.e., $\mbox{Wo}= \sqrt{\varpi/\nu}H$. Figure~\ref{fig:5} shows a comparison of the velocity profiles obtained using the RC-LBM for simulation of Womersley flow at ${Wo}=4.25$, $F_m=1 \times 10^{-5}$, $T=40,000$ and using the relaxation time $\tau=1.0$. We considered a grid resolution of 400 grid points along the wall normal direction choosing $H=50$, which means the rectangular grid has an aspect ratio of $a=0.125$. Since it was found that the results of RNR-LBM are visually indistinguishable from those obtained using RC-LBM, only the latter results are presented in this figure. Clearly, the spatio-temporal variations in the
velocity profiles at various instants within the time period shown by the analytical solution are well reproduced by our new rectangular LB scheme based on central moments.

\subsection {Lid-driven cavity flow}
Finally, we will now present simulations of the flow inside a square cavity driven by the shear from the motion of the top lid. It is a standard flow problem for benchmarking new CFD methods by going beyond the use of analytical solutions and involves flow patterns characterized by the presence of a main or primary vortex around center of the cavity and accompanied by various secondary vortices around the corners, whose sizes and locations depend on the Reynolds number $\mbox{Re}$. Prior rectangular LB schemes based on either SRT or orthogonal MRT formulations have been used to perform simulations of this problem~\cite{bouzidi2001lattice,zhou2012mrt,zong2016designing,peng2016hydrodynamically,peng2016lattice}, which have reported results for only relatively low $\mbox{Re}$ as they were often subjected to numerical instability issues when the Reynolds number was increased to even moderate values. Hence, it would be interesting to study the performance of the present RNR-LBM and RC-LBM for this case study. For a square cavity of side $H$ whose top lid is set into a uniform motion at a velocity $U$, the characteristic Reynolds number can be defined as $\mbox{Re}=UH/\nu$. We carried out simulations of this flow problem using both RNR-LBM and RC-LBM involving rectangular lattices with grid aspect ratios of $a=4, 2, 0.8, 0.5, 0.25$, and $0.2$ at a wide range of Reynolds numbers of 100, 400, 3200, 5000 and 7500 covering those reported in a prior work providing benchmark numerical solutions based on a NS solver~\cite{ghia1982high}. The choices of the various parameters considered for these simulations are given in Table~\ref{tab:3}. The half-way bounce back scheme was used to impose the no-slip conditions on all the bounding walls, with momentum correction being added to those for the top wall via Eqs.~(\ref{eq:57})-(\ref{eq:59}) constructed for use with the rectangular lattice.
\begin{table}[H]
\small
\centering
\captionsetup{justification=centering}
\caption{Parameters used in the simulation of lid-driven cavity flow.}
\begin{tabular}{c c c c c c c}
\hline
\mbox{Re} & $a$ &  $N_x \times N_y$   & U &  $c_s^2$  & $\nu$ & $\tau$ \\
\hline
\hline
\multirow{3}{4em}{100}
& 2 & $100\times 50$ & 0.1 & 0.3333 & 0.099 & 0.797\\
& 4 & $200 \times 50$ & 0.1 & 0.3333 & 0.199 & 1.097 \\
& 0.8 & $100\times 125$ & 0.1 & 0.3333 & 0.099 & 0.797\\
& 0.5 & $100 \times 200$ & 0.04 & 0.18 & 0.042 & 0.736\\
& 0.25 & $ 100 \times 400$ & 0.02 & 0.04 & 0.02 & 0.995\\
& 0.2 & $100 \times 500$ & 0.02 & 0.02 & 0.02 & 1.49\\
\hline
\multirow{2}{4em}{400} & 0.25 & $100 \times 400$ & 0.02 & 0.04 & 0.0049 & 0.623\\
& 0.2 & $100 \times 500$ & 0.02 & 0.02 & 0.0049 & 0.747\\
\hline
\multirow{1}{4em}{1000} & 0.5 & $150 \times 300$ & 0.02 & 0.04 & 0.0003 & 0.5745\\

\hline
\multirow{1}{4em}{3200} & 0.5 & $ 150 \times 300$ & 0.02 & 0.02 & 0.0009 & 0.5468\\
\hline
\multirow{1}{4em}{5000} & 0.5 & $150 \times 300$ & 0.02 & 0.02 & 0.0006 & 0.53\\
\hline
\multirow{1}{4em}{7500} & 0.5 & $150 \times 300$ & 0.05 & 0.03 & 0.0099 & 0.533\\
\hline
\end{tabular}
\label{tab:3}
\end{table}

\begin{figure}[H]
\centering
\advance\leftskip-1.7cm
    \subfloat[] {
        \includegraphics[scale=0.6] {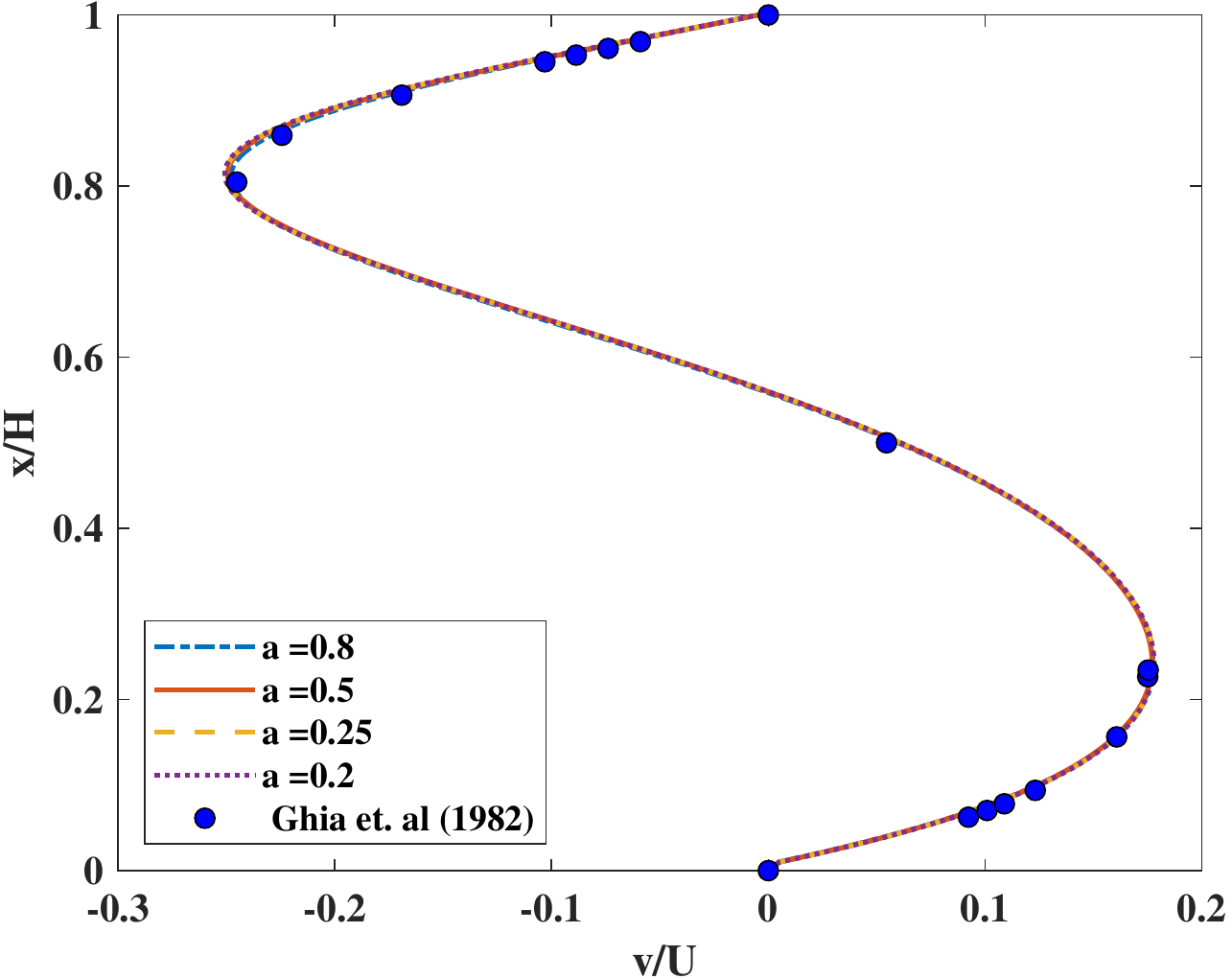}
        \label{fig:6a} } 
    \subfloat[] {
        \includegraphics[scale=0.6] {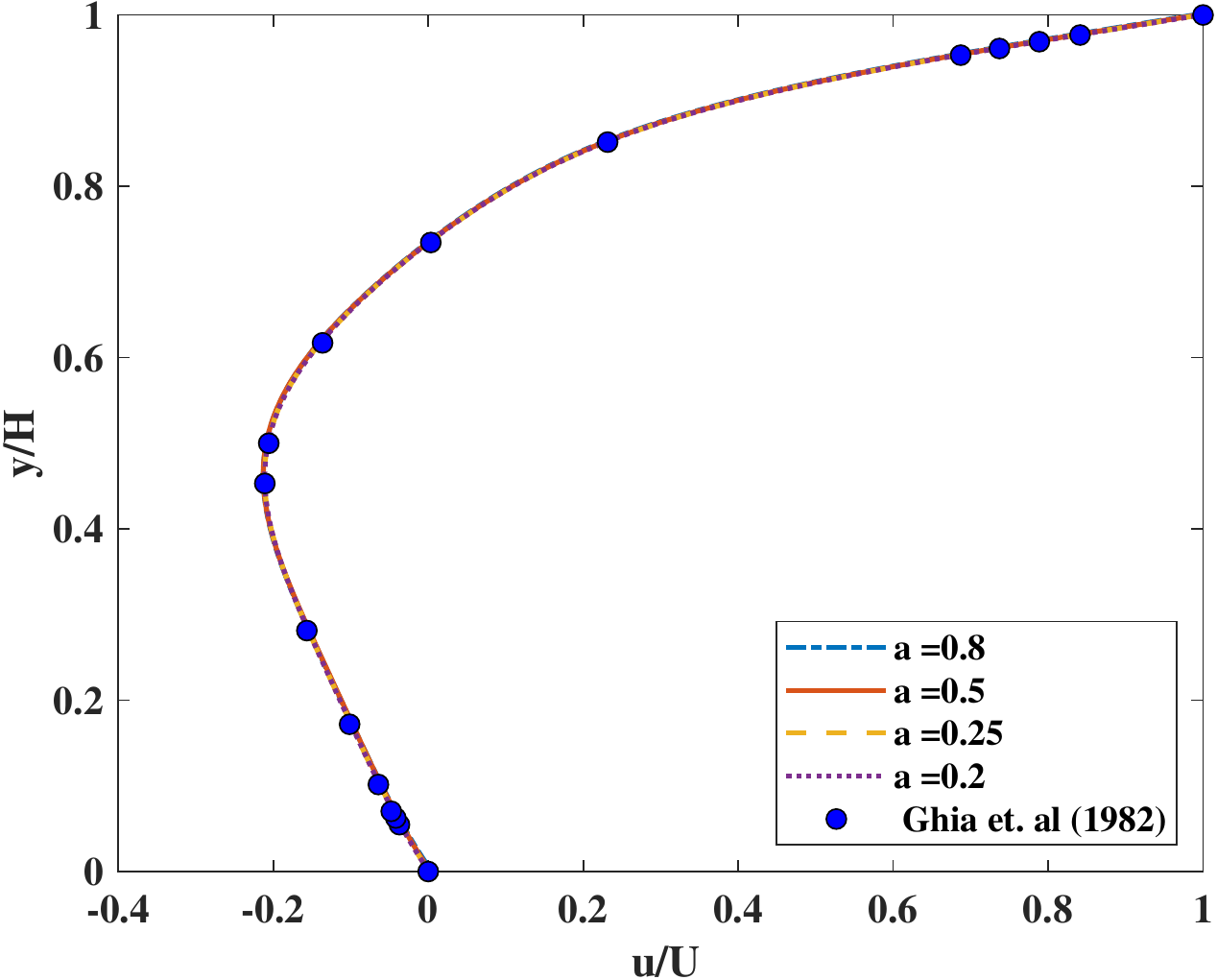}
        \label{fig:6b} } \\
                \advance\leftskip0cm
    \caption{Velocity profiles (a) along the vertical centerline of the lid driven cavity flow for the $u$ component and (b) along the horizontal centerline for the $v$ component computed using RNR-LBM with grid aspect ratios of $a=0.8, 0.5, 0.25$ and $0.2$ compared with the benchmark solution of Ghia \emph{et al}~\cite{ghia1982high} (symbols) at $\mbox{Re}=100$.}
    \label{fig:6}
\end{figure}

\begin{figure}[H]
\centering
\advance\leftskip-1.7cm
    \subfloat[] {
        \includegraphics[scale=0.6] {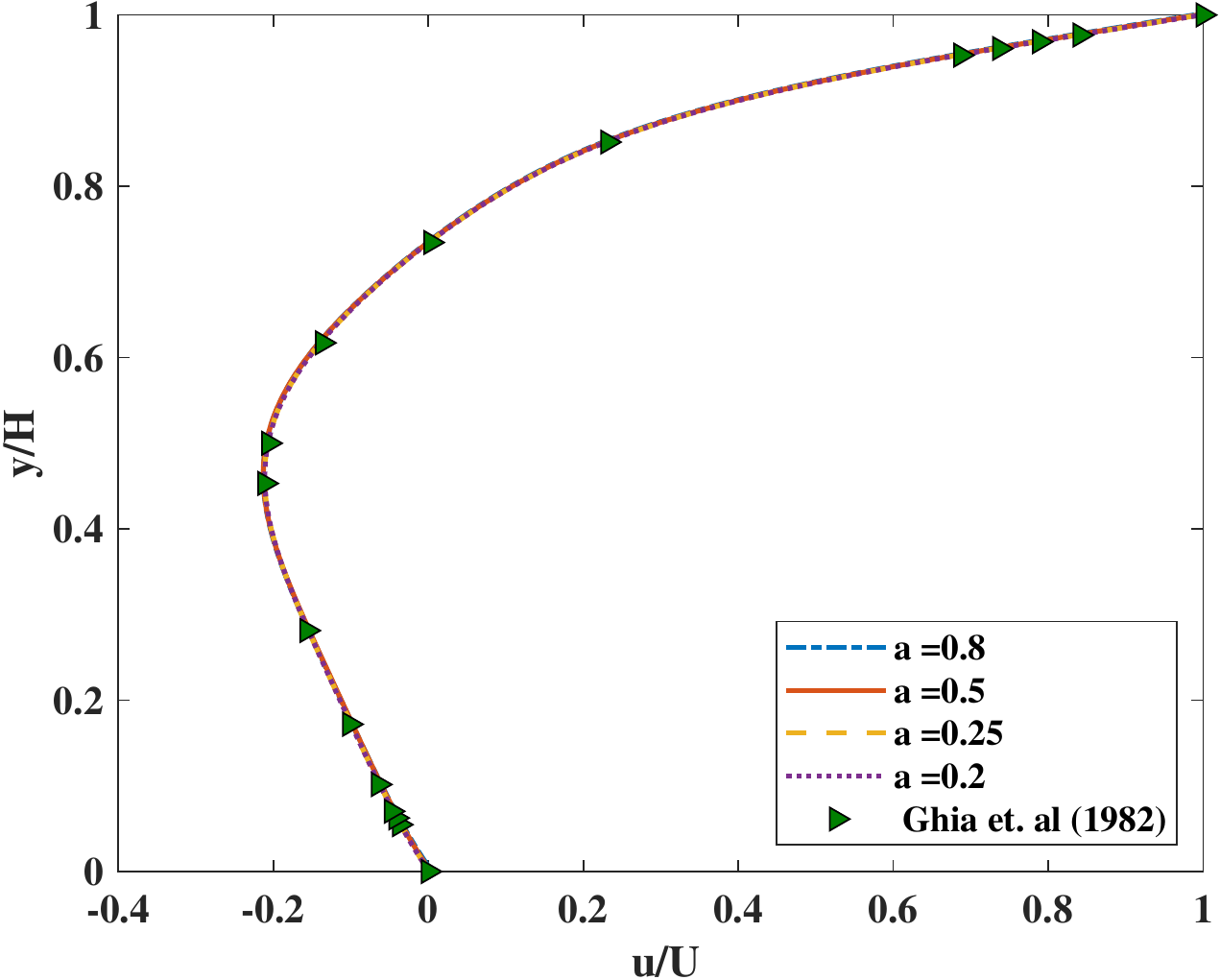}
        \label{fig:7a} } 
    \subfloat[] {
        \includegraphics[scale=0.6] {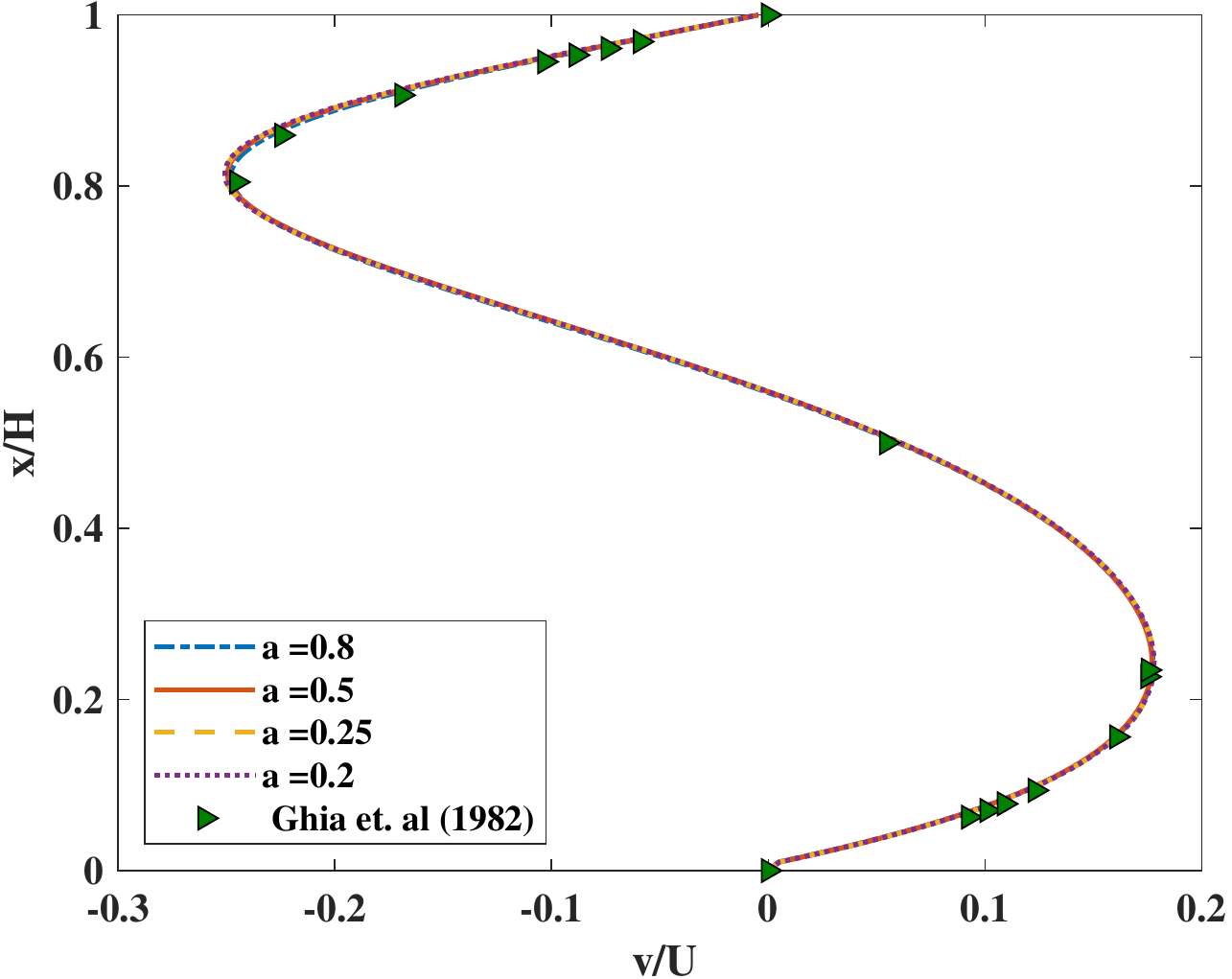}
        \label{fig:7b} } \\
                \advance\leftskip0cm
    \caption{Velocity profiles (a) along the vertical centerline of the lid driven cavity flow for the $u$ component and (b) along the horizontal centerline for the $v$ component computed using RC-LBM with grid aspect ratios of $a=0.8, 0.5, 0.25$ and $0.2$ compared with the benchmark solution of Ghia \emph{et al}~\cite{ghia1982high} (symbols) at $\mbox{Re}=100$.}
    \label{fig:7}
\end{figure}

\begin{figure}[H]
\centering
\advance\leftskip-1.7cm
    \subfloat[] {
        \includegraphics[scale=0.6] {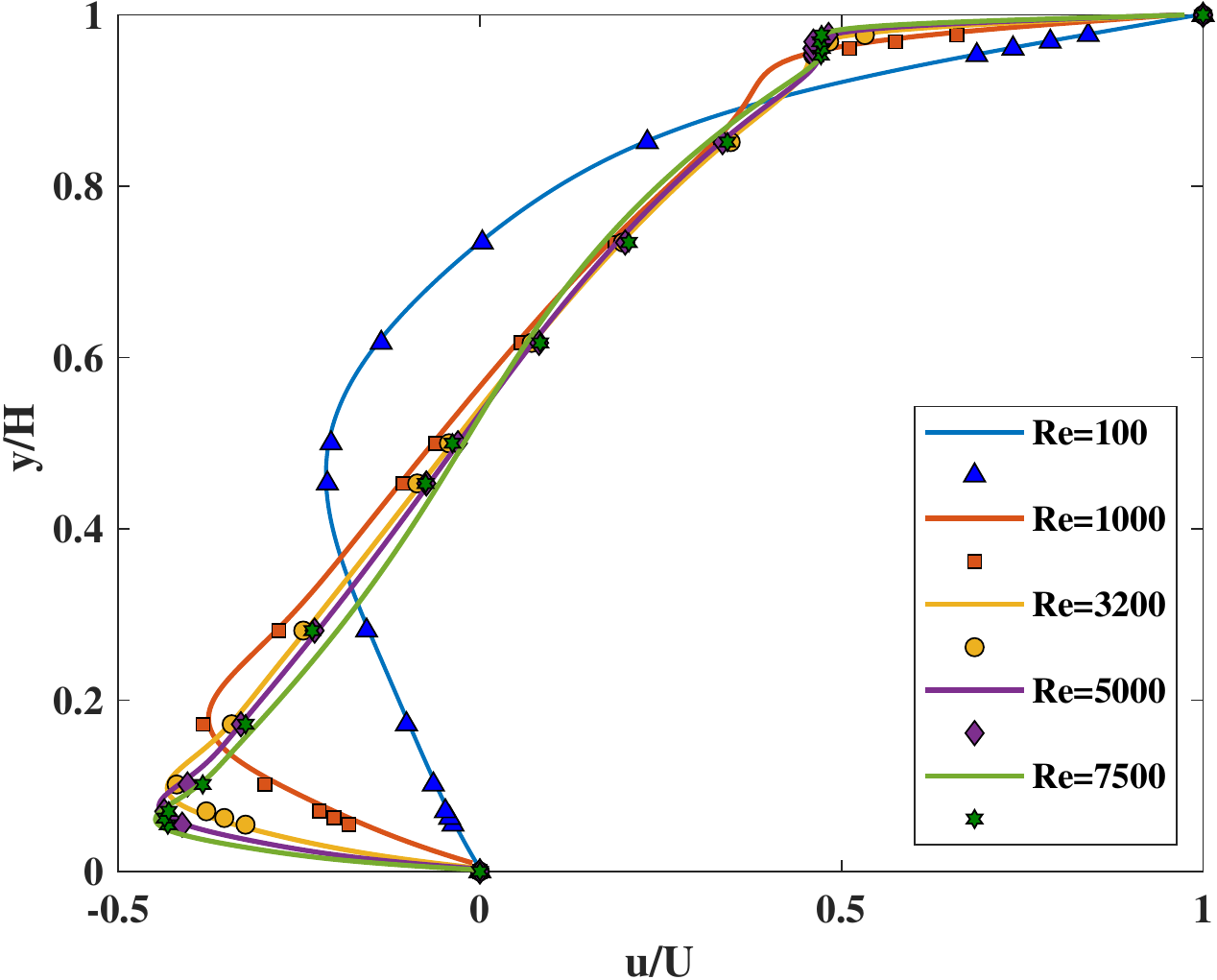}
        \label{fig:8a} } 
    \subfloat[] {
        \includegraphics[scale=0.6] {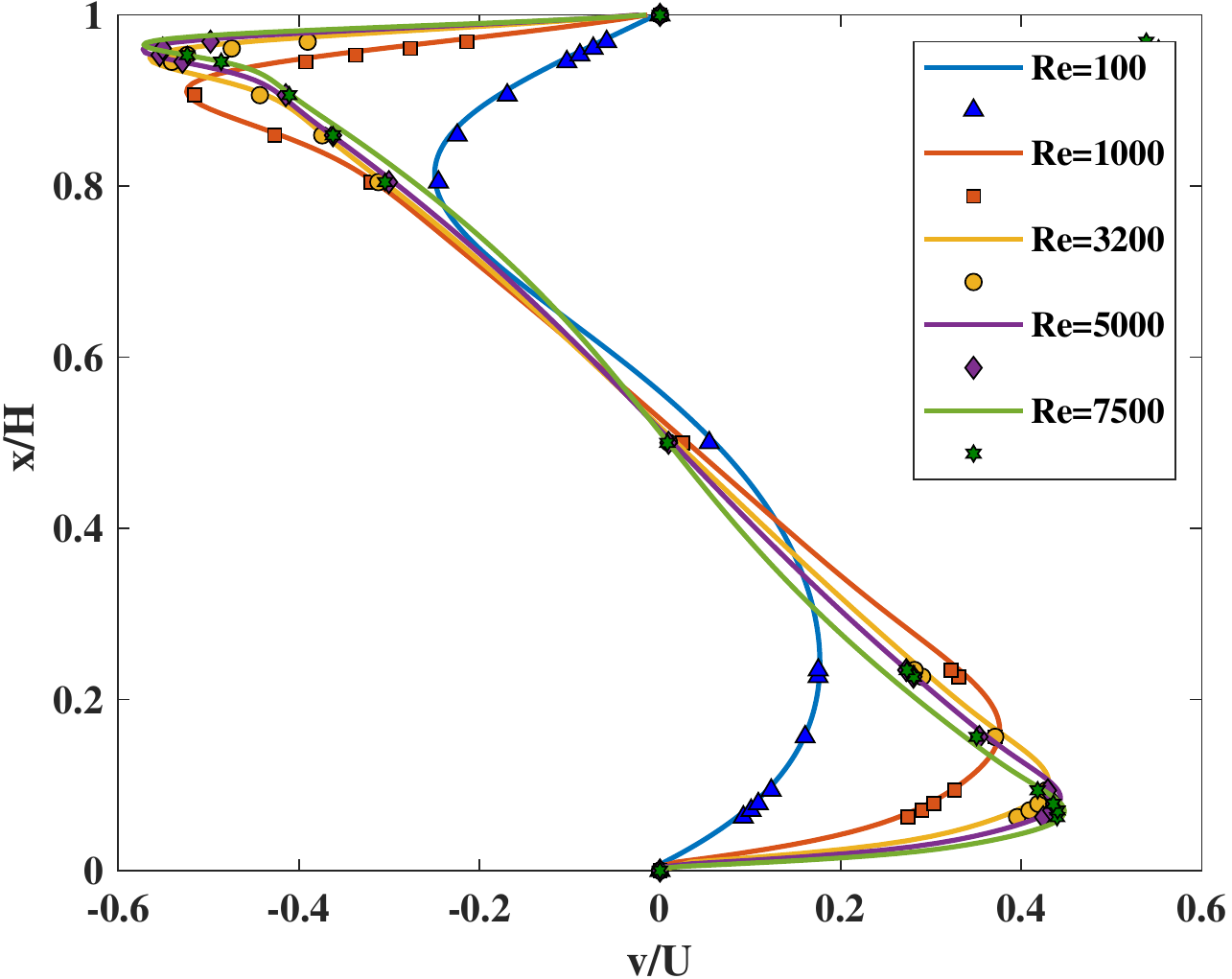}
        \label{fig:8b} } \\
                \advance\leftskip0cm
    \caption{Velocity profiles (a) along the vertical centerline of the lid driven cavity flow for the $u$ component and (b) along the horizontal centerline for the $v$ component computed using RNR-LBM with a grid aspect ratio of $a=0.5$ at different Reynolds numbers of $\mbox{Re} = 100, 1000, 3200, 5000$ and $7500$ compared with the benchmark solution of Ghia \emph{et al}~\cite{ghia1982high} (symbols).}
    \label{fig:8}
\end{figure}

\begin{figure}[H]
\centering
\advance\leftskip-1.7cm
    \subfloat[] {
        \includegraphics[scale=0.6] {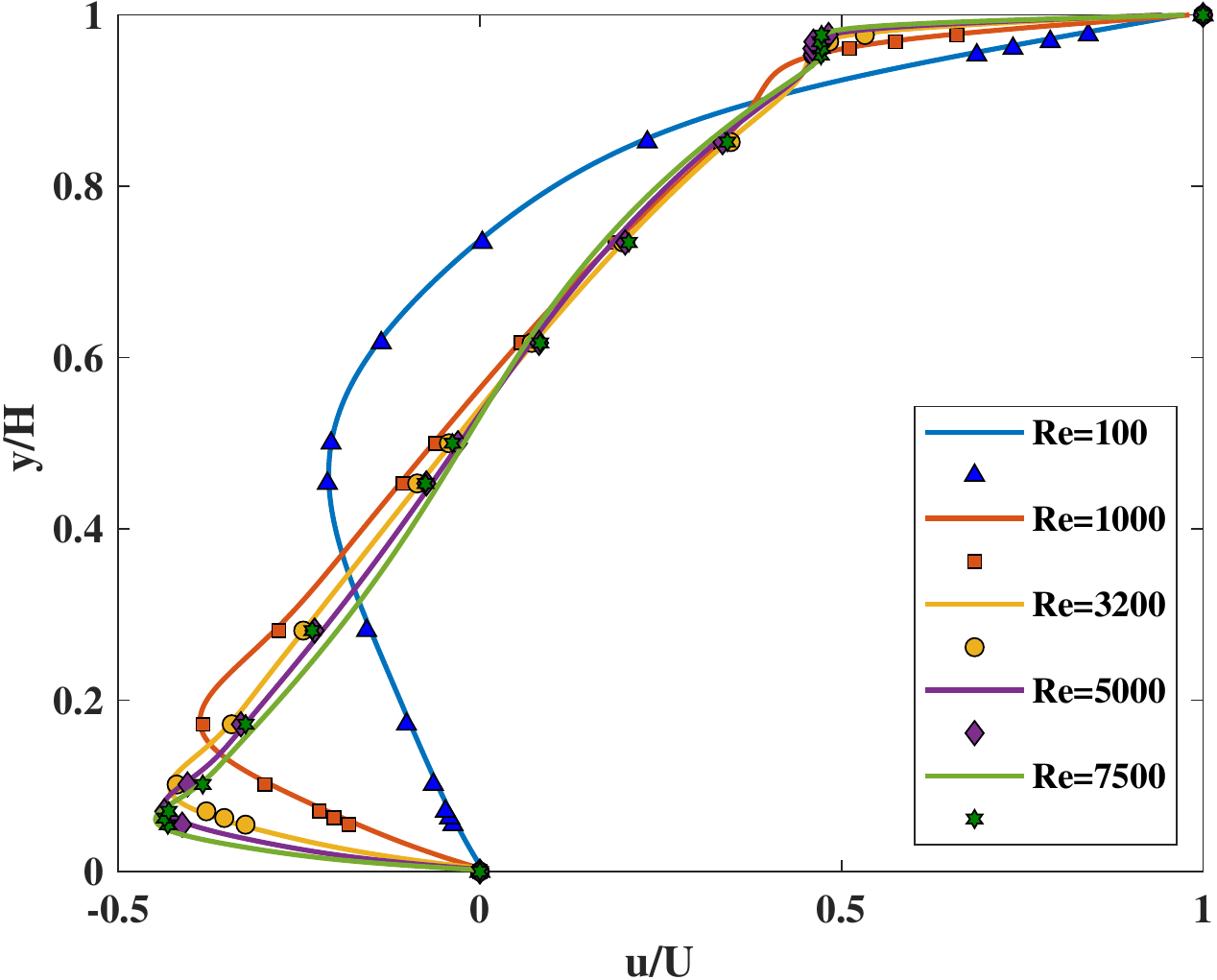}
        \label{fig:9a} } 
    \subfloat[] {
        \includegraphics[scale=0.6] {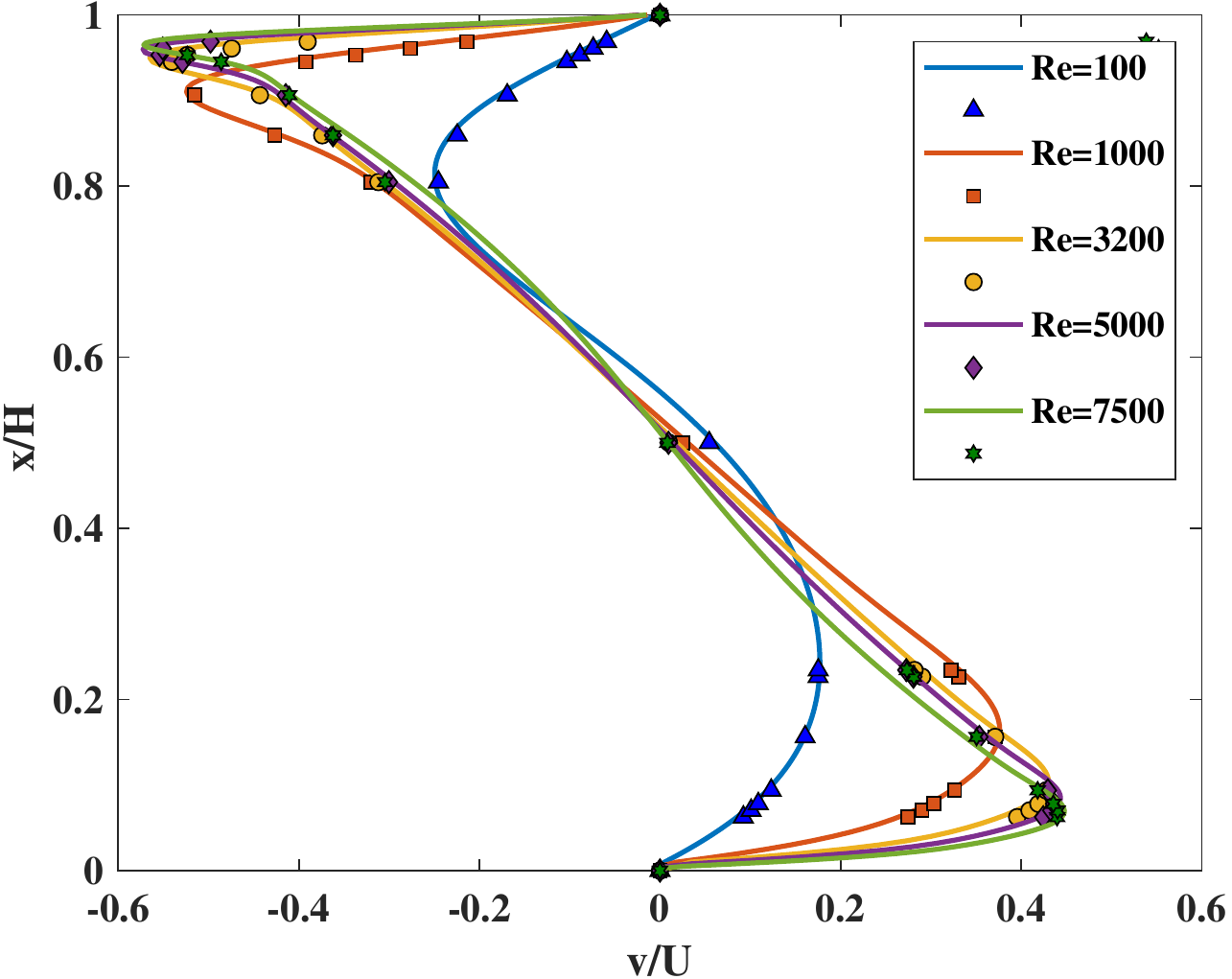}
        \label{fig:9b} } \\
                \advance\leftskip0cm
    \caption{Velocity profiles (a) along the vertical centerline of the lid driven cavity flow for the $u$ component and (b) along the horizontal centerline for the $v$ component computed using RC-LBM with a grid aspect ratio of $a=0.5$ at different Reynolds numbers of $\mbox{Re} = 100, 1000, 3200, 5000$ and $7500$ compared with the benchmark solution of Ghia \emph{et al}~\cite{ghia1982high} (symbols).}
    \label{fig:9}
\end{figure}

\begin{figure}[h]
\centering
\advance\leftskip-1.7cm
    \subfloat[] {
       \includegraphics[scale=0.6] {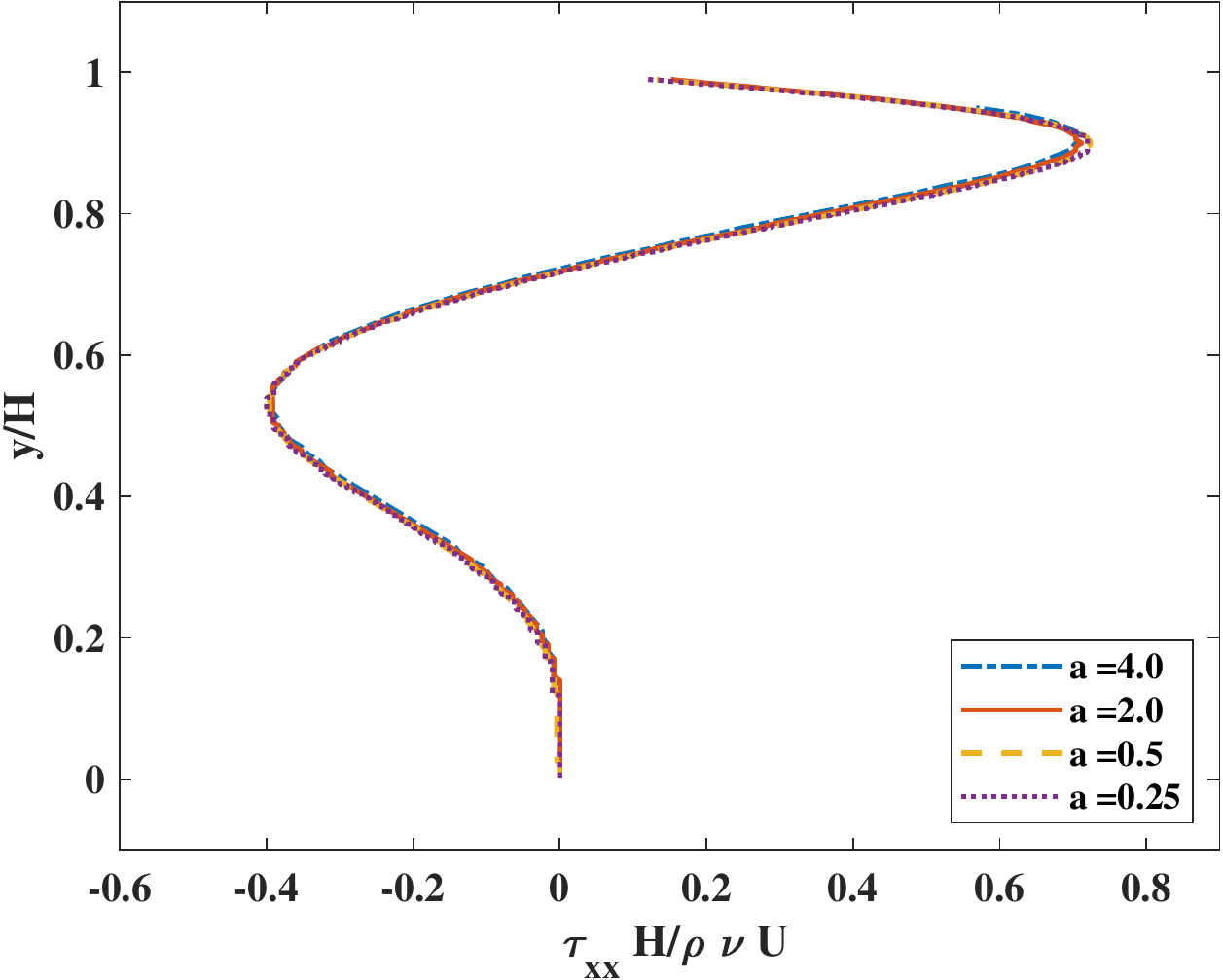}
        \label{fig:10a} } 
    \subfloat[] {
        \includegraphics[scale=0.6] {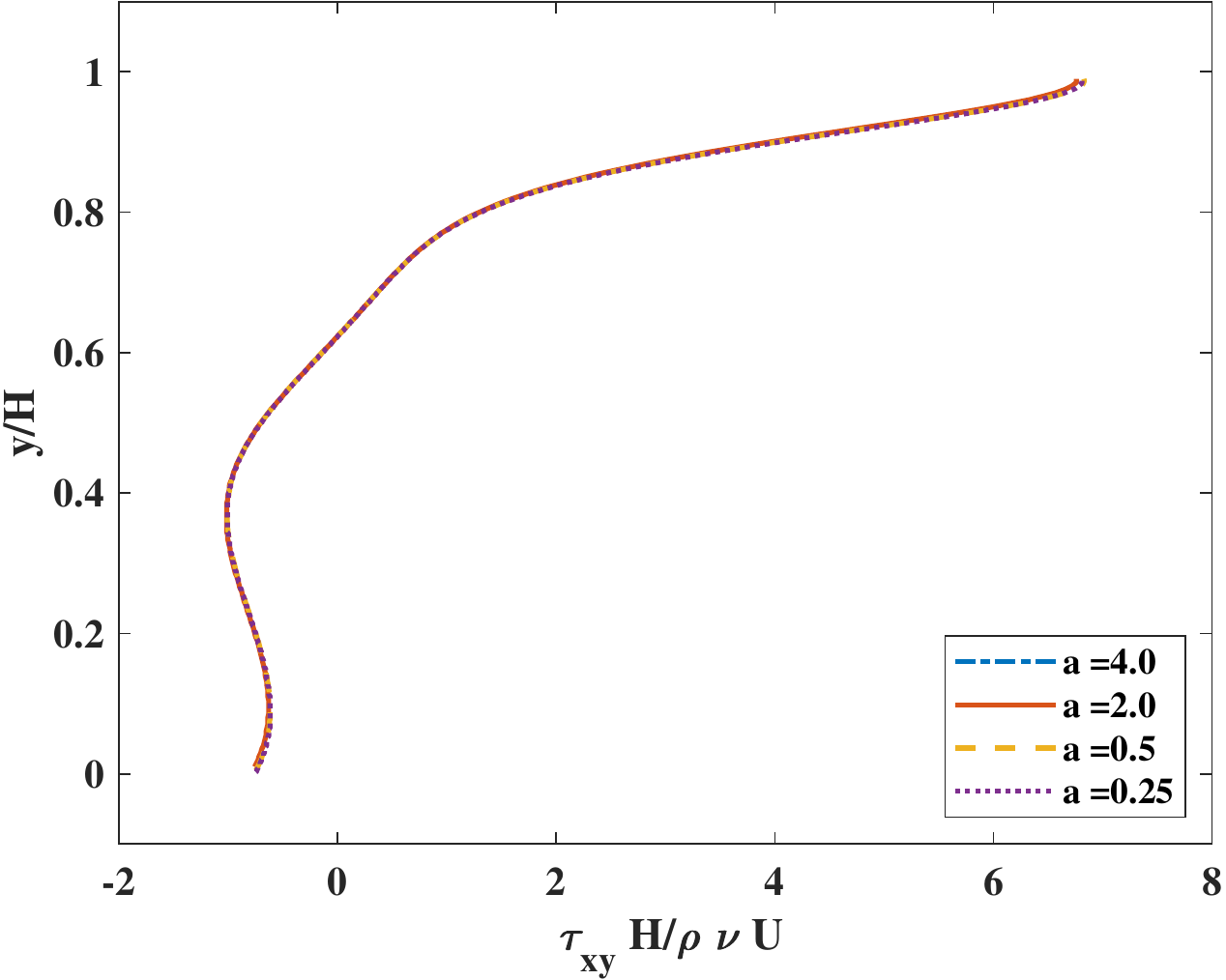}
        \label{fig:10b} } \\
                \advance\leftskip0cm
    \caption{The normal stress profile (a) $\tau_{xx}$ and shear stress profile (b) $\tau_{xy}$ along the vertical centerline of the lid driven cavity flow computed using RC-LBM with grid aspect ratios of $a =4.0, 2.0, 0.5$, and $0.25$ at $\mbox{Re}=100$.}
    \label{fig:10}
\end{figure}

Figures~\ref{fig:6} and \ref{fig:7} show the velocity profiles along the vertical and horizontal centerlines computed using
RNR-LBM and RC-LBM, respectively, with $a=0.8, 0.5, 0.25$ and $0.2$ at $\mbox{Re}=100$. Good agreement with the benchmark results~\cite{ghia1982high} can be seen for all the choices of the grid aspect ratio.
On the other hand, next we fix the grid aspect ratio and vary the Reynolds number. Figures~\ref{fig:8} and \ref{fig:9} present the velocity profiles along the vertical and horizontal centerlines obtained using RNR-LBM and RC-LBM, respectively, with $a=0.5$ at $\mbox{Re} = 100, 1000, 3200, 5000$ and $7500$. The computed results again match well with those provided in Ref.~\cite{ghia1982high} for all the $\mbox{Re}$ tested. By contrast, it may be noted that Ref.~\cite{zong2016designing} reported that the use of a prior rectangular MRT-LB scheme~\cite{bouzidi2001lattice} became unstable for $\mbox{Re}>1000$, while the more recent rectangular MRT-LB formulations showed results for $\mbox{Re}=100$ in Ref.~\cite{peng2016hydrodynamically} and up to $\mbox{Re=3200}$ in Ref.~\cite{zong2016designing}. All these prior schemes used orthogonal moment basis. Furthermore, a more recent rectangular SRT-LBM also showed results only up to $\mbox{Re}=1000$~\cite{peng2016lattice}. On the other hand, our RNR-LBM and RC-LBM can reach significantly higher values of $\mbox{Re}$, including $7500$, the largest value for which the benchmark data involving steady state results for making comparisons are available~\cite{ghia1982high}. Moreover, in the next section involving a numerical stability study, simulations with even higher Reynolds numbers will be reported.
For further assessment, we investigate the ability of our rectangular LB schemes to evaluate the components of the viscous stress tensor locally. We compute the normal stresses $\tau_{xx}$ and $\tau_{yy}$, and the shear stress $\tau_{xy}$ based on the strain rate tensor components given in Eqs.~(\ref{eq:37})-(\ref{eq:39}) involving the non-equilibrium moments and parameterized by the grid aspect ratio $a$. As illustrated in Fig.~\ref{fig:10} makes a comparison of the normal and shear viscous stress profiles obtained using RC-LBM at $\mbox{Re}=100$ for different choices of the grid aspect ratio, i.e., with $a =4.0, 2.0, 0.8, 0.5$, and $0.25$. The results show that they are consistent to each other for a wide range of the grid aspect ratio.

\begin{figure}[H]
\centering
\advance\leftskip-1.7cm
    \subfloat[Re=100] {
        \includegraphics[width=.40\textwidth] {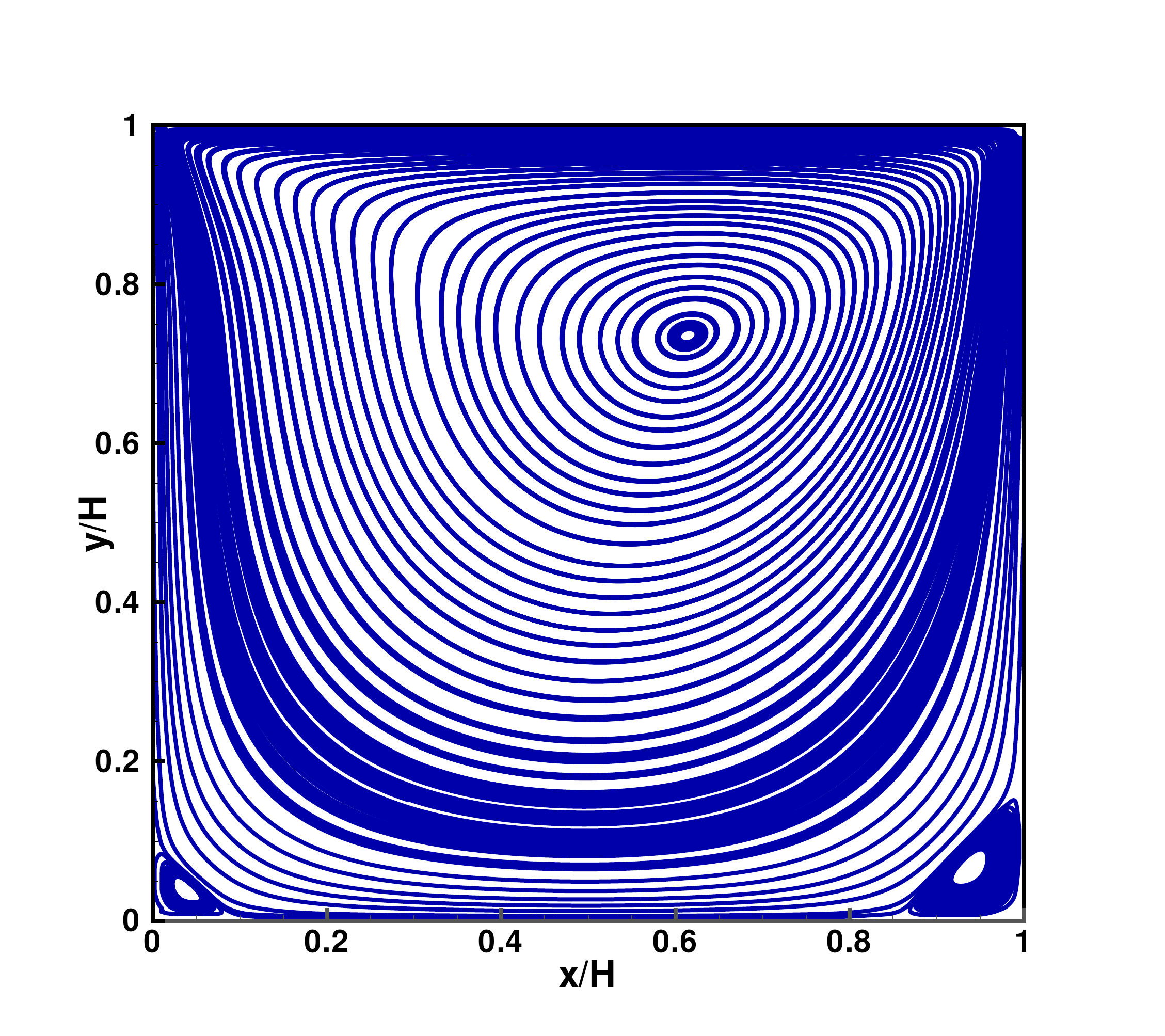}
        \label{fig:12a} } 
    \subfloat[Re=400] {
        \includegraphics[width=.40\textwidth] {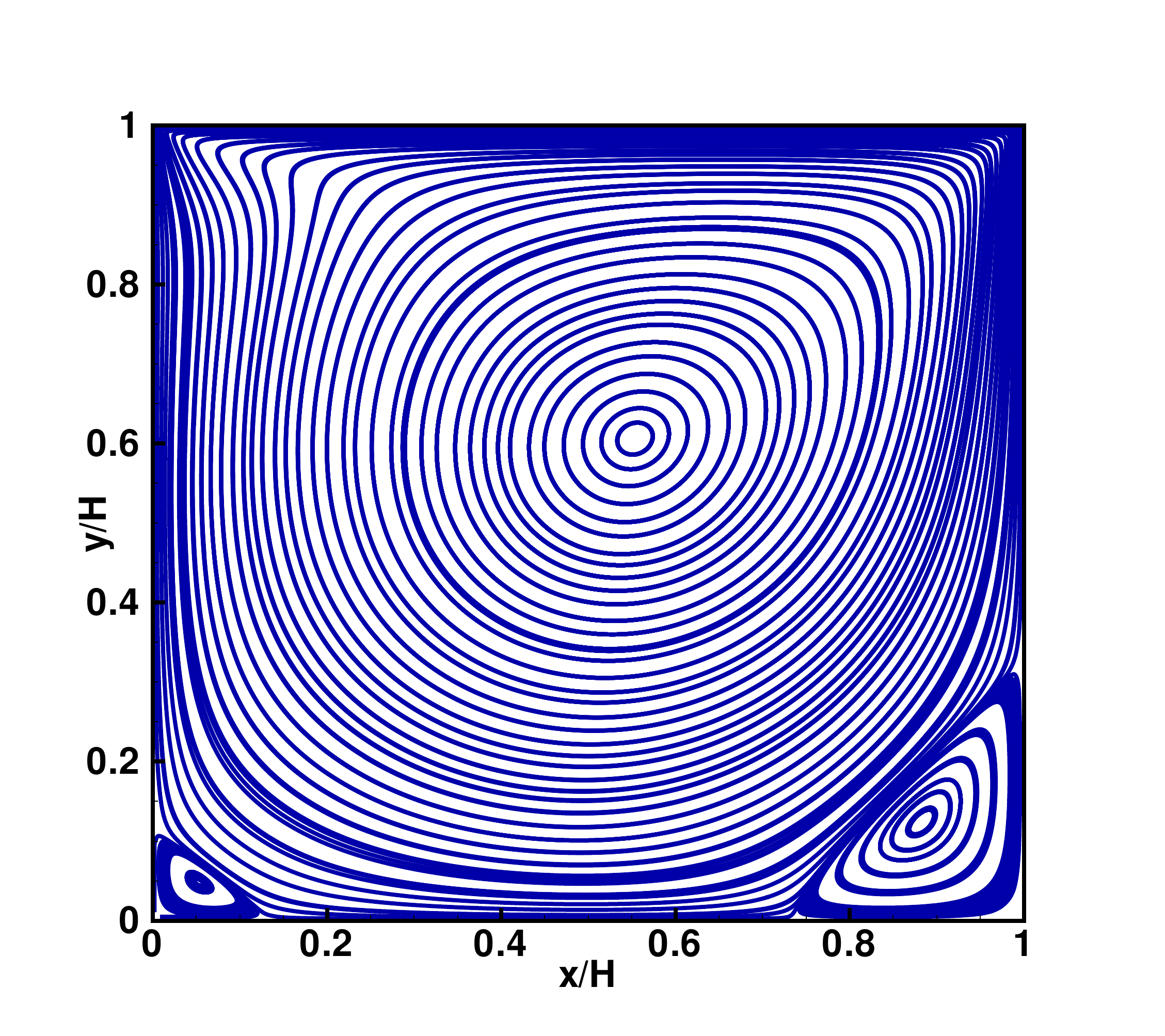}
         \label{fig:12b} } \\

          \subfloat[Re=1000] {
        \includegraphics[width=.40\textwidth] {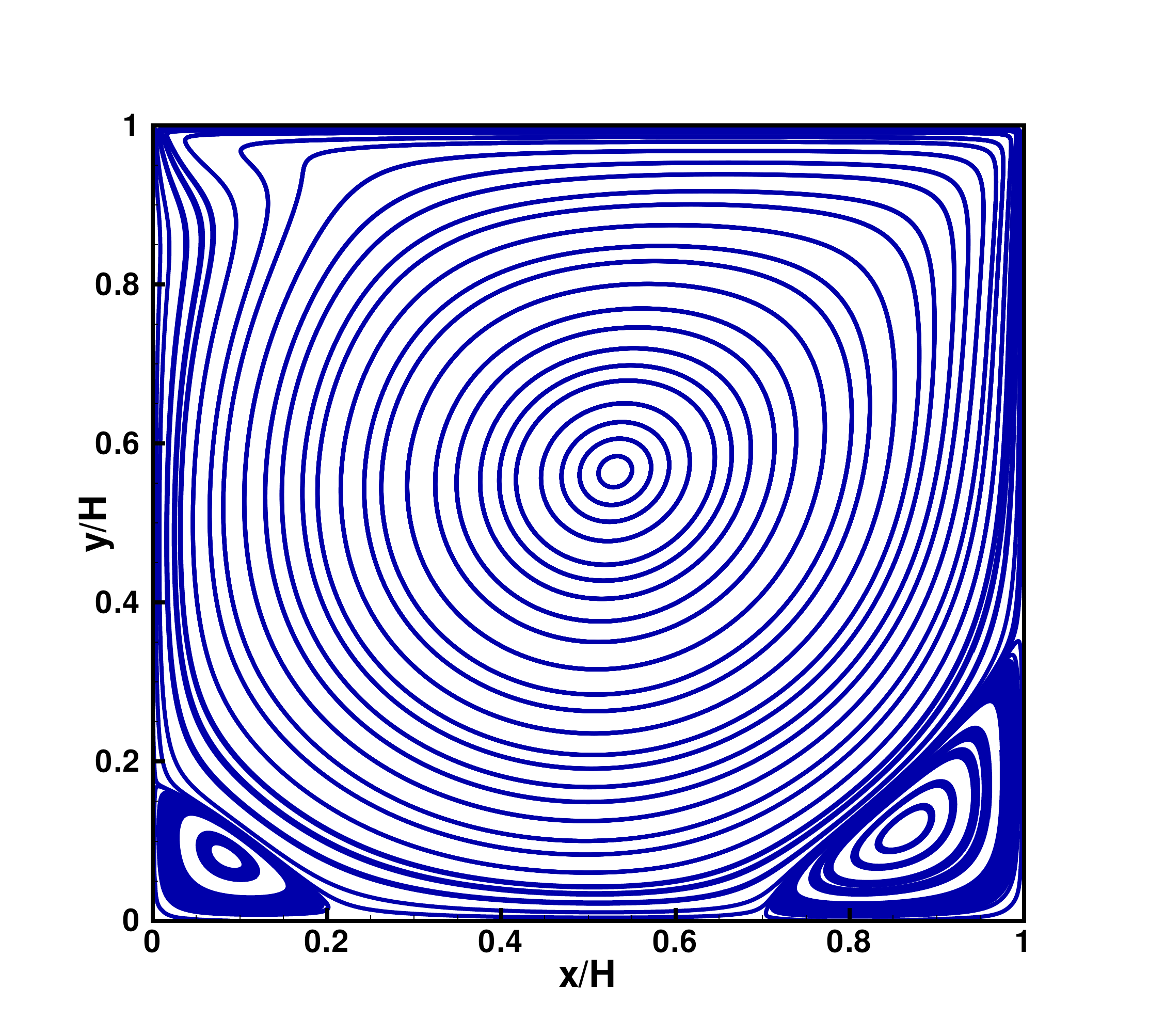}
        \label{fig:12c} } 
    \subfloat[Re=3200] {
        \includegraphics[width=.40\textwidth] {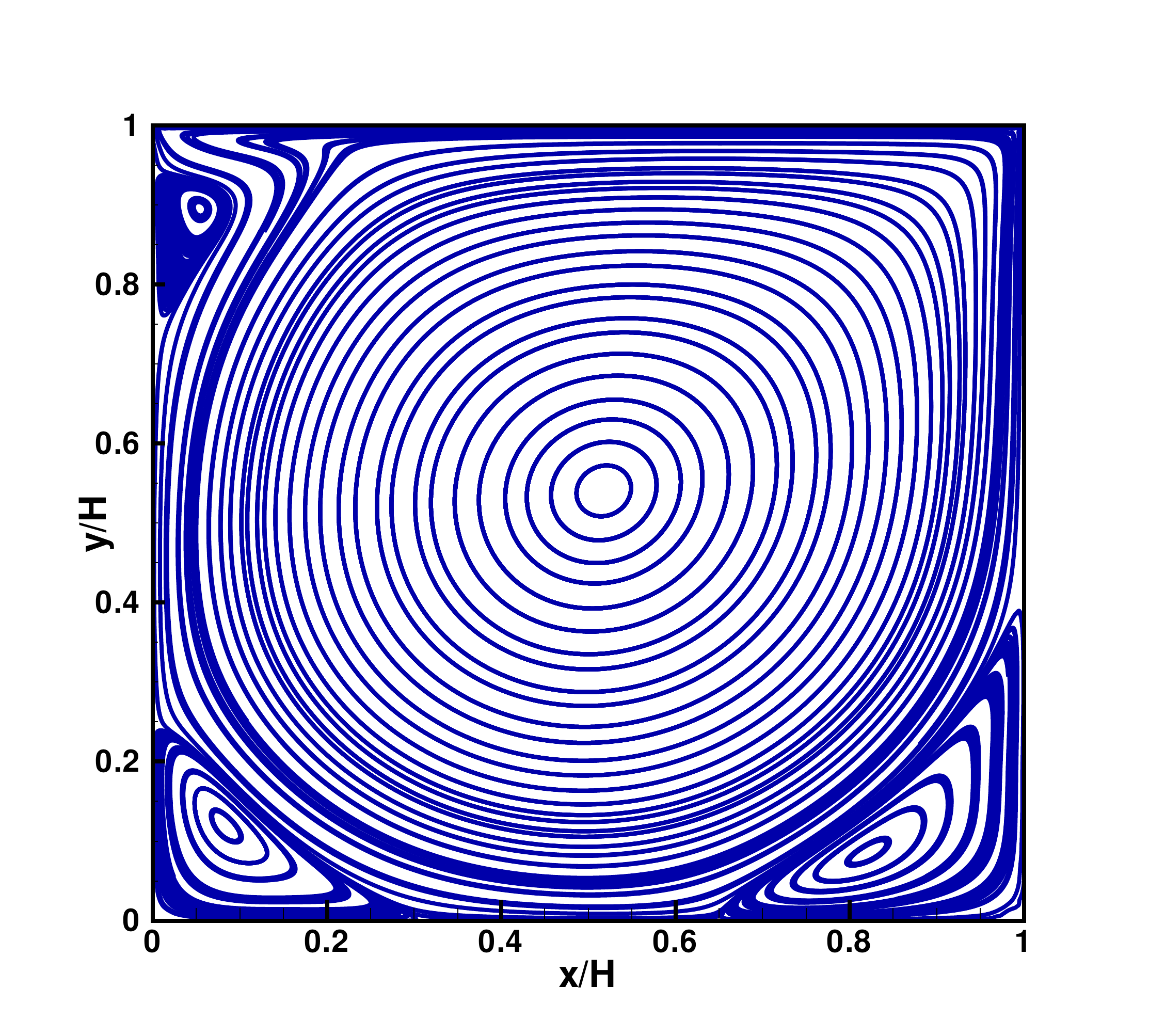}
        \label{fig:12d} } \\
        \subfloat[Re=5000] {
        \includegraphics[width=.40\textwidth] {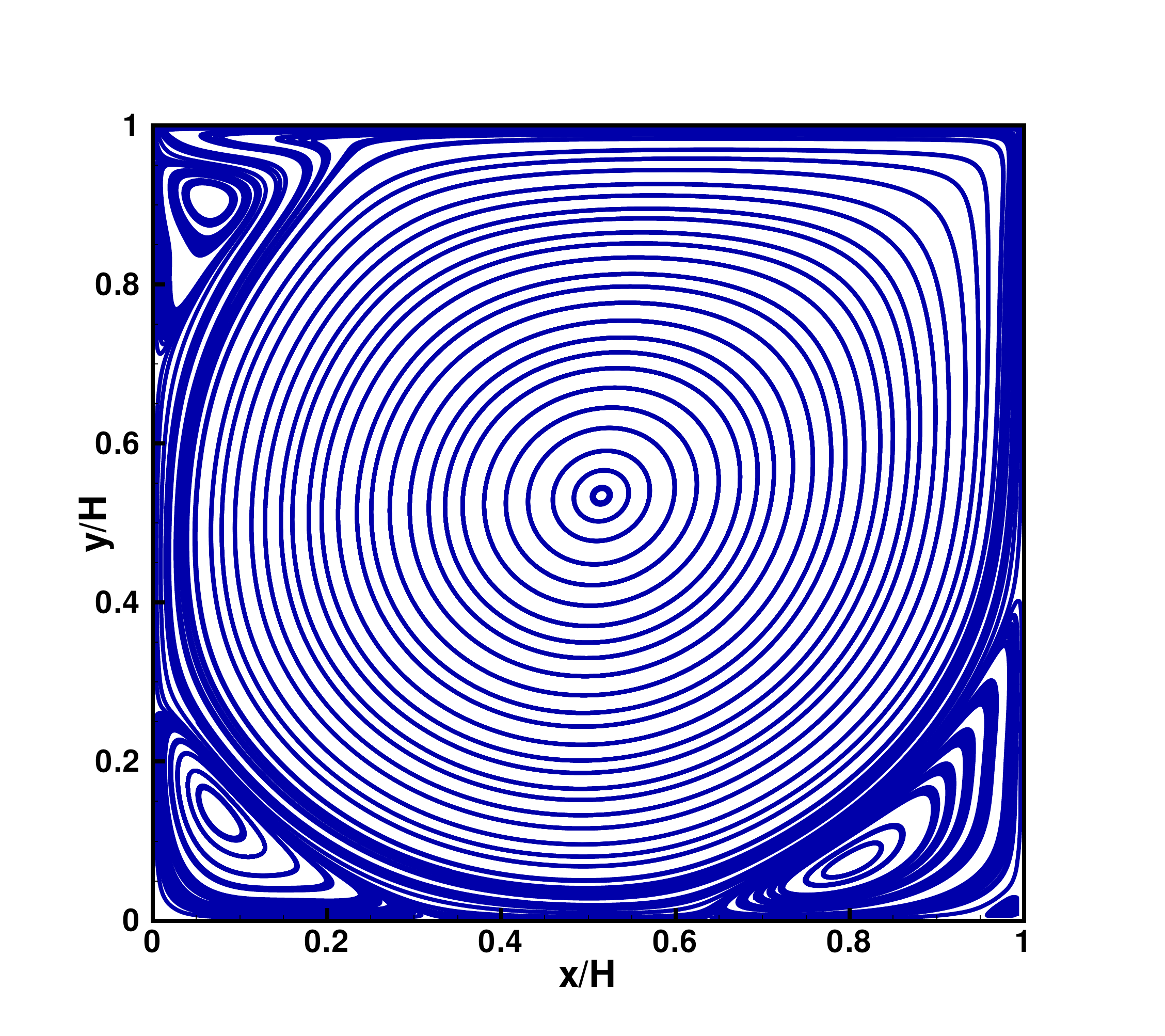}
        \label{fig:12e} } 
    \subfloat[Re=7500] {
        \includegraphics[width=.40\textwidth] {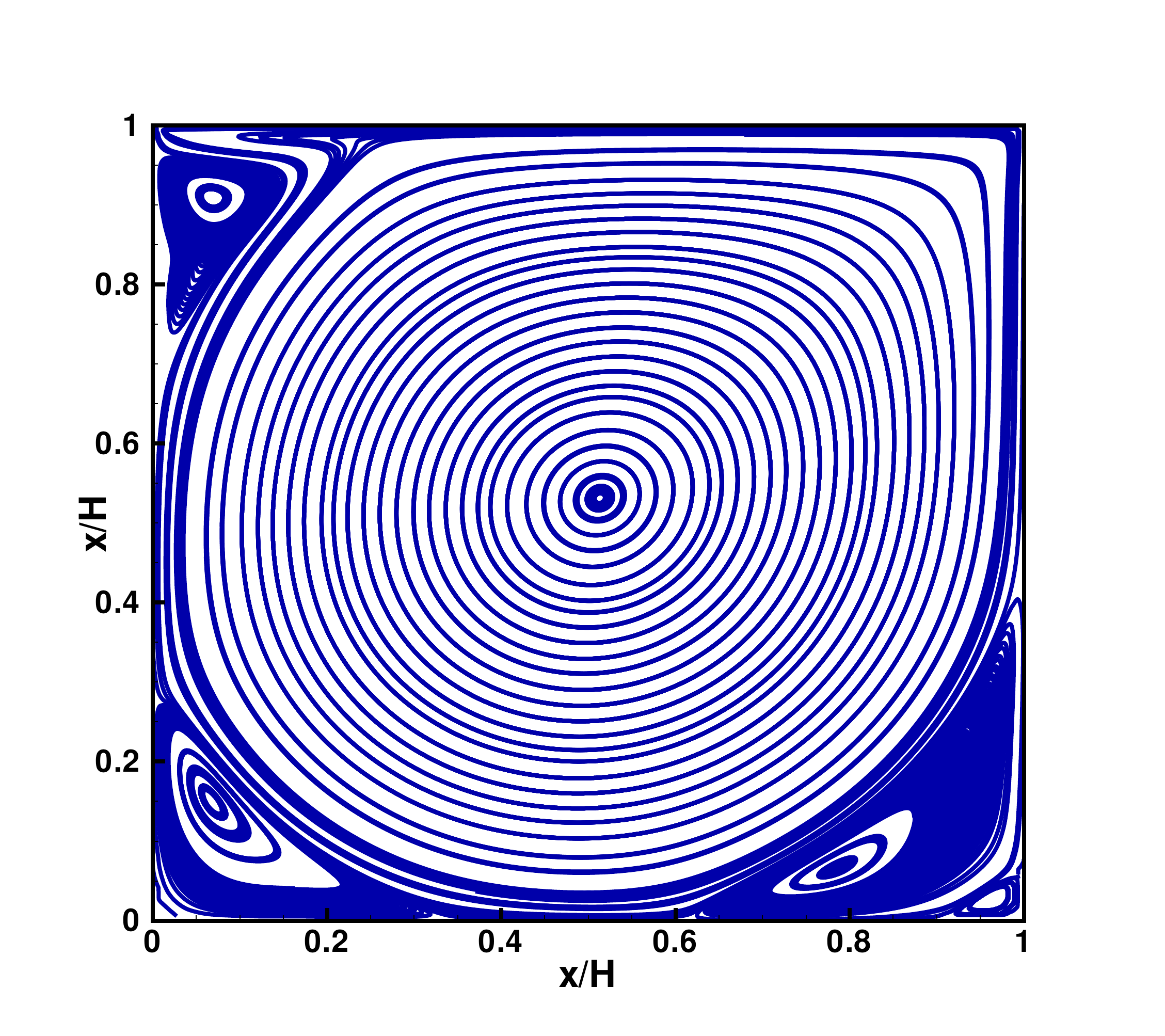}
        \label{fig:12f} } \\
    \caption{Streamline contours of the flow field in a 2D lid driven cavity computed by RC-LBM on a rectangular lattice with the grid aspect ratio of $a=0.5$ at different Reynolds numbers (a) $\mbox{Re}=100$, (b) $\mbox{Re}=400$, (c) $\mbox{Re}=1000$, (d) $\mbox{Re}=3200$, (e) $\mbox{Re}=5000$ and (f) $\mbox{Re}=7500$.}
    \label{fig:12}
\end{figure}

\begin{figure}[ht]
\centering
\captionsetup{justification=centering}
\includegraphics[scale=0.7]{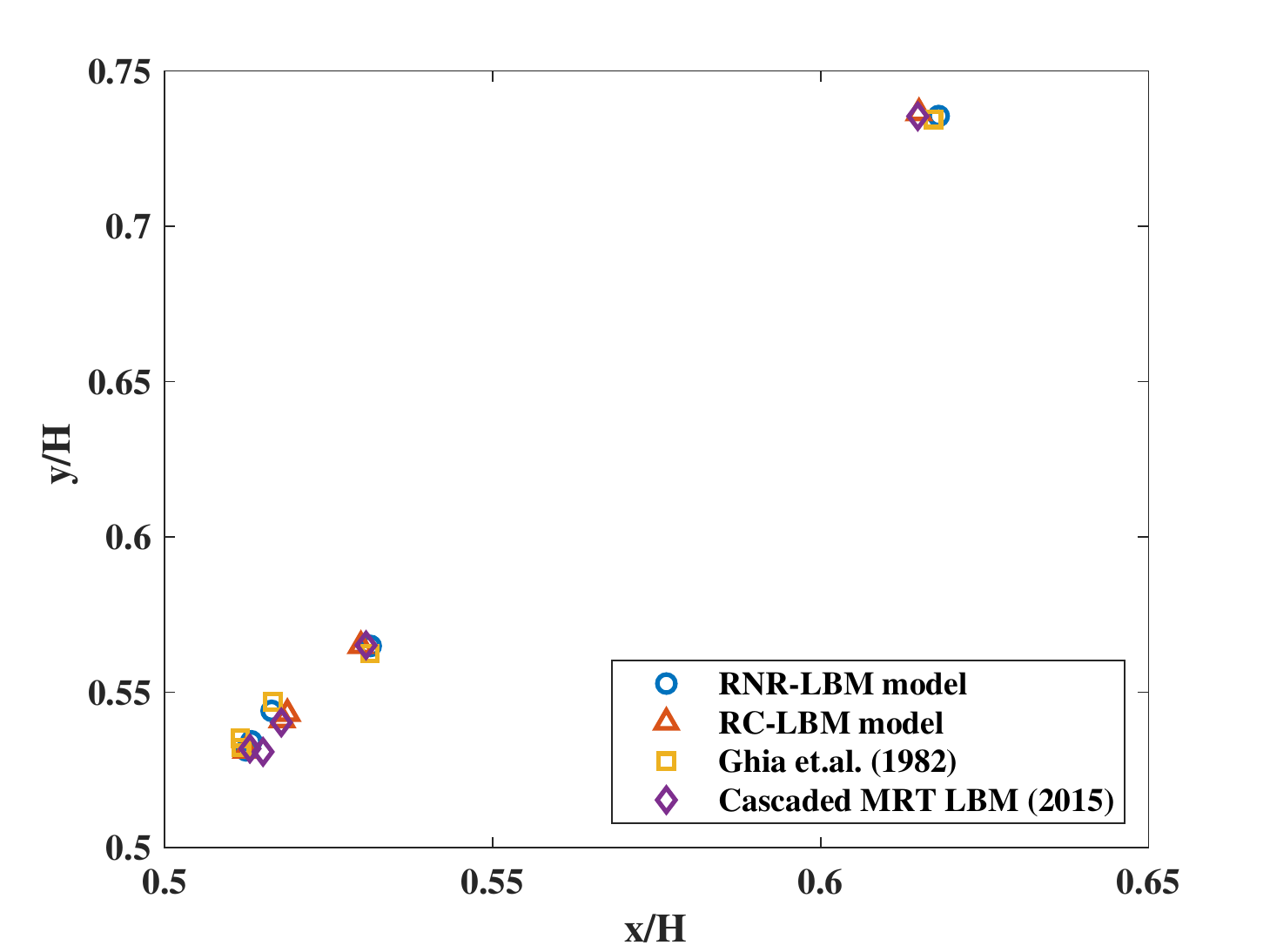}
\caption{Comparison of the computed results of the location of the primary vortex at Reynolds numbers $\mbox{Re} = 100, 1000, 3200, 5000$ and $7500$ using RNR-LBM and RC-LBM with the grid aspect ratio of $a = 0.5$ with the benchmark numerical results of Ghia \emph{et al}~\cite{ghia1982high}.}
\label{fig:13}
\end{figure}

Next, Fig.~\ref{fig:12} shows streamline patterns at Reynolds numbers of 100, 400, 1000, 3200, 5000 and 7500 computed using RC-LBM with a grid aspect ratio of $a=0.5$. The center of the primary vortex is seen to move towards the middle of the cavity as the Reynolds number increases. This can also be more clearly observed from Fig.~\ref{fig:13} which plots the coordinate locations of this vortex at different $\mbox{Re}$, which match well with those given in Ref.~\cite{ghia1982high}. Meanwhile, additional secondary vortices emerge and grow in a counter-clockwise direction at the right and left of the bottom wall. At Reynolds number above 3200 (Figure \ref{fig:12d}), a secondary vortex appears on the upper left corner while the secondary vortices at the bottom corner become relatively larger. Furthermore, a second secondary vortex emerges at the right bottom when $\mbox{Re}$ reaches a values above 5000. All these features are consistent with those presented in the benchmark results~\cite{ghia1982high} and confirm the ability of our rectangular LB formulation to reproduce physically correct complex vortical flow patterns well without any spurious grid anisotropy effects that limited some of the prior rectangular LB schemes.
\newpage
In addition, in order to provide some quantitative comparisons, the locations of the primary vortices as well as those for various secondary vortices are presented in the form of tabulated data in Tables~\ref{tab:44} and \ref{tab:4}, respectively, at various Reynolds numbers. The numerical results obtained using both the RNR-LBM and RC-LBM at a grid aspect ratio $a=0.5$ are compared with those given in the benchmark paper~\cite{ghia1982high} as well as those based on the recent cascaded LBM~\cite{ning2016numerical}. Very good quantitative agreement seen between our rectangular LB formulations and these earlier investigations provide further evidence to their validity in computing physically consistent and accurate solutions of this flow problem for a wide range of Reynolds numbers.
\begin{table}[h!]
\small
\centering
\captionsetup{justification=centering}
\begin{tabular}{l c c c c c }
\hline
 \multicolumn{6}{c}{Primary Vortex} \\
 \hline
 \hline
 Method & Re=100   & Re=1000 &  Re=3200&  Re=5000 & Re=7500\\
 \hline
RNR-LBM  & (0.6182,0.7355)  & (0.5314,0.5649)  & (0.5132,0.5342)& (0.5132,0.5342) & (0.5124, 0.5313) \\
error(\%)&  (< 0.0016)      &   (< 0.004)     &   (< 0.02)     &   (< 0.0033)     &   (< 0.0016) \\
RC-LBM   & (0.6151,0.7365)  & (0.5299,0.5649)& (0.5179,0.5411)& (0.5151, 0.5337) & (0.5122,0.5321)\\
error(\%)&  (< 0.0034)      &   (< 0.004)     &   (< 0.01)     &   (< 0.007)     &   (< 0.0009) \\
Cascaded LBM~\cite{ning2016numerical} & (0.6148,0.7354)  &  (0.5307,0.5651)  & (0.5177,0.5402)& (0.5149,0.5352) & (0.5129, 0.5318) \\
error(\%)&  (< 0.0038)      &   (< 0.0046)     &   (< 0.012)     &   (< 0.006)     &   (< 0.002) \\
Ghia \emph{et al}~\cite{ghia1982high}    & (0.6172,0.7344)  & (0.5313,0.5625)&(0.5165,0.5469) &(0.5115,0.5352) & (0.5117,0.5322)\\
\hline
\end{tabular}
\caption{Location of primary vortices in a 2D lid-driven cavity flow at different Reynolds numbers obtained using RNR-LBM and RC-LBM with rectangular lattice grid aspect ratio $a=0.5$ and compared with the results of Ghia \emph{et al}~\cite{ghia1982high} based on a NS solver and cascaded LBM~\cite{ning2016numerical}.}
\label{tab:44}
\end{table}

\begin{table}[h!]
\small
\centering
\captionsetup{justification=centering}
\begin{tabular}{l c c c c c }
\hline
 \multicolumn{6}{c}{Re=100} \\
 \hline
 Method & \multicolumn{3}{l}{First Secondary Vortex}&  \multicolumn{2}{l}{Second Secondary Vortex} \\
   & Top vortex & Bottom Left & Bottom Right & Bottom Left & Bottom Right\\
 \hline
RNR-LBM   &NA & (0.0383,0.0382)  & (0.938,0.0652)& NA & NA\\
RC-LBM    &NA& (0.0373,0.0392)& (0.9381,0.0657)& NA & NA\\
Cascaded LBM~\cite{ning2016numerical} & NA& (0.0387,0.0387)  & (0.9383,0.0658)& NA & NA\\
Ghia \emph{et al}~\cite{ghia1982high}    & NA& (0.0313,0.0391)&(0.9453,0.0625) &NA & NA\\
\hline
\multicolumn{6}{c}{Re=1000} \\
 \hline
RNR-LBM   & NA& (0.0860,0.0778) & (0.8581,0.1147) &NA & (0.9905,0.0065)  \\
RC-LBM    & NA& (0.0842,0.0764)  & (0.8612,0.1121)&  NA & (0.9912,0.0075)\\
Cascaded LBM~\cite{ning2016numerical} & NA& 0.0842,0.0791) & (0.8631,0.1128)&NA & (0.9923,0.0076) \\
Ghia \emph{et al}~\cite{ghia1982high}   & NA & (0.0859,0.0781) & (0.8594,0.1094)&NA & (0.9922,0.0078)\\
\hline
\multicolumn{6}{c}{Re=3200} \\
 \hline
RNR-LBM   & (0.0536,0.8966) & (0.0836,0.1191)  &  (0.8206,0.0859)& (0.0069,0.0089) & (0.9838,0.0094)  \\
RC-LBM    & (0.0551,0.8962) & (0.0839,0.1192)  &  (0.8202,0.0857)& (0.0077,0.0068) & (0.9871,0.0103) \\
Cascaded LBM~\cite{ning2016numerical} & (0.0547,0.8976)&  (0.0821,0.1207)  &  (0.8229,0.0853)& (0.0075,0.0075) & (0.9875,0.0113) \\
Ghia \emph{et al}~\cite{ghia1982high} &(0.0547,0.8984)  & (0.0859,0.1094)  &  (0.8125,0.0859)& (0.0078,0.0078) & (0.9844,0.0078)     \\
\hline
\multicolumn{6}{c}{Re=5000} \\
 \hline
RNR-LBM   & (0.0649,0.9062)  &  (0.0795,0.1341)& (0.8029,0.0728) & (0.0123,0.0073)&  (0.9883,0.0127) \\
RC-LBM    & (0.0641,0.9073) & (0.0765,0.1349)  &  (0.8059,0.0749)& (0.0097,0.0061) &  (0.9796,0.0165)  \\
Cascaded LBM~\cite{ning2016numerical} & (0.0644,0.9081)&  (0.0740,0.1378)  &  (0.8037,0.0739)& (0.0075,0.0075) & (0.9775,0.0200) \\
Ghia \emph{et al}~\cite{ghia1982high}   & (0.0625,0.9102)& (0.0703,0.1367)  &  (0.8086 ,0.0742)& (0.0117,0.0078) & (0.9805,0.0195)     \\
\hline
\multicolumn{6}{c}{Re=7500} \\
 \hline
RNR-LBM  & (0.0653, 0.9105)  & (0.0635,0.1532)  &(  0.7804,0.0612)& (0.0123,0.0123) & (0.9608,0.0260)\\
RC-LBM   & (0.0672,0.9108)  & (0.0672,0.1501)  &  (0.7846,0.0636)& (0.0130,0.0117) & (0.9542,0.0370) \\
Cascaded LBM~\cite{ning2016numerical} & (0.0676,0.9102) &  (0.0654,0.1536)& (0.7892,0.0663)   & (0.0125,0.0125)&(0.9508,0.0429)  \\
Ghia \emph{et al}~\cite{ghia1982high}   & (0.0664, 0.9141)  &  (0.0645,0.1504) &(0.7813,0.0625) & (0.0117,0.0117)& (0.9492,0.0430) \\
\hline
\end{tabular}
\caption{Location of secondary vortices in a 2D lid-driven cavity flow at different Reynolds numbers obtained using RNR-LBM and RC-LBM with rectangular lattice grid aspect ratio $a=0.5$ and compared with the results of Ghia \emph{et al}~\cite{ghia1982high} based on a NS solver and cascaded LBM~\cite{ning2016numerical}.}
\label{tab:4}
\end{table}
\subsection{Illustration of computational advantages of using rectangular lattice over square lattice: Shear flow in a shallow rectangular cavity\label{sec:rectVssquare}}
While the previous examples validated the accuracy of our rectangular LB formulations against benchmark solutions, we will now present a case study that demonstrates the computational advantages of employing the rectangular lattice in lieu of the square lattice. In particular, when the spatial gradients in the flow field in one of the coordinate directions are significantly larger than those in the other direction, such as in inhomogeneous shear flows, the rectangular LB schemes are expected to be more efficient. In order to emphasize this numerically, we will now consider the shear flow in a shallow rectangular cavity of width $L$ and height $H$, where $L$ is significantly larger than $H$, driven by the motion of the top lid at a velocity $U$ along the $x$ direction. Specifically, we choose $H/L=0.25$ with the Reynolds number, defined by $\mbox{Re}=UL/\nu$, to be 100. In this flow configuration, the gradients are dominant in the direction normal to the shearing lid at the top, i.e., the $y$ direction. The use of the uniform square lattice would require considerably larger computational resources as it does not exploit the inhomogeneous features inherent in such flows. If $N_x$ and $N_y$ are the number of grid nodes along $x$ and $y$ directions, respectively, the grid spacings in the respective directions for this problem are $\Delta x = L/N_x$ and $\Delta y = H/N_y$. In the case of the square lattice, since $\Delta x = \Delta y$, we require $N_x/N_y = L/H$. If the grid resolution normal to the top lid is resolved with $100$ nodes, i.e., $N_y =100$, for $H/L=0.25$ this implies that the number of grid nodes in the other direction to be $N_x=400$.

\begin{figure}[ht]
\centering
\advance\leftskip-1.4cm
    \subfloat[] {
        \includegraphics[scale=0.6] {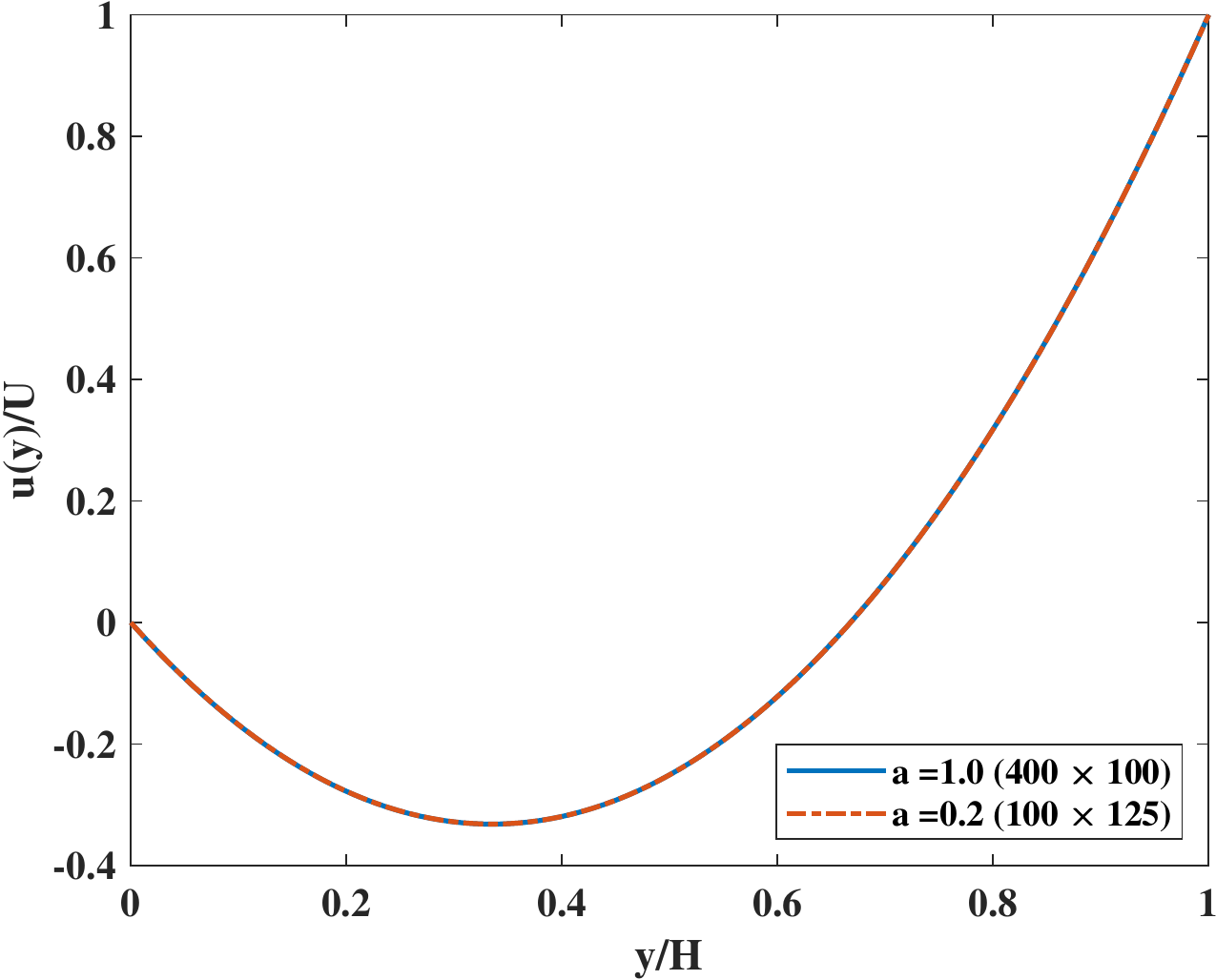}
        \label{fig:14a} } 
    \subfloat[] {
        \includegraphics[scale=0.6] {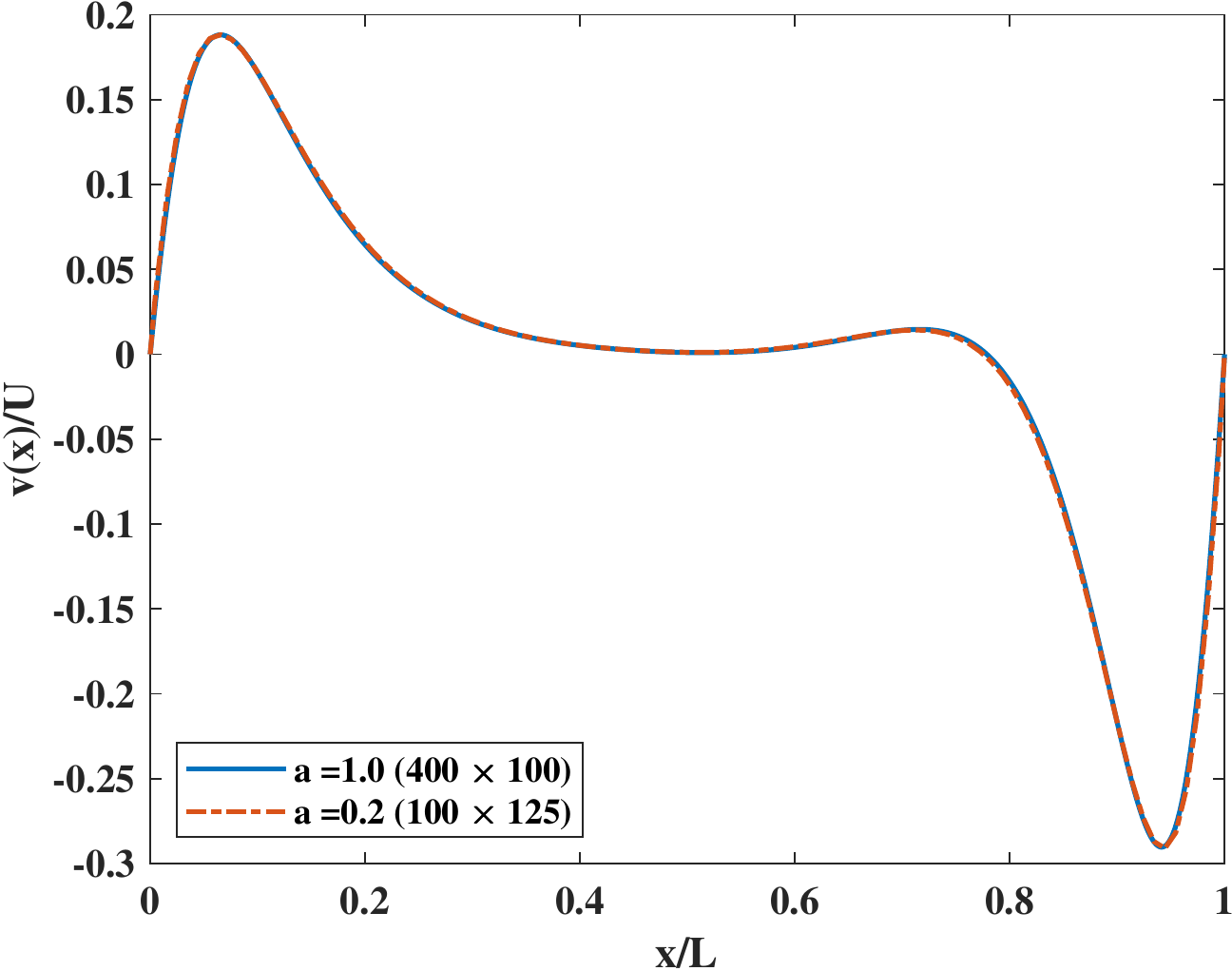}
        \label{fig:14b} } \\
                \advance\leftskip0cm
    \caption{Comparison of the velocity profiles along the centerlines of a shallow rectangular lid driven cavity of aspect ratio $H/L=0.25$ at a Reynolds number $\mbox{Re}=100$ computed using RC-LBM with square lattice ($a=1.0$) and rectangular lattice of grid aspect ratio of $a=0.2$.  (a) $u$ component along the vertical centerline, and (b) $v$ component along the horizontal centerline}
    \label{fig:14}
\end{figure}

\begin{figure}[h!]
\centering
\advance\leftskip-1.7cm
    \subfloat[] {
        \includegraphics[scale=0.35] {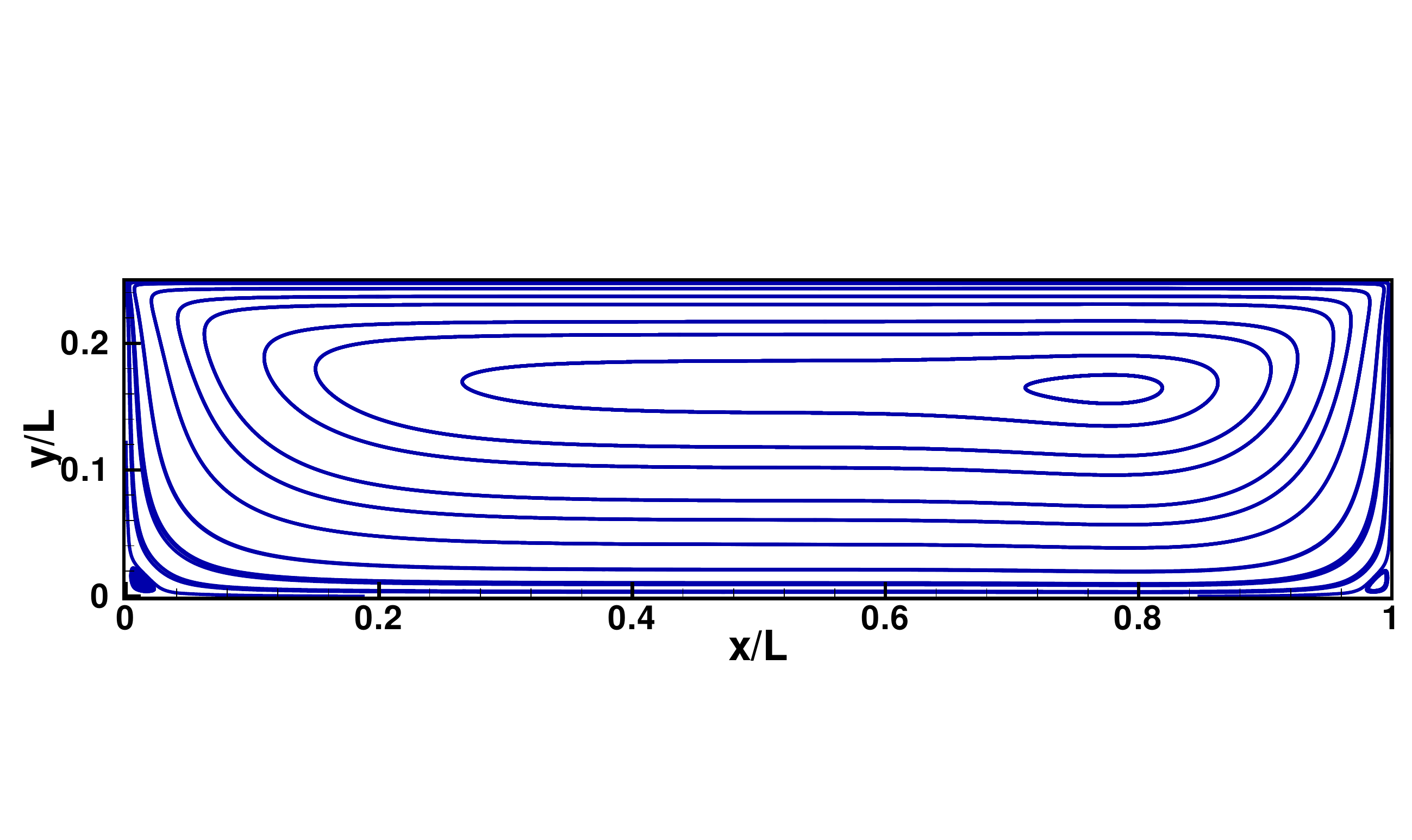}
        \label{fig:15a} } 
    \subfloat[] {
        \includegraphics[scale=0.35] {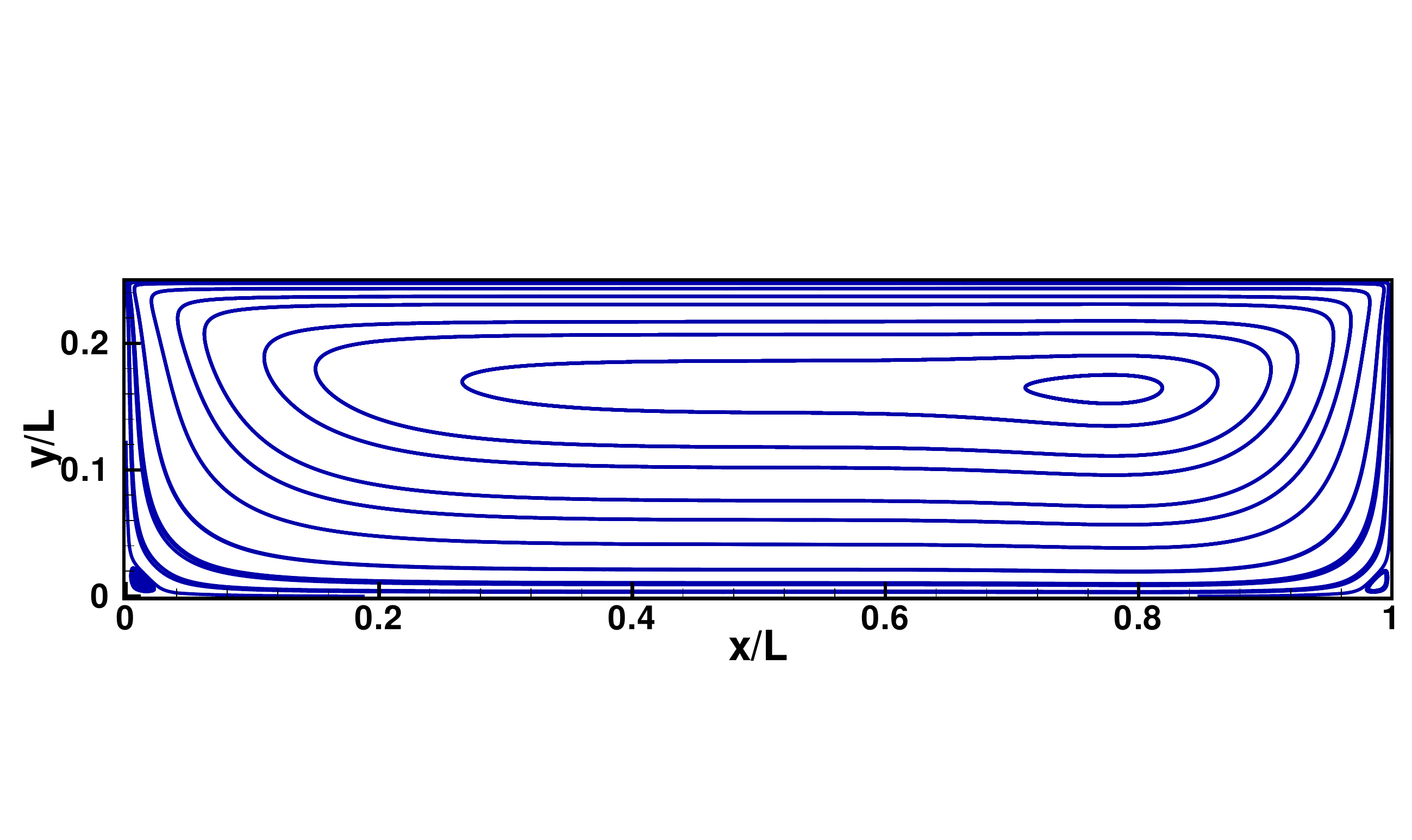}
        \label{fig:15b} } \\
                \advance\leftskip0cm
    \caption{Comparison of the streamline patterns in a shallow rectangular cavity of aspect ratio $H/L=0.25$ at a Reynolds number $\mbox{Re}=100$ computed using RC-LBM with (a) square lattice ($a=1.0$) using a grid resolution of 400$\times$ 100 and (b) rectangular lattice ($a=0.2$) using a grid resolution 100$\times$ 125.}
    \label{fig:15}
\end{figure}

On the other hand, in the case of the rectangular lattice, based on the characteristic of this flow, we could choose $\Delta y \ll \Delta x$. Since, by definition, the grid aspect ratio is $a=\Delta y/\Delta x$, from the above it follows that $N_x/N_y= a(L/H)$. Thus, even if we choose $N_y=125$ (i.e., larger than that considered for the square lattice to resolve the flow better in the dominant gradient direction) and by taking $a=0.2$, the number of grid nodes in the other direction $N_x$ is only $100$ in the case of the rectangular lattice. We will now compare the flow fields computed using the RC-LBM with the square lattice ($a=1$) considering $400 \times 100$ grids nodes and the rectangular lattice ($a=0.2$) considering a grid resolution of $100 \times 125$. The results for the velocity profiles along the vertical and horizontal centerlines are presented in Fig.~\ref{fig:14}, while the streamline contours within the shallow rectangular cavity  at $\mbox{Re}=100$ are shown in Fig.~\ref{fig:15}. It is evident that the results of the rectangular LB scheme, which uses considerably fewer grid nodes adapted to reflect the spatial variations in the flow, are in excellent agreement with the those for the obtained for the square lattice. The use of fewer grid nodes in the case of the rectangular LB scheme results in considerable savings in memory as well as reduction in the simulation turnaround time, by a factor of about 3 in this case. Thus, this demonstrates that the rectangular LBM provides a flexible and computationally efficient approach for resolving inhomogeneous shear flows.

\section{Numerical stability tests: Comparisons between RNR-LBM and RC-LBM}  \label{sec:6}
From the previous section, it was clear the other existing rectangular LB schemes~\cite{bouzidi2001lattice,zhou2012mrt,zong2016designing,peng2016hydrodynamically,peng2016lattice} are limited to relatively low or moderate Reynolds number simulations due to numerical stability issues. This is generally due to their choice of orthogonal moment basis and construction of equilibria and the correction terms involving cumbersome specifications of several model parameters that limited the possible ranges of variation of their transport coefficients. These aspects have been avoided in the present work that uses a non-orthogonal moment basis, a matching principle to construct the equilibria directly from the Maxwell distribution function, and simpler expressions for tuning the transport coefficients and specifying the correction terms to restore isotropy. As a result, the RNR-LBM and RC-LBM developed here represent as significant improvements over the prior rectangular LB formulations. Now, between these two options, the simulations carried out earlier (see Table~\ref{tab:2}) showed that the latter can deliver smaller global relative errors compared to the former in a body force driven flow. Besides such accuracy improvements, we will now clarify the utility of performing the collision step in the local moving frame of reference in the case of RC-LBM in improving the robustness of computations. In this regard, we will now perform two different types of numerical stability tests involving the shear flow generated within a square cavity due to the motion of the lid that compares the RNR-LBM and RC-LBM. Such systematic numerical stability investigations of rectangular LB formulations are lacking in the literature.

\begin{figure}[h!]
    \centering
    \captionsetup{justification=centering}
   \includegraphics[scale=0.7]{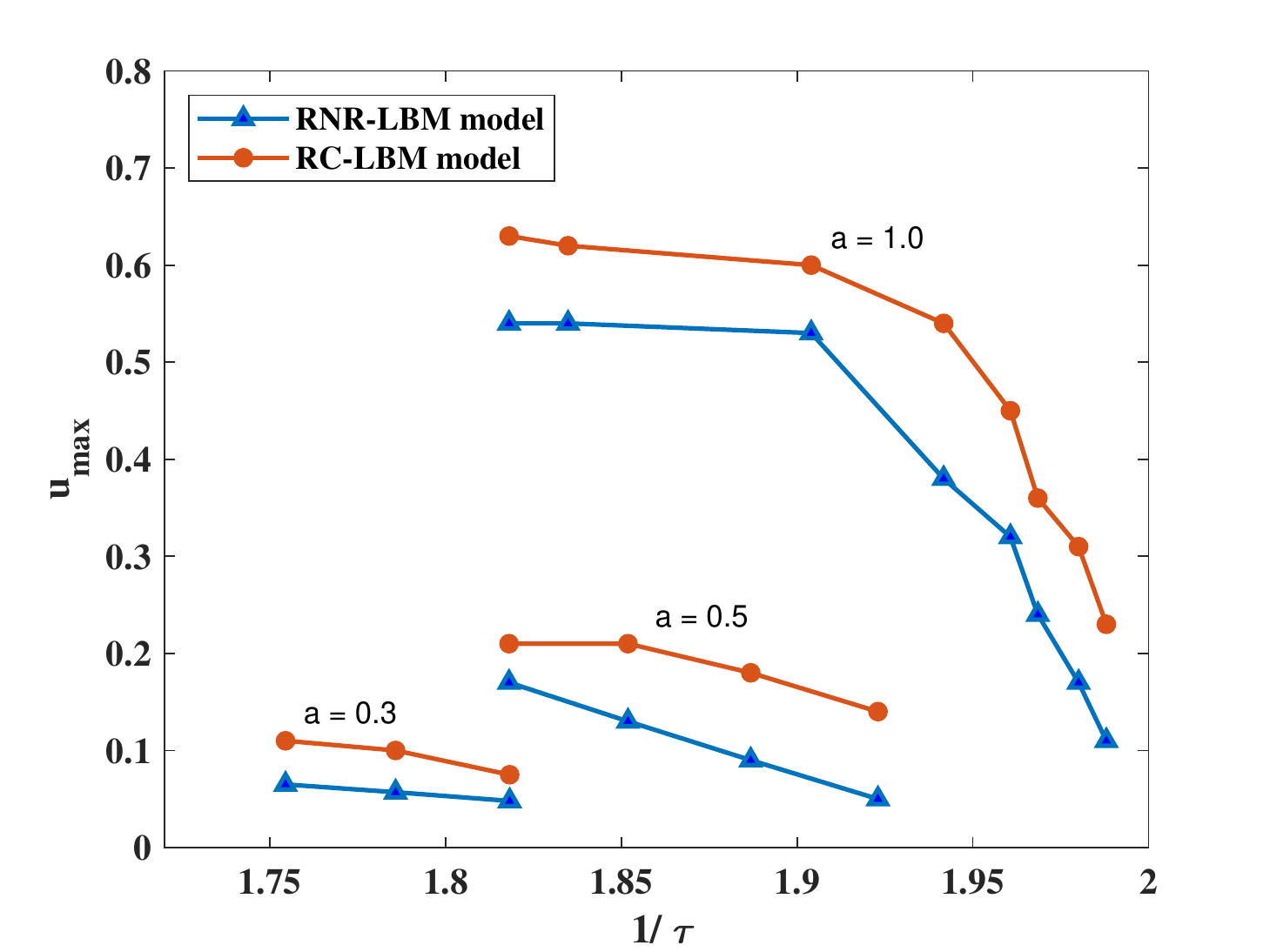}
    \caption{Numerical stability test results showing the maximum threshold velocity of the lid $U$ in a 2D lid driven cavity flow at different values of the relaxation parameter controlling the shear viscosity. Comparisons are made between the RNR-LBM and RC-LBM using rectangular lattice with grid aspect ratios of $a= 1.0, 0.5$, and $0.3$.}
    \label{fig:17}
\end{figure}

\begin{figure}[H]
    \centering
    \captionsetup{justification=centering}
  \includegraphics[scale=0.7]{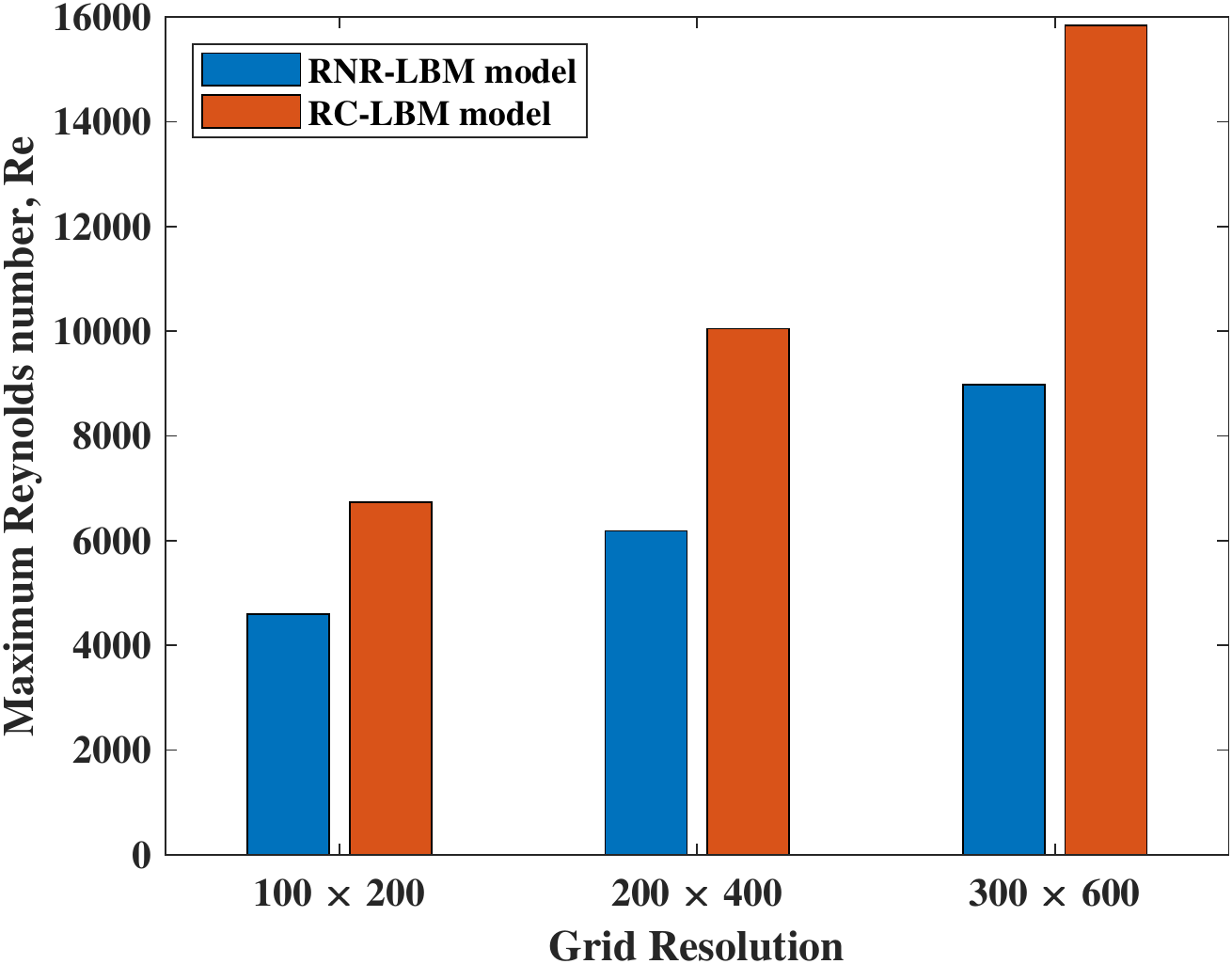}
    \caption{Comparison of the maximum Reynolds number for numerical stability of RNR-LBM and RC-LBM at different mesh resolution with a grid aspect ratio of $a=0.5$.}
    \label{fig:16}
\end{figure}

In the first cast study, we determine the maximum threshold velocity of the top plate $U$ in a lid-driven cavity flow at various relaxation times $\tau$ for RNR-LBM and RC-LBM using the rectangular lattice with grid aspect ratios of $a= 1.0, 0.5$, and $0.3$. These aspect ratios correspond to choosing fixed
coarse grid resolutions of $21\times 21$, $21\times 41$ and $21\times 61$, respectively. Following a strategy similar to Refs.~\cite{luo2011numerics,ning2016numerical}, we evaluate the maximum lid velocity which maintains stable simulations for 100,000 time steps for the rectangular LB formulations. Figure~\ref{fig:17} shows the stability regime results for RNR-LBM and RC-LBM for different choices of $a$. It can be seen that, in general, as the grid aspect ratio decreases, characterized by greater geometric anisotropy of the lattice, the numerically stable regime becomes narrower. However, in all cases, the RC-LBM based on central moments is found to be significantly more stable compared to the RNR-LBM based on raw moments, with the former generally taking a relatively small additional computational overhead of about $25\%$ when compared to the former.
As a second type of numerical stability test, we perform simulations to investigate the maximum Reynolds number sustained by each of the two rectangular LB schemes at a fixed grid aspect ratio of $a=0.5$, while maintaining a constant lid velocity at $U=0.2$ and $c_s^2=0.1$ and reducing the shear viscosity of the fluid to a smallest possible value for which the computations remains numerically stable. In this regard, the tests are carried out for
grid resolutions of $100\times200$, $200\times400$, $300\times600$ and, under the above conditions in each case, the relaxation time $\tau$ is decreased gradually until the computations become unstable. The results are tabulated in Table~\ref{tab:6} and illustrated in Fig.~\ref{fig:16}. It can be seen that the RC-LBM is found to be significantly more stable when compared to the latter. Further improvements may be possible by adjusting the relaxation times for the higher order moments and the speed of sound.
\begin{table}[H]
\small
\centering
\captionsetup{justification=centering}
\caption{The maximum Reynolds number for numerical stability of RNR-LBM and RC-LBM at different mesh resolution with a grid aspect ratio of $a=0.5$.}
\begin{tabular}{c c c}
\hline
\hline
 Grid resolution &   RNR-LBM  &  RC-LBM \\
 \hline\hline
 $100\times200$ & 4591 &6733 \\
 $200\times400$ & 6185 & 10050 \\
 $300 \times600$& 8985 & 15842 \\
\hline
\end{tabular}
\label{tab:6}
\end{table}

Here, it should be mentioned that the rectangular central moment LB formulation, while better than its raw moment counterpart, has a reduced stability range when compared to the square lattice based central moment LBM (see e.g.,~\cite{ning2016numerical}). This is due to the fact that while the correction terms eliminate the truncation errors arising from the use of the rectangular lattice in the second order moments and recover the desired viscous stress tensor and hence the Navier-Stokes equations, the effects of grid anisotropy and the associated non-GI terms remain in the higher order (kinetic) moments that influence such numerical behavior. However, this stability reduction can be compensated for by the flexibility available in the rectangular LBM in choosing the grid sizes that reflect the nature of the flow being resolved, such as in inhomogeneous shear flows, which then would result in significant improvements in computational efficiency as demonstrated in Sec.~\ref{sec:rectVssquare}. The key aspect in such cases is the careful selection of the grid aspect ratio in the range that maintains numerical stability while delivering reductions in the overall computational cost, which depends on the type of flow problem to be simulated. We believe that further improvements are possible by the development of a multiblock rectangular central moment LBM, where instead of resolving the entire domain using a single grid aspect ratio with a relatively low value, appropriate grid aspect ratios can be utilized in few selected zones of the flow domain in such a way that the multiblock interfaces are constrained to match the hydrodynamics from the respective zones. The construction of such general forms of rectangular lattice based LBM will be addressed in a future work.

\section{Summary and Conclusions} \label{sec:7}
In this paper, we presented two new rectangular LB schemes based on raw moments and central moments, designated as RNR-LBM and RC-LBM, respectively, where the collision step involves relaxation of the corresponding moments, each with its own individual rates. Unlike the other existing schemes, they are constructed using a non-orthogonal moment basis and the moment equilibria are directly obtained via matching with those of the continuous Maxwell distribution function, thereby involving higher order velocity terms and without the use of many free parameters. By using a Chapman-Enskog analysis, correction terms are derived to eliminate the grid anisotropy introduced on the viscous stress tensor arising from the use of the rectangular lattice and the cubic non-Galilean invariant terms due to aliasing effects on the standard D2Q9 lattice. Such correction terms, which are incorporated via extending the equilibria for the second order moments, along with the expressions for the transport coefficients have simpler functional relationships involving the grid aspect ratio, the speed of sound and the diagonal components of the velocity gradient tensor. Formulas are derived to compute the latter locally based on non-equilibrium moments. Furthermore, the attendant transformation matrices that map between the distribution functions and the moments and parameterized by the grid aspect ratio are also considerably simplified owing to the use of non-orthogonal moment basis. All these considerations result in more robust and efficient implementations of the proposed RNR-LBM and RC-LBM when compared to the other existing rectangular LB formulations. These two schemes are validated against a variety of benchmark flow problems yielding accurate solutions for a wide range of grid aspect ratio. Furthermore, simulations demonstrate improvements in accuracy and significantly greater numerical stability regime for shear driven flows with the use of the RC-LBM when compared to RNR-LBM. Moreover, the effectiveness of using of our rectangular LB scheme in lieu of that based on the square lattice in reducing the computational cost is shown. The present rectangular central moment LB formulation can be extended to a 3D cuboid LB approach, which will be reported in the near future. The approach presented here involving the RNR-LBM and RC-LBM is for athermal flows. It can also be extended to include temperature variations either by using extended lattice sets or using dual distribution functions-based formulations in a rectangular lattice for efficient simulations of flow with heat transfer. Furthermore, while the present approach allows local variations in the viscosity via the relaxation parameters, a pressure-based rectangular central moment LB formulation involving variations in various fluid properties can be constructed using a modified kinetic equation and equilibria via a transformation similar to that presented for the square lattice in Ref.~\cite{hajabdollahi2021central}. Moreover, the development of multiblock rectangular/cuboid lattice based central moment LB schemes represents another interesting area. These are important topics for practical applications and are subjects for future investigations.

\section*{Acknowledgements}
Parts of this work were presented at the 70th Annual Meeting of the APS Division of Fluid Dynamics (DFD), Denver, Colorado, Nov. 2017~\cite{yahiaAPSDFD2017} and the 71st Annual Meeting of the APS DFD, Atlanta, Georgia, Nov. 2018~\cite{yahiaAPSDFD2018}. The first author thanks the Graduate School of the University of Colorado Denver for the travel grants to make these presentations. The second author would like to acknowledge the support of the US National Science Foundation (NSF) under Grant CBET-1705630.

\appendix
\section{Inverse of the transformation matrix for mapping raw moments to distribution functions}\label{sec:appendix1}
The transformation from the raw moments to the distribution functions $\tensor{T}^{-1}$ is obtained by inverting Eq.~(\ref{eq:4}) for the rectangular lattice and can be explicitly written as
\begin{equation}\label{eq:Tinverse}
\tensor{T^{-1}}=
  \begin{bmatrix}
  1 &  0 &  0 &  r_1 &  r_2 &  0 &  0  &  0  &  \frac{-1}{a^2}\\[4pt]
  0  &  \frac{1}{2} &  0 &  \frac{1}{4} &  \frac{1}{4} &  0  &  0  & -\frac{1}{2a^2} &  -\frac{1}{2a^2}\\[4pt]
  0  &  0  & \frac{1}{2a}  & \frac{1}{4a^2}  &  -\frac{1}{4a^2}   & 0 & -\frac{1}{2a} & 0 &   -\frac{1}{2a^2} \\[4pt]
  0  &  -\frac{1}{2} &  0 &  \frac{1}{4} &  \frac{1}{4} &  0  &  0  & \frac{1}{2a^2} &  -\frac{1}{2a^2}\\[4pt]
  0  &  0  & -\frac{1}{2a}  & \frac{1}{4a^2}  &  -\frac{1}{4a^2}   & 0 & \frac{1}{2a} & 0 &   -\frac{1}{2a^2} \\[4pt]
  0  &  0  & 0  & 0  &  0   & \frac{1}{4a}  & \frac{1}{4a} & \frac{1}{4a^2} &  \frac{1}{4a^2} \\[4pt]
  0  &  0  & 0  & 0  &  0   & -\frac{1}{4a}  & \frac{1}{4a} & -\frac{1}{4a^2} &  \frac{1}{4a^2} \\[4pt]
  0  &  0  & 0  & 0  &  0   & \frac{1}{4a}  & -\frac{1}{4a} & -\frac{1}{4a^2} &  \frac{1}{4a^2} \\[4pt]
  0  &  0  & 0  & 0  &  0   & -\frac{1}{4a}  & -\frac{1}{4a} & \frac{1}{4a^2} &  \frac{1}{4a^2}
 \end{bmatrix},
\end{equation}
where $r_1=-\frac{1}{2} \left(1+\frac{1}{a^2}\right)$ and $r_2=-\frac{1}{2} \left(1-\frac{1}{a^2}\right)$, which is parameterized by the grid aspect ratio $a$. Note that the use of a non-orthogonal moment basis leads to a simpler mapping matrix with several zero elements allowing a more efficient implementation in their component form.

\section{Frame transformation matrix and its inverse for mapping between central moments and raw moments}\label{Sec:appendix2}
The elements of the frame transformation matrix $\tensor{\mathcal{F}}$ that maps from raw moments to central moments follow from enumerating the components of the binomial transforms written at different orders, which read
\begin{equation} \label{eq:53}
  \tensor{\mathcal{F}} =
  \begin{bmatrix}
   1 &  0 &  0 &  0 &  0 &  0 &  0  &  0  &  0\\[4pt]
  -u_x  &  1 &  0 &  0 &  0 &  0 &  0  &  0  &  0\\[4pt]
  -u_y  &  0  & 1  & 0  &  0   & 0 & 0 & 0 &   0 \\[4pt]
  u_x ^2 +u_y ^2  &  -2 u_x  & -2 u_y  & 1   &  0   & 0 & 0 & 0 &   0 \\[4pt]
  u_x ^2 -u_y ^2   &  -2 u_x  & 2 u_y & 0   & 1   & 0 & 0 & 0 &   0
  \\[4pt]
  u_x u_y  &  -u_y  & -u_x  & 0  &  0   & 1 & 0 & 0 &  0 \\[4pt]
  -u_x ^2 u_y    &  2 u_x u_y  & u_x ^2   & - \frac{1}{2} u_y  &  - \frac{1}{2} u_y  & -2 u_x & 1 & 0 &  0 \\[4pt]
  -u_x u_y ^2   & u_y ^2   & 2 u_x u_y  & - \frac{1}{2} u_x  &  \frac{1}{2} u_x  & -2 u_y & 0 & 1 &  0 \\[4pt]
  u_x^2 u_y ^2   & -2 u_x u_y ^2   & -2 u_x^2 u_y  &  \frac{1}{2} (u_x^2+u_y^2)  &  \frac{1}{2} (u_y^2 -u_x^2)  & 4 u_x u_y & -2 u_y & -2 u_x &  1\\.
  \end{bmatrix}
\end{equation}
On the other hand, the elements of the transformation from central moments to raw moments denoted by $\tensor{\mathcal{F}}^{-1}$ can be obtained directly from those of $\tensor{\mathcal{F}}$ without needing to perform an explicit matrix inversion by replacing the signs of $u_x$ and $u_y$ (i.e., $u_x \leftrightarrow -u_x$ and $u_y \leftrightarrow -u_y$) based on an interesting property of binomial transforms that immediately follows from their generating function representation. In other words, if $\tensor{\mathcal{F}}=\tensor{\mathcal{F}}(u_x,u_y)$, then $\tensor{\mathcal{F}}^{-1}=\tensor{\mathcal{F}}(-u_x,-u_y)$. Thus, we have
\begin{equation} \label{eq:55}
  \tensor{\mathcal{F}^{-1}} =
  \begin{bmatrix}
   1 &  0 &  0 &  0 &  0 &  0 &  0  &  0  &  0\\[4pt]
  u_x  &  1 &  0 &  0 &  0 &  0 &  0  &  0  &  0\\[4pt]
  u_y  &  0  & 1  & 0  &  0   & 0 & 0 & 0 &   0 \\[4pt]
  u_x ^2 +u_y ^2  &  2 u_x  & 2 u_y  & 1   &  0   & 0 & 0 & 0 &   0 \\[4pt]
  u_x ^2 -u_y ^2   &  2 u_x  & -2 u_y & 0   & 1   & 0 & 0 & 0 &   0
  \\[4pt]
  u_x u_y  &  u_y  & u_x  & 0  &  0   & 1 & 0 & 0 &  0 \\[4pt]
  u_x ^2 u_y    &  2 u_x u_y  & u_x ^2   &  \frac{1}{2} u_y  &   \frac{1}{2} u_y  & 2 u_x & 1 & 0 &  0 \\[4pt]
  u_x u_y ^2   & u_y ^2   & 2 u_x u_y  & \frac{1}{2} u_x  & -\frac{1}{2} u_x  & 2 u_y & 0 & 1 &  0 \\[4pt]
  u_x^2 u_y ^2   & 2 u_x u_y ^2   & 2 u_x^2 u_y  &  \frac{1}{2} (u_x^2+u_y^2)  &  \frac{1}{2} (u_y^2 -u_x^2)  & 4 u_x u_y & 2 u_y & 2 u_x &  1\\.
  \end{bmatrix}
 \end{equation}


\end{document}